\documentclass[10pt,a4paper]{report}

\usepackage{amsmath,epsfig}
\usepackage{amssymb,amsfonts}
\usepackage{latexsym}
\usepackage{xcolor}
\usepackage{setspace}
\usepackage{feynmf}
\usepackage{booktabs}
\usepackage{tensor}
\usepackage{ulem}
\usepackage{verbatim}

\usepackage{hyperref}
\relax
\renewcommand{\theequation}{\arabic{chapter}.\arabic{equation}}
\def\be{\begin{equation}}
\def\ee{\end{equation}}
\def\bea{\begin{eqnarray}}
\def\eea{\end{eqnarray}}
\def\bs{\begin{subequations}}
\def\es{\end{subequations}}

\newcommand{\HRule}{\rule{\linewidth}{0.5mm}}
\numberwithin{equation}{chapter}
\usepackage{graphicx}
\usepackage{color}
\usepackage[left=3.2cm,right=3.2cm,top=2.8cm,bottom=3cm]{geometry}
\usepackage{titlesec}
\titleformat{\chapter}[display]
  {\bfseries\huge}
  {\filleft\Large\chaptertitlename~\thechapter}
  {3ex}
  {\titlerule\vspace{1.5ex}\filright}
  [\vspace{1ex}\titlerule]
\usepackage{fancyhdr}
\newcommand\fverb{\setbox\pippobox=\hbox\bgroup\verb}
\newcommand\fverbdo{\egroup\medskip\noindent%
                        \fbox{\unhbox\pippobox}\ }
\newcommand\fverbit{\egroup\item[\fbox{\unhbox\pippobox}]}

\newcommand{\bear}{\begin{eqnarray}}

\newcommand{\eear}{\end{eqnarray}}

\newbox\pippobox

\def\6{\partial}

\def\a{\alpha}

\def\pa{\partial}

\def\e{\epsilon}

\def\sp{\;\;\;,\;\;\;}

\def\sq
\def\a{\alpha}
\def\b{\beta}

\def\hri#1#2{\href{http://arxiv.org/abs/#1}{[ArXiv:#1]#2}}

\def\na{\nabla}

\def\e{\epsilon}

\begin{document}

\begin{titlepage}
\begin{center}

\textsc{\Huge University of Crete}\\[0.5cm]
\textsc{\LARGE Department of Physics}\\[2.0cm]
\textsc{\Large Master Thesis}\\[0.2cm]
\textit{\large Submitted in partial fulfillment\\ of the requirements for the degree of\\ Master of Science in Advanced Physics}\\[1.1cm]

\begin{figure}
\begin{center}
\includegraphics[width=0.25\textwidth]{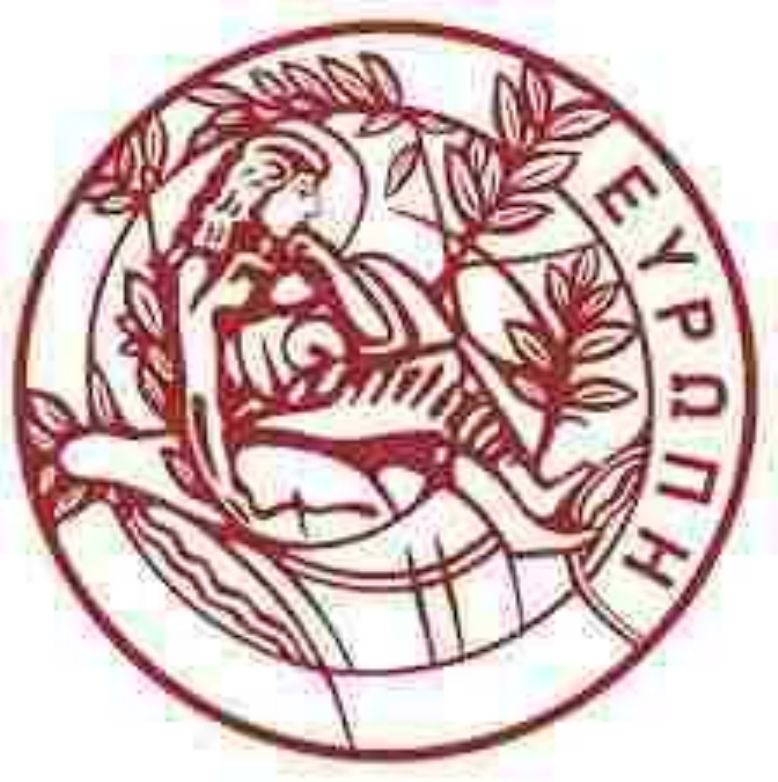}
\end{center}
\end{figure}

\HRule \\[0.5cm]

{ \huge \bfseries Renormalization Group Flows in Non-Relativistic Holographic Effective Field Theories}\\[0.6cm]

\HRule \\[1.5cm]
\begin{minipage}{0.4\textwidth}
\begin{flushleft} \large
\emph{Author:}\\
Georgios M. \textsc{Koutentakis}
\end{flushleft}
\end{minipage}
\begin{minipage}{0.4\textwidth}
\begin{flushright} \large
\emph{Supervisor:} \\
Prof.  ~Elias \textsc{Kiritsis}
\end{flushright}
\end{minipage}

\vfill\Large{Heraklion}
\\
 August 2016
 \\
CCTP-2016-21
\newpage
\pagenumbering{roman}

\end{center}
\end{titlepage}
\newpage
\begin{center}
{\Large{\textbf{Acknowledgments}}}
\begin{quote}
It is a pleasure for me to thank my supervisor, Prof.  Elias Kiritsis, for his support and continuous discussions during my studies and, also, for tolerating my numerous wrongdoings. It is also important to thank Prof. Dr. Peter Lambropoulos for taking the time to support me all of the times I needed his guidance.\\
\indent I would also like to acknowledge Prof. Dr. F. Diakonos, Prof. Dr. A. Donos and Dr. L. Katsimiga for helpful remarks during the course of this work, as well as, N. Angelinos and A. Papanikolaou for proofreading parts of this thesis.
Finally, I would also like to extend my thanks to all of my colleagues and friends for their moral and scientific support.~\\
\end{quote}
\end{center}
\newpage

\newpage
\begin{abstract}

We develop a formalism with two different UV cutoff scales, one for space and one for time, appropriate for the richer structure of non-Lorentz invariant quantum field theories.
In this formalism there are two different $\beta$ functions for each coupling constant, arising from independent variations of the energy or momentum cutoffs.
For holographic non-relativistic theories with rotational invariance, we develop the technique to calculate such $\beta$-functions using a generalization of the superpotential formalism developed in \cite{Kiritsis:2012ma}.
We then proceed and compute the $\beta$ function around a Lifshitz critical point, as well as for general Lifshitz-invariant theories with hyperscaling violation.
Finally, we do a similar computation in a weakly coupled Lifshitz invariant QFT.
\end{abstract}
\newpage

\newpage
\pagenumbering{arabic}

\tableofcontents

\newpage

\chapter{Introduction}

\indent
The purpose of this thesis is to introduce a new formalism
with two different UV cutoff scales for energy and momentum, appropriate for the richer structure of non-Lorentz invariant quantum field theories.
This choice introduces two distinct $\beta$ functions for each coupling constant, arising from independent variations of the energy or momentum cutoffs.
Most of our analysis will be done using holographic ideas as they are very powerful in providing insight on many interacting non-relativistic scale-invariant quantum field theories. In the current chapter we explain the previously mentioned terms and motivate the importance of the proposed construction.

Quantum Field Theory \cite{Ryder,Peskin,Zee} is a theoretical framework for constructing quantum mechanical models of interacting particles (or quasi-particles) and fields. The fundamental difference between QFT and Quantum Mechanics is that the former treats particles as excited states of an underlying field and, consequently, particle interactions are described as interactions between the corresponding fields. A QFT can be made consistent with special relativity and is able to describe systems where the number of particles is not conserved or particles change from one kind to another. A QFT that is invariant under conformal transformations\footnote{A transformation is called conformal at a point P if it preserves oriented angles between curves through P, with respect to their orientation.} is called a conformal field theory (CFT). CFTs are important because they describe the high or low energy limits of QFTs.

 Non-relativistic (also known as non-Lorentz invariant) field theories are especially interesting because they can describe various condensed matter systems and can be used to investigate the possible breaking of Lorentz invariance in high-energy physics. An example for the latter is the newly developed Ho{\v r}ava Lifshitz gravity \cite{HL-orig,HL-add,HL-app}. Non-relativistic theories may exhibit anisotropic scaling in space and time in their corresponding scaling limit\footnote{The high or low energy limits where the field theory becomes scale invariant}, such as:
\be
x \to \lambda x \sp  t \to \lambda^z t\;,
 \ee
 which is referred to as Lifshitz scaling with scaling exponent $z$ . It is expected that Lifshitz invariant field theories with $z=2$ appear in two-dimensional rotationally invariant condensed matter systems equipped with an additional $U(1)$ symmetry \cite{lifshitz}.

Another important aspect observed in the scaling limits of non-Lorentz invariant field theories is the violation of hyperscaling. Within the framework of a $d+1$ dimensional hyperscaling-violating field theory, the thermodynamic quantities scale as if the number of spatial dimensions was $(d-\theta)$, where $\theta$ is the hyperscaling violation exponent.

Although all of the weakly-coupled known Lifshitz field theories are free (an example being the scaling theory stemming from the anisotropic next nearest neighbor Ising --ANNNI-- model \cite{ANNNI}), condensed matter systems such as strange metals \cite{the_strange,exp_strange} are believed to be described by strongly coupled non-relativistic field theories (see \cite{cp_strange}), which motivates the study of strongly coupled Lifshitz models. Therefore, non-relativistic field theories provide interesting perspectives on the understanding of condensed matter systems or even realizing new physics.
 The aforementioned scale invariant theories naturally arise in the study of quantum phase transitions.

The concept of phase transitions is of great importance, as many systems are shown to transit from a particular ordered state to a disordered one for different external conditions. The most well-known example of a phase transition is the boiling of water, where above a critical temperature the liquid water (ordered phase) is transformed into gaseous water vapor (disordered phase). Such phase transitions are referred to as thermal phase transitions (or classical phase transitions) as the tuning parameter, i.e. the external cause that drives the change of phase, is the temperature. Quantum phase transitions occur at zero temperature and are not driven by thermal fluctuations but by the competition of internal interactions and the effect of an external probe. The boundary between the interaction dominated phase (ordered phase) and the phase dominated by the effect of the external probe (disordered phase) is called a quantum critical point. The dynamics in the region near the quantum critical point are described by scale-invariant theories which, as stated before, can be Lorentz invariant or non-relativistic. The effect of a quantum phase transition is observable for finite temperature as a quantum critical region. The importance of that region is that, unlike the ordered and disordered regions, its long-time dynamics cannot be described in terms of thermally excited quasi-particles corresponding to the low-lying excitations. The physics of those regions is tied with the properties of the corresponding critical point, which can be adequately described by Lifshitz scaling QFTs.

Quantum phase transitions are important for understanding the behavior of condensed matter systems that exhibit highly desired properties such as the aforementioned strange metals. Consequently, quantum criticality constitutes an active topic of ongoing research. For more information on the topic of quantum criticality we refer to the chapter \ref{cpt:QPTs}.

Renormalization \cite{Ryder,Peskin,Zee} is the process of extracting finite values from infinities that arise in the calculations of observables, when an infinite range of short-distance scales are integrated out. There are two steps for treating such infinities: the step of regularization and the step of renormalization. The standard process of regularization, in the well established case of relativistic QFTs, is based on the introduction of a $(d+1)$-momentum cutoff, $\Lambda$, which sets the maximum values of the momenta up to which the physical quantities are evaluated. Manifestly, regularization introduces a cutoff dependence to all quantities, but their value is now finite, since momenta larger than the cutoff are excluded. However, low energy measurable quantities are depended on bare couplings and the cutoff.

The definition of low energy couplings, involves a novel scale, $\mu$ known as the renormalization group (RG scale). It is the scale at which the measurable quantities are defined.
A change of the cutoff scale can be compensated by a scale of the RG scale so the theory remains invariant.
The $\beta$-function encodes how a coupling runs with the renormalization scale, defining a RG flow. The $\beta$-functions allow us to identify the values of the momentum for which the QFT becomes strongly coupled (the associated coupling runs to infinity) or nearly free (the associated coupling runs to zero), which is important for the perturbative treatment of the QFT, as perturbation theory fails for strongly coupled theories. Lifshitz invariant field theories are expected not to be stable under generic perturbations and that eventually the perturbed theory will end up in a Lorentz-invariant CFT ($z=1$ see also discussion in \cite{lifshitz}) or in another Lifshitz invariant theory. Therefore, the running of the couplings is expected to be more elaborate than the relativistic case and thus a different scheme than the standard (sketched above), needs to be developed to describe the emerging RG flow.

The holographic correspondence, sometimes referred to as gauge/gravity correspondence, stems from a more general idea called the Holographic Principle. The Holographic Principle conjectures that all of the information in a region of space can be perceived as encoded on a boundary of the region. In this sense, the space and its boundary are related much like holograms and the corresponding film tapes in optical holography. This conjecture was first proposed by 'd Hooft \cite{largeN} and later elaborated upon by Susskind \cite{sus.hol}. The holographic principle follows from the argument (first proposed by Bekenstein, in \cite{bekens}) that the maximal entropy of a region of space with boundary of area $A$ is proportional to $A/G_N$ (where $G_N$ the Newton's constant). This argument seems not to be in agreement with QFT, according to which the number of degrees of freedom in a region of spacetime should scale with its volume rather than the surface of its boundary. However, it is expected that a successful quantum gravity theory should satisfy the Holographic Principle. The most successful realization of holographic principle to date is the AdS/CFT correspondence \cite{mald,ads2,ads3}.

The AdS/CFT correspondence was first proposed by Maldacena in \cite{mald}. In the same paper, it was shown that a strongly coupled, $\mathcal{N}=4$ supersymmetric $U(N)$ Yang-Mills theory in 3+1 dimensions, in the 't Hooft limit\footnote{Roughly speaking, the limit of the Yang-Mills theory, in which the number of colors tends to infinity, $N \to \infty$.}, \cite{largeN} is dual to  IIB string theory in 9+1 dimensions. It is conjectured that any string theory with gravity in AdS space-time is exactly equivalent to a CFT living on its boundary, as every field in the AdS theory can be translated to an operator in the CFT and vice-versa \cite{ads2,ads3}. An important aspect of AdS/CFT is that a low energy string theory is dual to a strongly coupled CFT. This weak/strong correspondence allows us to examine the properties of strongly coupled CFTs, in terms of low energy gravity or string theory in AdS space, which is easier to handle. Therefore, it is rather obvious that this kind of holographic field theory-gravity dualities are of great theoretical interest, as they seem to have a wide range of applications \cite{Erlich,Nishioka,Gursoy:2010fj}. Further clarifications and analysis on the topic of AdS/CFT correspondence are provided in the section \ref{sec:AdS_CFT_correspondence} of the present thesis.

A generalization of the AdS/CFT correspondence has been developed that includes the cases of Lifshitz scaling field theories without \cite{Kachru2008,Taylor:2008tg} or with hyperscaling violation \cite{cgkkm,gk1,Dong_hv_sol}. Non-relativistic holography, in contrast to standard weakly-coupled field theory methods, has successfully implemented interacting Lifshitz theories. Thus, non-relativistic holography seems adequate for the analysis of the quantum critical points that are conjectured to be described by strongly coupled non-relativistic field theories. Furthermore, non-relativistic holography has been successful in calculating physical observables such as the conductivity \cite{cgkkm}, that allows for direct comparison with the experiment.

In the case of relativistic QFTs, the way that the $\beta$-functions are extracted from the dual gravity was studied in
\cite{BoerVer,GursoyKir,Ramallo,Bourdier:2013axa}. In the non-relativistic case the momentum and energy are not related by Lorentz symmetry and are allowed to vary in different ways. This allows for non-relativistic scaling symmetries at the fixed points of the RG flow. The scale invariance at the fixed point adds a constraint to the scaling of energy and momentum but in contrast to the relativistic case that constraint is lifted whenever the system is driven away from the fixed point. Consequently, the RG flow is expected to be more involved and its analysis requires a different prescription than the Lorentz invariant case. Indeed, the additional  freedom introduced by the lifting of Lorentz symmetry can be readily identified by inspecting the form of the domain wall frame ansatz for a non-Lorentz invariant gravity theory
\be
ds^2=dr^2 -e^{2 A(r)} dt^2 +e^{2 B(r)} dx_i dx^i, \label{domain.ansatz}
\ee
where the dependence of the proper length, $ds$, on two distinct functions $A(r)$ and $B(r)$ is manifested. Such metrics cannot be realized for relativistic holography as Lorentz symmetry imposes the constraint $A(r)=B(r)$. According to the holographic prescription (see also chapter \ref{sec:AdS_CFT_correspondence}) the metric of the gravity theory (eq. \ref{domain.ansatz}) corresponds to a RG flow in the dual field theory and variations of the radial coordinate correspond to different values of the renormalization scale that interpolate between the IR (horizon) and UV (boundary) fixed points of the RG flow. Since, the functions $A(r)$ and $B(r)$ determine how the proper length depends on time and space, we can interpret them as the energy and momentum scales of the dual field theory respectively. We define the energy and momentum $\beta$ functions for a coupling $\phi$ (see chapter \ref{cpt:setup}) as
\be
\beta_E(\phi) \equiv \frac{d \phi}{d A} \sp \beta_P(\phi) \equiv \frac{d \phi}{d B}. \label{beta.intro}
\ee

We were able to evaluate within the holographic framework the two aforementioned $\beta$-functions in terms of the superpotential in the generic case of EMD gravity (section \ref{sec:running_phi}). We have found that the dependence of the $\beta_P$-function on the superpotential is similar to that in the relativistic case, but the dependence of the $\beta_E$ function is more complicated and shifts away from the relativistic value $\beta_E=\beta_P$ in terms of the superpotential.
 Therefore, a desired property for the two $\beta$ functions $\beta_E$ and $\beta_P$ is that they run to the corresponding value of the Lorentz-invariant $\beta$ function when the system becomes Lorentz invariant, we have shown in section \ref{sec:red_to_rel_b} that the definition (eq. \ref{beta.intro}) captures this property. We also comment on the behavior of the couplings near the boundary (UV fixed point) of the asymptotic solutions of the EMD gravity studied in \cite{cgkkm,gk1} (section \ref{sec:asympotic_EMD}). Then we continue with the running of the UV relevant operator near a Lifshitz fixed point by perturbing the Lifshitz solution of gravity with massive gauge field (first found in \cite{Taylor:2008tg}) with a UV relevant scalar field (section \ref{hp}). By this process we are able to identify the leading quantum corrections to the corresponding $\beta$-functions. In order to verify our holographic results, we evaluate the energy and momentum $\beta$-functions in the case of Lifshitz scalar $\phi^4$ theory using the standard field theoretic prescription (section \ref{sec:field_theory_treatment}). Finally in chapter \ref{cpt:conclusions}, we summarize our results and provide the outlook of the present thesis.

\chapter{Quantum Criticality} \label{cpt:QPTs}
\indent In this section we briefly review the basics about classical and quantum phase transitions, via some illustrative examples.

\section{Setup}

\subsection{Classical phase transitions}

Statistical mechanics studies the properties of physical systems consisting of a large (often intractable) amount of constituents by employing probability theory. A basic notion is the macroscopic and microscopic state. The microscopic state of a system is defined by knowing the velocity, position and internal state of each individual constituent. The corresponding dynamics is dominated by a microscopic Hamiltonian $H$. It is clear that in the case of typical atomic matter the knowledge of the microscopic state implies that the knowledge of $\sim 10^{22}$ time evolving quantities per $cm^3$, in order to track the state of each individual atom. To characterize the state of the system in a tractable way, the notion of a statistical ensemble is introduced. A statistical ensemble $XYZ$ is the collection of all distinct microscopic states yielding the same macroscopic observables\footnote{Macroscopic observables are measures of the average behavior of the system instead of each individual constituent.} $X$, $Y$ and $Z$. For instance, the canonical ensemble or $NVT$ ensemble is the collection of all microscopic states that yield a given value for the particle number $N$, the volume $V$ and the temperature $T$. In the above mentioned example the canonical ensemble can be realized by enclosing $N$ atoms in a impenetrable container of volume $V$ and assuming thermal equilibrium with a heat bath of temperature $T$. The characteristic constant of the ensemble is the Helmholtz free energy $F=F(N,V,T)$. The rest of the macroscopic quantities (eg. the pressure $P$, chemical potential $\mu$ and entropy $S$) are related with the $N$, $V$ and $T$ by their corresponding definitions which are characteristic for the canonical ensemble
\be
\langle P \rangle \equiv - \frac{\partial F}{\partial V} \sp \langle \mu \rangle \equiv \frac{\partial F}{\partial N} \sp \langle S \rangle \equiv - \frac{\partial F}{\partial T}. \label{definitions}
\ee
In order to connect the macroscopic observables $N$, $V$, $T$, $P$, $\mu$, $S$ with the corresponding thermodynamic quantities we should impose the thermodynamic limit that reads
\be
N \to \infty \sp V \to \infty \sp n=\frac{N}{V}=\text{finite}.
\ee
the thermodynamic limit also cures the ill definition of specific thermodynamic quantities such as the chemical potential $\mu$ (see eq. \ref{definitions}).
In order to connect the microscopic with the macroscopic description we should consider an isolated system where the microscopic description is well established. Obviously, the aforementioned example system is not isolated since it is in thermal contact with a heat bath, but the composite system consisting of the contained atomic matter plus the heat bath is indeed isolated. We assume that each of the distinct microscopic states $i$ of the composite system has the same total energy $E^{\text{comp}}$ and that the bath is weakly interacting with the system so that we can obtain the total energy by the summation of energies of each subsystem
\be
E^{\text{comp}} = E^{\text{bath}}_i +E_i, ~\forall i,
\ee
where $E_i$ and $E^{\text{bath}}_i$ are the energies of the contained atomic matter and the heat bath in the microscopic state $i$.
By using those assumptions the probability distribution for each of the microscopic states $i$ is
\be
P_i=e^{\frac{F-E_i}{K_B T}},
\ee
where the Helmholtz free energy $F$ reads
\be
F= - K_B T \ln \left[ \sum_i e^{- \frac{E_i}{K_B T}} \right], \label{free-energy}
\ee
where $K_B$ refers to the Boltzmann constant. Equation \ref{free-energy} connects the spectrum of the microscopic Hamiltonian $H$ with the Helmholtz free energy, $F$, which is used to define the macroscopic description of the system.
The macroscopic state of the system is defined by the three quantities that define the ensemble $N$, $V$, $T$ or any three quantities via which the value of the $F$ can be unambiguously determined. If those parameters are determined the rest of the macroscopic quantities can be evaluated. Besides the canonical $NVT$ ensemble other ensembles can be employed for the identification of macroscopic states, examples being the microcanonical $NVE$ ensemble and the grand canonical $\mu VT$ ensembles.

Before discussing phase transitions an understanding of what is the context of the term ``phase'' should be established. The concept of phase is tied with the concept of ``order''. As an example we consider the solid, liquid and gas phases of atomic matter. In the solid phase each of the atoms is localized at a specific position and is performing a small-amplitude oscillation around that position. In the liquid phase the atoms are also closely packed, but not localized as they can move across the volume of the liquid, as long as, there are other neighboring atoms in their vicinity. Finally, in the case of a gas the atoms are free to move throughout the volume of the gas almost independently. Therefore, whenever matter is in a solid phase the microscopic state of the atoms has to comply with a large amount of constraints. In the case of a liquid the number of constraints is smaller and, finally, in the case of a gas there are almost no constraints whatsoever. Consequently, we can claim that a solid is more ordered than a liquid and a liquid is more ordered than a gas. In a phase transition one of the involved phases is the ordered phase (more constraints) and the other phase is the disordered phase (less constraints). The concept of order is quantified by the order parameter. In the aforementioned case the most natural choice for an order parameter is the entropy, $S$ which measures the number of different microscopic configurations which may give rise to the same macroscopic state. Manifestly, for equal number of molecules more constraints correspond to less available microscopic configurations and therefore, $S_s < S_\ell < S_g$, where $s$, $\ell$, $g$ denote the solid, liquid and gas phases respectively. Therefore, the value of entropy can be used to quantify the ``order'' of the system and identify its phase. However, the order parameter is not unique. Indeed, different quantities such as the mass or number density could be used to identify those phases and thus are valid order parameters. Distinct macroscopic states yield different values for the order parameter and therefore we conclude that the phase of the system depends on the values of those parameters, which are referred to as tunable parameters.

We are finally able to define the term phase transition. $1$-st order phase transition is defined as the discontinuous change of the order parameter as one (or more) tunable parameters crosses a specific value which is referred to as a critical point. Except the $1$-st order phase transitions $n$-th order phase transitions can be defined, in this case the $(n-1)$-th derivative of the order parameter is discontinuous at the critical point.

To review the aforementioned concepts we rephrase the familiar effect of the boiling of water at atmospheric pressure by using the relevant jargon. We define the macroscopic state of the system at the $\mu p T$ ensemble and the mass density $\rho$ is introduced as an order parameter. In the ordered (liquid) phase the density of water is $\rho_\ell \sim 1000 ~kg/m^3$ while in the disordered (gas) phase the value of the density is $\rho_g \sim 0.8 ~kg/m^3$. At constant pressure $p=1 ~atm$ and at $T_c = 100 ~{}^oC= 373 ~K$ and unconstrained value of the chemical potential $\mu$\footnote{Remember that the boiling point of the liquid is does not depend on its quantity (intensive property).} the value of the order parameter changes abruptly from $\rho \sim \rho_\ell$ to $\rho \sim \rho_g$ signifying a $1$-st order phase transition. Such phase transitions that occur at non-zero temperature because of the variation of thermodynamic quantities are called classical phase transitions.

\subsection{Ising spin chain under a transverse magnetic field}

Quantum phase transitions are phase transitions that occur at zero temperature as an effect of the variation of non-thermodynamic parameters. In order to understand their emergence we review the example of the ferromagnetic Ising $S=1/2$ spin-chain under transverse magnetic field. Ferromagnetic Ising spin-chain is an one-dimensional array of spins (spanning along the $x$-axis) that interact in such a way, that they prefer to be aligned along one of the axes perpendicular to the array (for instance the $z$-axis). Transverse magnetic field is a homogeneous magnetic field parallel to the axis that spin chain extends ($x$-axis). The Hamiltonian for such a system reads
\be
\hat{H}= - J \sum_{i=1}^N \hat{\sigma}_i^z \hat{\sigma}_{i+1}^z - h \sum_{i=1}^N \hat{\sigma}_i^x, \label{transverse_ising}
\ee
where $J$ measures the strength of the spin-spin interaction (since we assumed ferromagnetic interactions $J \geq 0$) and $h$ measures the interaction of each of the spins to the external magnetic field.

The ground states with no magnetic field ($h=0$) are the states in which all of the spins are aligned either in the $+z$ or the $-z$ direction
\be
| \Uparrow \rangle \equiv \bigotimes_{i=1}^N | \uparrow \rangle_i \sp | \Downarrow \rangle \equiv \bigotimes_{i=1}^N | \downarrow \rangle_i, \label{ground_state1}
\ee
where $| \uparrow \rangle_i$ ($| \downarrow \rangle_i$) denotes that the spin in the $i$-th position is found in the state $|S=1/2,m_s=1/2 \rangle$ ($|S=1/2,m_s=-1/2 \rangle$).
On the contrary, if a strong transverse (along the $x$-axis) magnetic field is applied, such that the Ising interaction is negligible ($J \sim 0$), the ground states are the states in which all the spins are aligned in the $+x$ or $-x$ directions
\be
| \Rightarrow \rangle \equiv \bigotimes_{i=1}^N \frac{| \uparrow \rangle_i + | \downarrow \rangle_i}{\sqrt{2}}   \sp
| \Leftarrow \rangle \equiv \bigotimes_{i=1}^N \frac{| \uparrow \rangle_i - | \downarrow \rangle_i}{\sqrt{2}} . \label{ground_state2}
\ee
The ground states are competing, because the ground states for no magnetic field ($| \Uparrow \rangle$ and $| \Downarrow \rangle$) are highly excited states in the case of strong magnetic field and vice versa. The competition is manifested, when we consider the ground state in the case that neither the magnetic field nor the Ising interaction is negligible. In this case the answer is not as simple as in the cases of (eq. \ref{ground_state1}) and (eq. \ref{ground_state2}) but fortunately, the exact solution is known \cite{jordan-Wigner} (see section \ref{sec:exact_sol}). However, an exact solution is not known for a large class of problems and usually in order to study the behavior of such systems we have to rely on approximate schemes and/or numerical methods.

The notion of order is different in the case of zero-temperature. Ordered phase is the phase that is dominated by the internal interactions ($h=0$, eq. \ref{ground_state1}) and disordered is the phase that is dominated by the external fields ($h \to \infty$, eq. \ref{ground_state2}). A good choice for the order parameter\footnote{Strictly speaking, parameters that accumulate finite values for a system in the disordered phase and are equal to zero for a system in the ordered phase are referred to as disorder parameters.} in the given example is the magnetization along the $x$ axis
\be
M=\frac{1}{N}\sum_{i=1}^N s^x_i
\ee
and a good choice for the tuning parameter is $g=h/J$. The magnetization is equal to $M_{g=0} = 0$ in the case with zero magnetic field and $M_{g \to \infty} = \pm 1$ in the case of strong magnetic field. In order to identify the phase transition we should examine whether discontinuities in $M$ occur when $g$ is varied in the thermodynamic limit $N \to \infty$.

\section{Exact solution of the Ising spin chain under a transverse magnetic field} \label{sec:exact_sol}

The underlying idea behind the exact solution of the model described by the Hamiltonian (eq. \ref{transverse_ising}) is its mapping to a system of non-interacting spin-polarized fermions. Each spin in the state $\frac{| \uparrow \rangle_i + | \downarrow \rangle_i}{\sqrt{2}}$ is mapped to the absence of a fermion from the state $| \psi_i \rangle = \hat c_i^\dagger |0 \rangle$ ( $\hat c_i^\dagger$ the fermionic creation operator and $|0 \rangle$ the fermionic vacuum) and each spin in the state $\frac{| \uparrow \rangle_i - | \downarrow \rangle_i}{\sqrt{2}}$ to the presence of a fermion in the corresponding state $| \psi_i \rangle$. The correspondence between the spin operators $\hat \sigma^{\mu}_i$, $\mu \in \{x, y, z \}$ and the fermionic creation (annihilation) operator $\hat c^\dagger_i$ ($c_i$) reads
\be
\hat \sigma^x_m=2 \hat c^\dagger_m \hat c_m -1 \sp \hat \sigma^-_m= \frac{1}{2} \left( \hat \sigma^z_m - i \hat \sigma^y_m \right)= \hat c_m ~ \exp \left[ i \pi \sum_{n=1}^{m-1} \hat c^\dagger_n \hat c_n \right], \label{mapping}
\ee
where we denoted the site indices as $m$ and $n$ in order to avoid any confusion involving the imaginary unit $i^2=-1$. It can be shown that the definitions (eq. \ref{mapping}) comply with the aforementioned argumentation, as well as, the (anti) commutation relations
\be
\left[ \hat \sigma^{\mu}_m, \hat \sigma^{\nu}_n \right] =2 i \delta_{m,n} \epsilon^{\mu\nu\rho} \hat \sigma^{\rho}
\sp \left\{ \hat c_m, \hat c^\dagger_n \right\}= \delta_{m,n}
\sp \left\{ \hat c_m, \hat c_n \right\}=0.
\ee
By applying the mapping (eq. \ref{mapping}), the Hamiltonian (eq. \ref{transverse_ising}) reduces to
\be
\hat H= -J \sum_{n=1}^{N-1} \left( \hat c^\dagger_n \hat c_{n+1} + \hat c^\dagger_{n+1} \hat c_{n} +\hat c^\dagger_{n+1} \hat c^\dagger_n +\hat c_n \hat c_{n+1} \right) -h \sum_{n=1}^{N} \left( 2 \hat c^\dagger_n \hat c_n -1 \right),
\ee
this Hamiltonian is not-translationally invariant, this will introduce difficulties as the thermodynamic limit, $N \to \infty$, is taken. In order to fix that problem we impose periodic boundary conditions
\be
\hat H_{p}=\hat H -J \hat \sigma^z_N \hat \sigma^z_1 = \hat H + (-1)^{\sum\limits_{n=1}^{N} \hat c^\dagger_n \hat c_n} J \left( \hat c^\dagger_N \hat c_1 + \hat c^\dagger_1 \hat c_N - \hat c^\dagger_1 \hat c^\dagger_N - \hat c_N \hat c_1 \right),
\ee
the translational invariance implies that the quasi-momenta $k$ are good quantum numbers and thus $\hat c_k= \frac{1}{\sqrt{N}}\sum_{n+1}^N e^{-i k n a} \hat c_n$ is used to reduce the Hamiltonian $\hat H_p$ to
\be
\hat H_{p}= \sum_{k>0} -2 \left[ J \cos (k a) + h \right] \left(\hat c^\dagger_k \hat c_k - \hat c_{k} \hat c^\dagger_{-k} \right) +i 2 J \sin (k a) \left( \hat c^\dagger_k \hat c^\dagger_{-k} - \hat c_{-k} \hat c_{k} \right).
\ee
The Hamiltonian $H_p$ does not preserve the number of fermionic excitations, however since it is quadratic in the field operators it can be solved by introducing the Bogoliubov transformation
\be
\begin{split}
\hat d^\dagger_k=\sin \theta_k ~\hat c_k +i \cos \theta_k ~\hat c_{-k}^\dagger, \\
\hat d^\dagger_{-k}=\sin \theta_k ~\hat c_{-k} -i \cos \theta_k ~\hat c_{k}^\dagger,\\
\tan 2 \theta_k = - \frac{J \sin k a}{ J \cos k a + h}.
\end{split}
\ee
This concludes the mapping of (eq. \ref{transverse_ising}) to the Hamiltonian of free Fermions
\be
\begin{split}
\hat H_p = \sum_{k>0} \omega_k \left(d_{k}^\dagger d_{k} + d_{-k}^\dagger d_{-k} -1 \right),\\
\omega_k=2 \sqrt{ h^2 +J^2 +2 h J \cos(ka)}. \label{free_fermi}
\end{split}
\ee
The ground state of the system can be read from (eq. \ref{free_fermi}) as the state with no fermionic excitations (i.e. $d_k |GS \rangle=0$, $\forall k$) which reads
\be
| GS \rangle = \bigotimes_{k>0} \left( \cos \theta_k |0 \rangle + i \sin \theta_k | k, -k \rangle \right),
\ee
where $c_k |0 \rangle=0$ and $| k, -k \rangle=c^\dagger_{k} c^\dagger_{-k}|0 \rangle$, $\forall k$. Finally, the absolute value of magnetization, $|M|$ reads
\be
|M|= \left| \frac{2 a}{\pi} \int_{0}^{\frac{\pi}{a}} dk~ \sin^2 \left[ \frac{1}{2} \tan^{-1} \left( - \frac{\sin(k a)}{g+ \cos(k a)} \right) \right] -1 \right|,
\ee
where $\tan^{-1}$ denotes the arctangent. The value of this function is continuous as $g=h/J$ is varied. However, the value of its derivative
\be
\frac{d |M|}{d g}= \left| \frac{2 a}{\pi} \int_{0}^{\frac{\pi}{a}} dk~ \frac{ \sin \left( \tan^{-1} \frac{ -\sin(k a)}{g + \cos(k a)} \right)}{g^2 +2 g \cos(k a) +1} \right|
\ee
is discontinuous at $g=\pm 1$ signifying the existence of second order phase transition and showing that the points $g= \pm 1$ are quantum critical points. From the dispersion relation we can also verify that the system becomes gapless in the cases $g=1$ ($k=\pi/a$) and $g=-1$ ($k=0$). The gaplessness is important because it allows for the mapping of the system to a conformally invariant theory (in this case the theory of massless Majorana fermions).

\section{Emergent conformal field theory at the quantum critical point}

It is known that in the vicinity of a critical point the system is dominated by long-range correlations. Those correlations set a distance scale larger than the lattice spacing $a$. In this context it is reasonable to approximate the discrete Hamiltonian $\hat H$ with a continuous one $\hat H_c$ by introducing the spinor field $\hat \psi$
\be
\hat \psi(x_i)=\frac{1}{\sqrt{a}} \hat c_i. \label{psi}
\ee
We express the discrete Hamiltonian $\hat H$ (eq. \ref{transverse_ising}) in terms of the spinor field $\psi$ (eq. \ref{psi}) and then we take the continuous limit leading to the Hamiltonian
\be
\hat H_c= E_0 + \int dx~ \left[
 J a \left( \hat \psi^\dagger \frac{d \hat \psi^\dagger}{d x}
- \hat \psi \frac{d \hat \psi}{d x} \right)+2 \left| J-h \right| \hat \psi^\dagger \hat \psi \right].
\ee
Field theories are usually expressed in terms of a Lagrangian $\hat{\mathcal{L}}$ instead of a Hamiltonian $\hat{\mathcal{H}}$.  $\hat{\mathcal{H}}$ and $\hat{\mathcal{L}}$ are related via the Legendre transformation
\be
\hat{\mathcal{L}}=\int \hat P \frac{d \hat Q}{d t} -\mathcal{H},
\ee
where $\hat Q=\frac{\hat \psi +\hat \psi^\dagger}{\sqrt{2}}$ and $\hat P=\frac{\hat \psi -\hat \psi^\dagger}{i \sqrt{2}}$ are the generalized position and momentum respectively. The notion of Legendre transformation introduces a subtlety that is tied with the ordering of the generalized position and momentum. In order to properly define the Lagrangian density $\hat{\mathcal{L}}$ we demand that the partition function of the field theory equals the partition function obtained by the Hamiltonian.
\be
\mathcal{Z}=Tr e^{-\frac{H_C}{T}}=\int D \psi D \psi^\dagger e^{- \int_0^{1/T} d\tau dx \hat{\mathcal{L}} }, \label{partition}
\ee
where we have used that the partition function of the field theory corresponds to the appropriate path integral and that integration in imaginary time can be used to evaluate the effects of temperature. By using (eq. \ref{partition}) the Lagrangian of the field theory reduces to
\be
\hat{\mathcal{L}}  = -i \hat \psi^\dagger \frac{d \hat \psi}{d t} +J a \left( \hat \psi^\dagger \frac{d \hat \psi^\dagger}{d x} - \hat \psi \frac{d \hat \psi}{d x} \right)+2 \left( J-h \right) \hat \psi^\dagger \hat \psi. \label{majorana}
\ee
It can be checked that this Lagrangian is invariant under the conformal group defined in chapter \ref{subsec:ConfTrns}, in the vicinity of the quantum critical point, $J=h$, where the mass term $2 \left( J-h \right) \hat \psi^\dagger \hat \psi$ vanishes. In the case $J \neq h$, the aforementioned mass term breaks the scale invariance and introduces an energy gap in the field theory spectrum. The momentum dependence of the dispersion relation, $E \propto k^z$, defines the Lifshitz dynamical exponent, $z$. In the case of (eq. \ref{majorana}) the Lifshitz exponent is equal to $z=1$.

However, there are theories like the anisotropic next neighbor Ising (ANNNI) model \cite{ANNNI} that exhibit non-linear dispersion relations at criticality (for ANNNI $z=2$). Such theories exhibit a different type of scale invariance the Lifshitz invariance
\be
t \to \lambda^z t \sp x \to \lambda x.
\ee
The disappearance of the energy gap and the scaling invariance are characteristic properties of quantum critical points and are regularly used for their identification. The emergence of scaling invariant theories with $z \neq 1$ is especially interesting, as such field theories are non-Lorentz invariant. Gauge/gravity correspondence is a promising framework for explaining the behavior of such theories. AdS/CFT correspondence is the template of such theories which will be analyzed in the following.

\chapter{AdS/CFT correspondence} \label{sec:AdS_CFT_correspondence}

\indent In this chapter, we will briefly review the AdS/CFT correspondence based on the works \cite{stinnut,ads4,stinnut,Ramallo}. The AdS/CFT correspondence is the template for holographic theories. We begin by examining the connection between the perturbative structure of string theory and large-N gauge theories.

\section{Nambu-Goto equation and topological expansion of string theory}

Before discussing the dynamics of a string it is useful to review the dynamics of a point particle in the context of general relativity. The motion of such particle is described by its worldline $x^\mu=x^\mu(\tau)$. For a massive particle this worldline is the path with the least proper length that connects space time points, $x^m(\tau=\tau_i)$ and $x^m(\tau=\tau_f)$. Therefore, if the spacetime metric is $g_{\mu \nu}$ the relativistic action for a point particle with mass $m$ is the following (considering $-,+,+,+$ spacetime signature)
\be
S_0[x^\mu(\tau)]=-m \int ds=-m \int_{\tau_i}^{\tau_f} d\tau \sqrt{g_{\mu \nu} \frac{d x^\mu}{d\tau} \frac{d x^\nu}{d\tau} \;}.
\ee

A string is an extended one dimensional object, as a consequence its motion in spacetime is described by a two dimensional surface that we will call worldsheet $x^\mu=x^\mu(\tau,\sigma)$, where $\tau$ is the timelike coordinate embeded on the worldsheet, while $\sigma$ is the spacelike one, with $\sigma \in [ 0,\bar{\sigma} ]$. The quantity that corresponds to the mass of a point-particle is the string tension $T_s$, a dimensionfull parameter with $[T_s]=2$ in mass units. We can, also, introduce the string length, $\ell_s$ and string scale, $M_s$, as $T_s=\frac{1}{2 \pi \ell_s^{\;2}}=\frac{M_s^{\;2}}{2 \pi}$. In analogy with the point particle case, the worldsheet of a relativistic string is the surface with the minimal surface area that connects two string configurations, $x^m(\tau=\tau_i,\sigma)$ and $x^m(\tau=\tau_f,\sigma)$. This implies that the relevant action is the Nambu-Goto action \cite{stinnut}
\be
S_1[x^\mu(\tau,\sigma)]=-T_s \int dA ,
\ee

We assume that $\xi^0=\tau$ and $\xi^1=\sigma$ and $g_{\mu \nu}$ is the metric of the space where the string propagates (target space). The induced metric on the world-sheet, $\hat{G}_{\a \b}$ is
\be
ds^2=g_{\mu \nu} dx^\mu dx^\nu= g_{\mu \nu}  \frac{\pa x^\mu}{\pa \xi^\a} \frac{\pa x^\nu}{\pa \xi^\b} d\xi^\a d\xi^\b \equiv \hat{G}_{\a \b} d\xi^\a d\xi^\b \sp \hat{G}_{\a \b}\equiv g_{\mu \nu}  \frac{\pa x^\mu}{\pa \xi^\a} \frac{\pa x^\nu}{\pa \xi^\b}.
\ee

We rewrite the Nambu-Goto action as
\be
S_1[x^\mu(\tau,\sigma)]=-T_s \int \sqrt{- \det \hat{G}_{\a \b}} d^2 \xi = -T_s \int_0^{\bar{\sigma}} d\sigma \int_{\tau_i}^{\tau_f} d\tau  \sqrt{- \hat{G}}.
\ee

In order to obtain a unique solution from this variational principle we should impose boundary conditions. The boundary conditions depend on the kind of string theory we consider. In the case of closed strings, the worldsheet is a tube and therefore we should impose an additional periodicity condition in the $\sigma$ worldsheet coordinate, $x^\mu(\tau,\sigma+\bar{\sigma})=x^\mu(\tau,\sigma)$. For open strings the worldsheet is a strip. Two kinds of boundary conditions are used in this case
\be \text{Newmann: } \frac{\delta \mathcal{L}}{\delta x'^\mu} \bigg|_{\sigma=0 \text{ or } \bar{\sigma}}=0 \sp \text{Dirichlet: } \frac{\delta \mathcal{L}}{\delta \dot{x}^\mu} \bigg|_{\sigma=0 \text{ or } \bar{\sigma}}=0.\ee
 Where $x'^\mu\equiv \frac{\pa x^\mu}{\pa \sigma}$ and $\dot{x}^\mu\equiv \frac{\pa x^\mu}{\pa \tau}$.

The relativistic string can be quantized by standard methods of quantum field theory. The simplest way is canonical quantization, where we consider $x^\mu$ as operators and impose canonical commutation relations with the respective momenta, $[x^\mu,p^\mu \equiv \frac{\delta \mathcal{L}}{\delta \dot{x}^\nu}]=i \hbar \tensor{\delta}{^\mu_\nu}$. Although, we will not consider the details of the quantised string theory, its perturbative structure is of importance in the understanding of AdS/CFT correspondence.

As strings propagate through target space, they can undergo interactions. In closed string theory the basic interaction is that a string can split into two or two strings can merge into one. The corresponding worldsheet is the triple vertex of the theory. A loop can be obtained by combining two such vertices and the result is a Riemann surface of genus 1 (with a hole). Higher order terms in perturbation theory correspond to surfaces of higher genus. Therefore, the perturbative series in string theory is a topological expansion. The genus of each surface, $g$ in perturbation theory is equal to the number of string loops. If we weigh every vertex with a string coupling constant $g_s$ and each external line with $1/g_s$, we obtain that the string perturbative expansion is of the following form
\be
A= \sum\limits_{g=0}^\infty g_s^{-\chi} F_g \sp \chi=2(1-g), \label{amp.str}
\ee
where $\chi$ is the Euler number of the surface. This string expansion is identical to the large-N expansion of gauge fields (eq \ref{amp.thoo}), which we will consider in the next section, \ref{subsec:largeN}.

\section{Large N limit in gauge theories} \label{subsec:largeN}

We consider U(N) Yang-Mills theory with action
\be
S_{YM}=-\frac{1}{g_{YM}^2} \int d^4x \text{Tr}[ F_{\mu \nu} F^{\mu \nu}] \sp F_{\mu \nu}=\pa_\mu A_\nu -\pa_\nu A_\mu +[A_\mu , A_\nu],
\ee
where $A_\mu$ are $N \times N$ matrices and the YM coupling $g_{YM}$ can be expressed in terms of the t' Hooft coupling $\lambda$, as $g_{YM}^2=\lambda/N$. The 't Hooft large $N$ expansion is realized by keeping the coupling $\lambda$ fixed while performing an expansion of the amplitudes in powers of $N$. U(N) Yang-Mills theory has three and four point vertices that each brings a factor of $N/\lambda$ in the amplitude, the gauge propagator brings a factor of $\lambda/N$. Also each loop brings in a factor of $N$ because of the summation over $N$ colors.

We denote $L$ as the number of loops, $I$ as the number of internal propagators and $V$ the number of vertices. A typical diagram will be proportional to the following powers in $N$ and $\lambda$
\be
\left( \frac{\lambda}{N} \right)^I \left( \frac{N}{\lambda} \right)^V N^L=N^{L-I+V} \lambda^{I-V}. \label{pow.thoo}
\ee

\begin{figure}[h]
\begin{center}
\includegraphics[width=0.60\linewidth]{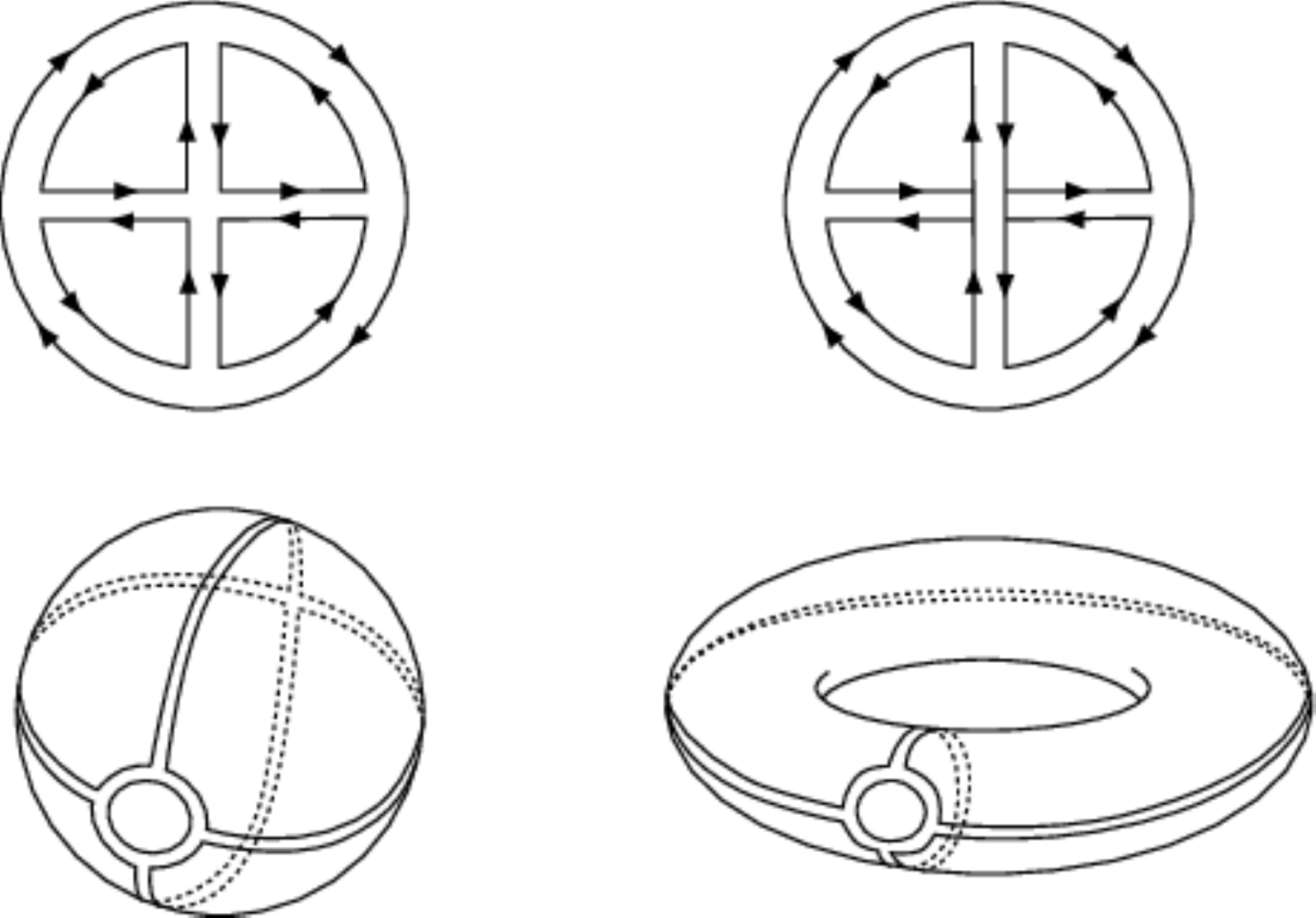}
\end{center}
\caption{In this figure we present a planar diagram on the left side that triangulates a closed surface of genus $g=0$ (sphere). While the non-planar diagram on the right triangulates a closed surface of genus $g=1$ (torus).} \label{fig:trian}
\end{figure}

It is useful to consider the double line 't Hooft notation, where we substitute each line in the Feynman diagrams of the theory with two lines of opposite orientation. One can then verify that a diagram of order $g$ triangulates a surface of genus $g$. Two such examples for $g=0$ and $g=1$ are presented in figure \ref{fig:trian}. The vertices on the 't Hooft diagram correspond to vertices (0-faces) on the surface, propagators correspond to edges (1-faces) and loops correspond to (2-)faces defined by the triangulation. Therefore, the combination that appears in the exponent of N in (eq. \ref{pow.thoo}) is nothing else than the Euler number $\chi=L-I+V$ of the surface, that the diagram triangulates. For compact closed surfaces $\chi=2(1-g)$, where $g$ is the genus of the surface (number of handles). Thus, we conclude that the standard perturbative expansion of an amplitude (for large $N$ and small $\lambda$) can be written as
\be
\sum\limits_{g=0}^\infty N^{2-2g} \sum\limits_{i=0}^\infty c_{i,g} \lambda^i \equiv \sum\limits_{g=0}^\infty N^{2-2g} Z_g(\lambda). \label{amp.thoo}
\ee

As a consequence, in the large N limit the result is dominated by the surfaces with smaller genus, $g$ (typically spheres). Diagrams with $g=0$ are called planar diagrams since they can be drawn on a plane (and if we add the point at infinity to a plane it becomes topologically equivalent to a sphere\footnote{We understand that by considering the usual stereographic projection of a sphere on a plane, if the south pole of the sphere touches the plane, all of the points on the sphere can be mapped on the plane except the north pole.}). The diagrams of higher genus are called non-planar and they are suppressed by negative powers of $N^2$, thus they become negligible in the large N limit.

The large N expansion is (in a sense) a topological expansion as we are expanding on diagrams that triangulate surfaces of increasing genus. The typical amplitude in this case (eq. \ref{amp.thoo}) is very reminiscent of the one obtained by topological expansion in string theory (eq. \ref{amp.str}), if we identify $g_s=N^{-1}$. Heuristically, one can claim that a gauge theory triangulates the worldsheets of an effective string theory. The AdS/CFT correspondence is a concrete realization of this connection.

\section{Dp-Branes}

Beside the perturbative structure of string theory, there is also a non-pertubative sector which is of major importance in connecting it with gauge theories. The most important (relevant for gauge-gravity correspondence) objects in the non-perturbative formulation of string theory are the Dp-branes. Dp-branes are extended objects in $p+1$ dimensions ($p$ spatial + time) that open strings are constrained to end on them by Dirichlet boundary conditions. Dp-branes act as sources for gravity fields and are, also, dynamical objects as their fluctuations are described by the dynamics of open strings that end on them.

\subsubsection{Field theories living in Dp-branes} \label{subsec:qftbrn}
There are two kinds of excitations that Dp-branes can exhibit, the first kind of which consists of rigid motions and deformations of their shape. If we assume a ten-dimensional target space (9 spatial directions + time), those degrees of freedom can be parametrized by $9-p$ coordinates\footnote{For later convenience we select $[\phi^i]=1$ in mass units.}, $\phi^i$ ($i=1,...,9-p$) that are transverse to the ($p+1$)-dimensional world-volume of the Dp-brane in the target space. Let $\xi^\mu$, with $\mu=0,1,...,p$, be the coordinates on the worldvolume of the Dp-brane and $G_{\a \b}$ the metric of the target space. If $X^\a(\xi^\mu)$ are the coordinates of the Dp-brane that describe its embedding in target space, then, $G_{\a \b}$ induces the metric:
\be
ds^2=G_{\a \b} dX^\a dX^\b = G_{\a \b} \frac{\pa X^\a}{\pa \xi^\mu} \frac{\pa X^\b}{\pa \xi^\nu} d\xi^\mu d\xi^\nu \equiv \hat{G}_{\mu \nu} d\xi^\mu d\xi^\nu.
\ee
If we consider Minkowski flat target space the induced metric in our case can be expressed, in the ``static gauge" (in the ``static gauge'' we consider the Dp-brane to be extended along $p$ spatial coordinates and be transverse to the rest $9-p$ coordinates) as
\be
\hat{G}_{\mu \nu}=\eta_{\mu \nu} + (2 \pi \ell_s^{\;2})^2 \pa_\mu \phi^i \pa_\nu \phi^i \label{Ind.metr}
\ee

In addition, Dp-branes, also, exhibit internal excitations; the endpoints of an open string are charges and therefore they source a gauge field, $F_{\mu \nu}$, on the world-volume. The action that takes into account these two kinds of excitations is called Dirac-Born-Infeld action and it can be written as:
\be
S_{DBI}=- T_{Dp} \int d^{p+1}x \sqrt{-det(\hat{G}_{\mu \nu}+ 2\pi \ell_s^{\;2} F_{mn})\;} \sp T_{Dp}=\frac{1}{(2\pi)^p g_s \ell_s^{\;p+1}}. \label{DBI}
\ee

If we expand the action (eq. \ref{DBI}) in powers of $F_{\mu \nu}$ and $\phi^i$, the quadratic terms in the expansion can be written as:
\be
S_{DBI}=-\frac{1}{g_{YM}^{\;2}} \int d^{p+1}x \left( \frac{1}{4}F_{\mu \nu} F^{\mu \nu} + \frac{1}{2} \pa_\mu \phi^i \pa_\nu \phi^i +... \right), \label{DBI.2}
\ee
they are the kinetic terms for a gauge field and $9-p$ scalar fields. The rest of the terms which are not presented in (eq.\ref{DBI.2}), are suppressed by additional factors of $\ell_s^{\;2}$. Therefore, in the low energy limit, the DBI action is equivalent to that of a U(1) gauge theory and $9-p$ scalar fields living in the world-volume of the Dp-brane. The Yang-Mills coupling, $g_{YM}$, in terms of string parameters $\ell_s$ and $g_s$, is expressed as:
\be
g_{YM}^{\;2}= 2 g_s (2 \pi)^{p-2} \ell_s^{\;p-3}. \label{gcoup.ym}
\ee

If we consider superstring theory, branes are also allowed to exhibit fermionic excitations, which correspond to fermionic fields in the low-energy expansion. Dp-branes can also be charged under p-forms which represent antisymmetric gauge fields, $A_{\mu_1,...,\mu_p}$. The most fascinating feature of Dp-branes, though, is the fact that they contain a gauge theory in their world-volume. In the case of N parallel Dp-branes it is expected that the fields $A_\mu$ and $\phi_i$ will be promoted to matrices transforming in the adjoint representation of $U(N)$. The elements of those matrices correspond to which of the branes are excited by those fields. In the system of N parallel coincident D3-branes\footnote{Those D3-branes are ``parallel" and ``coincident", meaning that they are extended along a (3+1)-dimensional hyperplane and they are located at the same point of the transverse six dimensional space.} the corresponding four-dimensional $U(N)$ gauge theory is a super Yang-Mills theory with four super symmetries ($\mathcal{N}=4$, sYM). This theory is an exact CFT. By utilizing the dynamical and gravitational descriptions of this system, we can explore the AdS/CFT correspondence.

\subsubsection{Langrangian description of N parallel D3-branes} \label{subsec:lagrbrns}

A system of N parallel coincident D3-branes would contain two types of excitations, open and closed strings. Open strings are excitations of the branes and closed strings are excitations of the bulk spacetime. Those two types of excitations interact. Therefore, we can schematically express the total action as:
\be S=S_\text{bulk}+S_\text{brane}+S_\text{interaction}. \ee

The low energy limit of $S_\text{bulk}$ is obtained by considering small deformations of an underlying 10D Minkowski space, $g_{\mu \nu}=\eta_{\mu \nu}+ \kappa h_{\mu \nu}$.
\be
S_\text{bulk} \sim \frac{1}{2 \kappa^2} \int d^{10}x ~ \sqrt{-g} R +... \sim \int d^{10}x ~ \left[ (\pa h)^2 + \kappa h (\pa h)^2 + ... \right],
\ee
where the gravity constant is related to string parameters as
\be 2 \kappa^2=(2 \pi)^7 g_s^{\;2} \ell_s^{\;8}. \ee

 We have not explicitly indicated all massless bulk fields and have suppressed Lorentz indices for simplicity. The crucial observation here is that interaction terms are proportional to positive powers in string length $\ell_s$ and thus for low energies ($\ell_s \to 0$) they become weaker. Thus, we conclude that in the low energy limit (or equivalently as $\ell_s \to 0$), $S_\text{bulk}$ describes a free bulk IIB supergravity.

Turning our attention to the $S_\text{brane}$ component of the action we recall that the DBI action reduces in the low energy limit to a $U(N)$ gauge field theory which is an exact CFT ($\mathcal{N}=4$ sYM), as stated in section \ref{subsec:qftbrn}.

We now consider the interaction component of the total action, $S_\text{interaction}$. With the expansion of the metric as $g_{\mu \nu}=\eta_{\mu \nu} + \kappa h_{\mu \nu}$ we can conclude that interaction terms are suppressed by positive powers of $\kappa$:
\be
S_\text{interaction}\sim\int d^4x~\sqrt{-g} \; \text{Tr}\left[ F^2 \right]+...\sim \kappa \int d^4x~h^{\mu \nu} \text{Tr}\left[ F^2_{\mu \nu} -\frac{\delta_{\mu \nu}}{4}F^2 \right],
\ee
where only the kinetic term for the gauge bosons was indicates for simplicity and the traces that appear in the expression are over colour indices. Therefore, in the low energy limit ($\ell_s \to 0$) all interaction terms in $S_\text{interaction}$ vanish.

In this limit, we end up with free bulk supergravity and $\mathcal{N}=4$, U(N) sYM theory not interacting with each other. This description, however, based on the Langrangian equations is not the unique way to describe the system of N parallel coincident D3-branes. In the following section \ref{subsec:grvbrns} we will consider an entirely gravitational description of this system.

\subsubsection{Gravitational description of N parallel D3-branes} \label{subsec:grvbrns}

String theory is a generalized gravity theory. Therefore Dp-branes, like other extended objects, should be solutions of the gravity sector of IIB string theory, the low energy limit of which is IIB supergravity. Since we are interested in D3-branes, we consider the following gravity action, involving the metric $g_{\mu \nu}$, the dilaton field $\Phi$ and the self-dual 4-form, $C_4$ under which D3-branes are charged with $F_5=\text{d} C_4$.
\be
S_3=\frac{1}{2 \kappa^2} \int d^{10}x \sqrt{-g}\left[ e^{-2 \Phi} \left[R+4(\na \Phi)^2 \right] -\frac{1}{240} F^{\;2}_5 \right].
\ee
The gravity constant is related to string parameters as
\be 2 \kappa^2=(2 \pi)^7 g_s^{\;2} \ell_s^{\;8}. \label{grav.con}\ee
We consider following D3-brane ansatz
\be
ds^2=H^{-1/2}(-dt^2+dx^i dx_i)+H^{1/2}(dr^2+r^2 d\Omega_5^{\;2}), \label{brn.metr}
\ee
where $i=1,2,3$ and $d\Omega_5$ is the solid angle in five dimensions. For constant dilaton field and $C_{0123}=1-1/H$, we obtain the solution
\be
H=1+\frac{L^4}{r^4} \sp L^4=4 \pi g_s \ell_s^{\;4} N. \label{L.gs}
\ee

Since the $g_{tt}$ metric component varies with $r$, the measured energy is $r$ dependent. If we measure energy $E_{r_0}$ at some point with $r=r_0$ then the energy that an observer at infinity would measure would be $E_\infty=H^{-1/4}(r_0) E_{r_0}$ due to the gravitational red-shift. In this context, there are two kinds of low energy excitations for an observer at infinity.

The first consists of massless large-wavelength excitations that can propagate in the whole bulk (i.e. they are not confined inside a brane but they are free to move in the 10D space) and therefore free to propagate far away from the brane where the space is essentially flat. Those excitations are described by the free limit of IIB supergravity.

The second low energy excitation mode consists of all kind of excitation that can approache $r=0$. In this case we can simplify the metric (eq. \ref{brn.metr}) taking the limit $r \to 0$. It is useful to express the metric in usual Poincar\'e coordinates by setting $u=L^2/r$ and taking the limit $u \to \infty$.
\bs \bear
ds^2=\frac{L^2}{r^2} dr^2+ \frac{r^2}{L^2} (-dt^2 +dx^i dx_i) + +L^2 d\Omega_5^{\;2}~~\text{for}~r \to 0 \label{corr.metr1}\\
ds^2=\frac{L^2}{u^2}(-dt^2 +du^2 +dx^i dx_i)+L^2 d\Omega_5^{\;2}~~\text{for}~u \to \infty. \label{corr.metr2}
\eear \es
The first part of the metric can be identified as Anti-de Sitter space\footnote{For more information, see appendix \ref{subsec:AdS}} in 5-dimensions while the second part is the 5-sphere. We conclude that such modes are described by IIB supergravity in an $AdS_5 \times S^5$ space. Those two modes are decoupled from each other in the low energy limit, as the coupling behaves as $\sim L^8 \omega^3$ \cite{stinnut} and therefore becomes negligible at low energies.

As a summary, we found that in the low energy limit the same system of $N$ parallel D3-branes can be described either as a free bulk IIB supergravity and a $\mathcal{N}=4$, sYM theory decoupled from one another or as a free bulk IIB supergravity and a IIB supergravity in $AdS_5 \times S^5$, also decoupled from one another. Since free bulk IIB supergravity is a component in both descriptions Maldacena, in \cite{mald}, conjectured that the remaining two theories, IIB supergravity in $AdS_5 \times S^5$ and $\mathcal{N}=4$ sYM, are equivalent. This conjecture is known as the AdS/CFT correspondence.

\section{Regime of validity for AdS$_5$/CFT$_4$ correspondence} \label{subsec:Val.cor}

We can find the limit that AdS/CFT holds by matching the parameters in the two sides of the correspondence. By combining equations (\ref{L.gs}) and (\ref{gcoup.ym}), we obtain the ratio between AdS radius $L$ and string length $\ell_s$.
\be
\left( \frac{L}{\ell_s} \right)^4=N g_{YM}^{\;2} \Leftrightarrow \left( \frac{\ell_s}{L} \right)^2 =\frac{1}{\sqrt{\lambda}}, \label{ls.con}
\ee
where we have used that 't Hooft coupling $\lambda$ is defined as $\lambda \equiv N g_{YM}^{\;2}$. In order to avoid stringy corrections due to the massive states of the string we should require that $\ell_s/L<<1$, which consequently by (eq. \ref{ls.con}) means that the 't Hooft coupling should be large $\lambda>>1$ and therefore the sYM theory strongly coupled. On the other hand, if sYM theory is weakly coupled, $\lambda<<1$ by the same equation, the string corections to the dual theory are large as $\ell_s >>L$.

This inverse proportionality of the string length $\ell_s$ and t' Hooft coupling, $\lambda$ is sometimes referred as strong/weak duality and is very important as it allows us to obtain information about strongly coupled dynamics from a dual weakly coupled theory.

The gravitational constant is expressed in terms of the Plank length, $\ell_P$ by the equation
\be 2 \kappa^2= 16 \pi \ell_P^{\;8}.\ee
By utilizing equations (\ref{ls.con}), (\ref{grav.con}) and (\ref{gcoup.ym}), we obtain the ratio between Plank length and AdS radius,as:
\be
\left( \frac{\ell_P}{L} \right)^8=\frac{\pi^4}{2 N^2}.
\ee
Therefore, in order to avoid quantum gravity corrections in the theory we should require that $\ell_P/L<<1$, which consequently means that $N>>1$, in the large N limit.

As a conclusion, we have found that the planar (large N limit) and strongly coupled (large 't Hooft coupling, $\lambda$) sYM can be described by a classical gravity on an AdS manifold. This is a weak form of the conjectured AdS/CFT correspondence. The strong form should be valid at any $N$ and $\lambda$. There is no contradiction to date with the strong form of the correspondence. We will continue by considering the symmetries that those theories share.

The $\mathcal{N}=4$ sYM is an exact 4-dimensional CFT. As a consequence, it is invariant under the conformal group $SO(2,4)$ (see appendix \ref{subsec:ConfTrns}). This is also  the symmetry group of $AdS_5 \times S^5$ (see appendix \ref{subsec:AdS}). The $\mathcal{N}=4$ sYM is a maximally supersymmetric theory. It has 32 conserved fermionic supercharges which correspond to the 32 Killing spinors of the dual supergravity. Finally the R-symmetry $SO(6)$ group of the CFT, which rotates the six scalar fields of the theory (see eq. \ref{DBI.2}) is identified with the symmetry of the five-sphere component of $AdS_5 \times S^5$. An interesting observation, is that the directions along the $S^5$ correspond to the scalar fields on sYM and that the isometries of the compact space correspond to internal rotations of scalar fields and supercharges.

\section{Physical significance of extra dimensions}

As stated before, the gravity theory lives in a 10D $AdS_5 \times S^5$ spacetime, while the dual field theory lives in the 4D boundary of the $AdS_5$. Since the two theories live in spacetimes of different dimensionality, the meaning of the extra dimensions in the gravity side of the correspondence should be explained. It is argued that the extra ``radial" dimension $r$ of the $AdS_5$, corresponds to the energy scale of the dual field theory, while the five-sphere component of $AdS_5 \times S^5$ is related to the scalar fields of the $\mathcal{N}=4$ sYM and their symmetries. The intention of the following section is to make those correlation clear.

\subsubsection{Radial coordinates $r$ and $u$} \label{subsec:radene}

To begin with, we review the role of the $r$ coordinate in the correspondence. We consider the gravitational redshift in the decoupling (low energy) limit $\ell_s \to 0$ in the near horizon area. The energy in string units ($\sim E_r \ell_s$) should be kept fixed in the limit $r \to 0$, thus the energy measured at infinity is
\be
E_\infty \approx E_r \frac{r}{\ell_s}=(E_r \ell_s) \frac{r}{\ell_s^{\;2}},
\ee
However, $E_\infty$ is the energy as measured in the field theory. We should keep it fixed in the same limit. We conclude that $U=r/\ell_s^{\;2}$ should be kept fixed in the $r \to 0$ limit and the radial coordinate is therefore proportional to the energy scale of the CFT in the near horizon limit. The near horizon metric in terms of $U$ is
\be
ds^2=\ell_s\left[ \frac{U^2}{\sqrt{4 \pi g_s N}} (-dt^2 +dx^i dx_i) +\sqrt{4 \pi g_s N} \left(\frac{dU^2}{U^2} + d\Omega_5^{\;2} \right) \right].
\ee
In these coordinates, it is manifest that if we ignore the 5-Sphere, that the AdS metric reduces to the CFT 4 dimensional metric with an extra coordinate that encodes the energy scale of the theory.

However, this argument works in the low energy, near horizon limit of AdS$_5$/CFT$_4$ correspondence. An argument with more extended validity is the following. Consider the $AdS_n$ metric in Poincar\' e coordinates
\be
ds^2=\frac{L^2}{u^2}(du^2-dt^2+dx^i dx^i)~~\text{with}~i=1,...,n-2.
\ee
This metric is invariant under scale transformations: $(u,t,x^i) \to (a u, a t, a x^i)$ (see appendix \ref{subsec:AdS}). We interpret this symmetry in the following way, if we scale up spatial coordinates $x^i$ we are, simultaneously, scaling up wavelengths, thus scaling down wavevectors in momentum space, therefore we scale down the energy of the particles and the system. However scaling up the coordinates means that we should also scale up $u$. Therefore, we conclude that $u$ is inversely proportional to the energy scale and consequently, the boundary at $u=0$ corresponds to the UV (high energy) limit of the theory, while the horizon at $u \to \infty$ corresponds to the IR (low energy) limit.

\subsubsection{Reduction on the $S^5$} \label{subsec:reducts}

We notice that any field living on $AdS_5 \times S^5$ space can be reduced to a tower of fields on a $AdS_5$ by expanding this field in terms of spherical harmonics. With this in mind, we will attempt to reduce the gravitational action to
\be
S=\frac{1}{2 \kappa_{(5)}^{\;2}} \int d^5x ~\sqrt{-g^{(5)}} \left[R^{(5)}- 2 \Lambda +\mathcal{L}_\text{matter} \right],
\ee
which describes gravity in 5 dimensions. The cosmological constant will be chosen in such way that a free gravity theory $\mathcal{L}_\text{matter}=0$, yields $AdS_5$ space as its solution. The appropriate choice is $\Lambda=-6/L^2$. The Plank scale in five dimensions will be obtained from the 10 dimensional one, by considering the integration on $d\Omega_5$
\be
\frac{1}{2 \kappa_{(10)}^2}\int d^5x d^5\Omega \sqrt{-g^{(10)}} R^{(10)} \to \frac{\pi^3 L^5}{2 \kappa_{(10)}^2}\int d^5x \sqrt{-g^{(5)}} R^{(5)}, \label{sphere.red}
\ee
where $\pi^3$ is the volume of unit $S^5$. Therefore, $\kappa_{(5)}^{\;2}=\kappa_{(10)}^2/(\pi^3 L^5)$, or by using the CFT parameters
\be \kappa_{(5)}^{\;2}=4 \pi^2 L^3 N^{-2}.\ee

We have verified that the 10 dimensional gravity component can be reduced to a five-dimensional one. We now examine whether the same is true for the $\mathcal{L}_\text{matter}$ term, we will consider the case of a massless scalar field for simplicity. The Klein-Gordon equation in this case is
\be
\na_{(10)}^2 \phi(x,\Omega)=0 \Rightarrow \left( \na_{AdS_5}^2+\na_{S^5}^2 \right) \phi(x,\Omega)=0
\ee
Where we have used the fact that since the metric factorizes into $AdS_5$ and $S^5$ parts, the D' Alambertian is the sum of those two parts. Furthermore, we know that the eigenfunctions of $\na_{S^5}^2$ are the 5-dimensional spherical harmonics defined by the eigenvalue equation
\be
\na_{S^5}^2 Y_\ell(\Omega)=-\frac{\ell(\ell+4)}{L^2} Y_\ell(\Omega), ~~~\ell=0,1,2,...
\ee
We conclude that the 5-dimensional scalar field $\phi_\ell(x)$, where $x$ are the $AdS_5$ coordinates, should satisfy a massive Klein-Gordon equation (remember that we are working with $-,+,+,+,+$ spacetime signature)
\be
(\na_{AdS_5}^2-m_\ell^2) \phi_\ell(x)=0, ~~~m_\ell^2=\frac{\ell(\ell+4)}{L^2}. \label{twr.flds}
\ee
Therefore, after the reduction on $S^5$, we obtain a tower of massive fields $\phi_\ell(x)$ (tower of fields means that we have an infinite sequence of them with ever increasing mass), with particular masses $m_\ell$, from a single massless field $\phi(x,\Omega)$. It can be shown that the same is also true for gauge and spinor fields that live on $AdS_5 \times S^5$.

\section{Bulk fields and boundary operators}

\subsubsection{Motivation}

In order to motivate the duality of bulk fields with boundary operators we examine the case of a non-gravitational system on a lattice with spacing $\a$ and hamiltonian given by
\be
H=\sum\limits_{x,i} J_i(x,\a) O^i(x), \label{ham.lat}
\ee
where $x$ denotes the different lattice positions and $i$ labels the different operators. $J_i(x,\a)$ are the coupling constants at a point $x$ of the lattice for a particular scale $\a$. In the Kadanoff-Wilson renormalization group approach we increase the lattice spacing by replacing multiple lattice sites with a single site. Each lattice variable of the new site is an average over the values of the replaced sites. In this process the Hamiltonian (eq. \ref{ham.lat}) retains its form but the couplings change in each step. The equation that determines how couplings change is the following
\be
u\frac{\pa}{\pa u} J_i(x,u)= \b_i \big(J(x,u),u \big) \label{beta},
\ee
where $\b_i$ are the so called $\b$ functions of the $i$-th coupling constant. At weak couplings, they are determined by perturbation theory. We observe that $u$ is a measure of the RG scale of the theory. We have already concluded in section \ref{subsec:radene}, that this scale is also related to the energy scale of the theory. The idea is to consider $u$ as an extra dimension. Therefore, the succession of lattices at different scale are considered as layers of new higher dimensional space, and coupling constants $J_i(x,u)$ are promoted to fields living in that space.

$\beta$-functions are also useful in the context of holography. A detailed study can be found in \cite{kir.beta}. The general prescription is to define the $\b$ function, as function of the fields $J_i$ in the gravity theory as presented in (eq. \ref{beta}), where $u$ is the energy scale of the dual field theory. Then by utilizing the equations of motion of the gravity theory, one can obtain and, a-priori, solve an equation solely for $\b (J_i)$. In \cite{kir.beta}, it has been also shown that Einstein-Dilaton theory yields non-linear first order differential equations for the holographic beta function.

\subsubsection{Mass scaling dimension of fields and dual operators} \label{subsec:mscldim}

We now consider a scalar field $\phi(u,x)$ in $AdS_5$ space, near the boundary. According to (eq. \ref{basy.scal}) the asymptotic behaviour is
\be
\phi(u,x)=u^{\Delta_-} \phi(x),~~~~ \Delta_\pm= 2\pm \sqrt{4+m^2L^2} \label{delta}
\ee
In order to identify the CFT source, $\varphi(x)$ with the boundary value of the field $\phi(u,x)$ we have to remove the trivial dimensionful part. Therefore we define $\varphi(x)$ as
\be
\varphi(x) \equiv \lim\limits_{z \to 0} u^{-\Delta_-} \phi(u,x). \label{philim}
\ee
In analogy with (eq. \ref{ham.lat}) we may consider the additional term in the gauge theory action for $\e \to 0$
\be
\int d^4x \sqrt{\gamma_\e} \phi(\e,x) O(\e,x) \xrightarrow{\e \to 0} L^4 \int d^4x \varphi(x) \e^{-\Delta_-} O(\e,x)
\ee
Where $\gamma_\e=(L/\e)^8$ is the induced metric at $z=\e$ and $O(\e,x)$ is the dual operator to the field $\phi(\e,x)$, we have also utilized (eq. \ref{philim}) and (eq. \ref{delta}). However the term should be independent of $\e$ in order to be finite at the boundary as a consequence we have to accept that the dual operator scales near the boundary as
\be
O(\e,x)=\e^{\Delta_+} O(x). \label{olim}
\ee
Thus from equations (\ref{philim}) and (\ref{olim}) we conclude that $\Delta_+$ is the mass scale dimension of the dual operator $O$ while $\Delta_-$ is the mass scale dimension of the bulk field $\phi$. If the BF bound (eq. \ref{BF.bound}) is satisfied scale dimensions are always real. For $m^2>0$, the dual operator is called irrelevant as it does not affect the IR of the theory. For $m^2=0$ the operator is called marginal and finally for $m^2<0$ the operator is called relevant as it affects the IR of the theory. Tachyonic scalars respecting the BF bound are acceptable as they do not lead to any bad instabilities.

\subsubsection{Field-Operator duality for a scalar field} \label{subsec:fld.opd}

We consider a massless scalar field $\Phi(u,x,\Omega)$ in $AdS_5 \times S^5$ space. As we have concluded in section \ref{subsec:reducts}, we can reduce the 10 dimensional gravity in $AdS_5 \times S^5$ space, coupled with a $\Phi(u,x,\Omega)$ to a 5 dimensional gravity in $AdS_5$ space which is now coupled with a tower of scalar fields $\Phi_\ell(u,x)$ of mass $m_\ell^2=\ell(\ell+4)/L^2$ with $\ell=0,1,2,...$ (see eq. \ref{twr.flds}). Therefore, according to the previous subsection (\ref{subsec:mscldim}), the field $\Phi_0$ is marginal ($m^2=0$) while the rest of the fields with $\ell \geq 1$ are irrelevant ($m^2>0$).

Since the field lives in $AdS_5 \times S^5$ space its dual operators should belong to a $\mathcal{N}=4$ sYM theory. The field $\Phi_0$ by utilizing equations (\ref{delta}) and (\ref{olim}), corresponds to a dual operator with mass scaling dimension $\Delta_+(\ell=0)=4$. Also remember that $\Phi_0$ corresponds to the spherical harmonic with $\ell=0$ which is singlet under the $S0(6)$ symmetry of the $S^5$. As a consequence the dual operator should also be singlet with respect to the dual R symmetry group. As a last observation it is obvious that since the field is scalar the dual operator should not have any indices. Therefore, the $\Phi_0(u \to 0,x)$ field corresponds to the coupling constant of the $\mathcal{N}=4$, sYM Lagrangian.

Scalar fields $\Phi_\ell$ with $\ell \geq 1$ have masses $m^2 L^2=\ell (\ell+4)$ (eq \ref{twr.flds}), therefore the mass scaling dimension of their dual operator will be $\Delta_+(\ell)=4+\ell$ and they will be irrelevant. $\Phi_\ell$ correspond to harmonics with $\ell \geq 1$, and thus the dual operator should transform under the appropriate representation of $S0(6)$. Then the dual operator is the $\mathcal{N}=4$, sYM Lagrangian
multiplied with the traceless symmetric product of $\ell$ scalar fields $\phi_{i}$ of $\mathcal{N}=4$ sYM, which is a marginal operator in four dimensions.

We have managed to identify the corresponding CFT operators of a dual field in AdS space. However, this identification would be in vain if there was not a way to calculate CFT quantities, the most important being correlation functions, in terms of the dual gravity theory.

\section{Correlation functions}

An important ingredient is the ability to compute CFT correlation functions of the type $\langle O(x_1)...O(x_1)\rangle$ from the dual gravity. This is achieved by utilizing the following postulate:
\be
\left\langle e^{\int d^4x \phi_0(x) O(x)} \right\rangle_\text{CFT}=Z_\text{string}\left[ \phi(x,u)|_{u=0} \to u^{4-\Delta} \phi_0(x) \right], \label{hl.grl}
\ee
where we have assumed that the operator $O$ has mass scale dimension $\Delta$ and is dual to the bulk field $\phi$.

In the rest of this section we will try to decipher what (eq. \ref{hl.grl}) means. On the left hand side of the equation, the expectation value is taken in the CFT and $\phi_0(x)$ plays the role of an arbitrary source for the operator $O$. On the right hand side, $Z_\text{string}$ is the generating functional of the string amplitudes in the $AdS_5 \times S^5$. In string theory, however we can only calculate amplitudes for a set of boundary data. That means that in order to calculate the string generating functional we should impose boundary conditions to the massless sting fields and then calculate the vacuum amplitudes as functions of those boundary conditions. In our case the boundary condition is $\phi(x,u)|_{u=0} \to u^{4-\Delta} \phi_0(x)$.

In order to use (eq. \ref{hl.grl}) since string theory on $AdS_5 \times S^5$ is not exactly solvable we need some approximation as we have argued in section \ref{subsec:Val.cor}, Strongly coupled planar $\mathcal{N}=4$ sYM theory is dual to IIB supergravity, thus in this limit the generating functional of string theory may be substituted for the IIB supergravity one, $Z_\text{string} \approx Z_\text{gravity}= e^{-I_\text{gravity}(\phi_0)}$, where $I_\text{gravity}(\phi_0)$ is the low energy (two derivative) IIB supergravity action.

The last important input (eq. \ref{hl.grl}) needed is the specific one-to-one correspondence between sYM operators and string theory fields, the methods of working out those correspondences were presented in section \ref{subsec:fld.opd}. The last complication with (eq. \ref{hl.grl}) is that we need both theories in the correspondence to be properly renormalized, on the field theory side renormalization and how to treat UV divergences is well understood (for more information see chapter \ref{sec:field_theory_treatment}), on the gravity side of the correspondence, divergences come from the neighbourhood of the boundary and they are long distance (IR) divergences that can be calculated. An interesting feature is that the UV counterterms that are introduced in CFT should be in one-to-one correspondence to the IR counterterms of the gravity theory. This is yet another manifestation of the strong/weak correspondence, we have observed in \ref{subsec:Val.cor} \cite{Henningson:1998gx,de Haro:2000xn,Skenderis:2002wp}.

\section{Holography, AdS/CFT and applications}

The AdS/CFT correspondence is the manifestation of the holographic principle. The holographic bound states that the entropy contained inside a closed region of space time with boundary area $A$ is less than or equal to $S_\text{max}=A/(4 G_N)$. This seems seemed to be in disagreement with QFT, as according to the latter we would expect entropy to scale with the volume of the region. However, in the context of AdS/CFT this disagreement is resolved. To show exactly this, we need to count the degrees of freedom in the CFT and the corresponding string/gravity theory.

In section \ref{subsec:radene}, we concluded that the radial coordinate in AdS corresponds to the energy scale. From (eq. \ref{AdS.metric}) for $n=5$ with the change of radial coordinate $\tilde{u}=tan (\rho/2)$, the metric becomes
\be
ds^2=L^2\left[- \left( \frac{1+\tilde{u}^2}{1-\tilde{u}^2} \right) d\tau^2+ \frac{4}{(1-\tilde{u}^2)^2} \left( d\tilde{u}^2 +\tilde{u}^2 d\Omega_3^2 \right) \right],
\ee
In these coordinates the boundary (extreme UV in the dual CFT) is at $\tilde{u}=1$, the horizon (extreme IR limit in the dual CFT) is at $\tilde{u}=0$ and the interior of the space at  $0<\tilde{u}<1$. We define a UV cutoff in $AdS_5$ at $\tilde{u}^2=1-\e$, where $\e\to 0^+$. According to the holographic bound the maximal entropy of $AdS_5$ is
\be
S_{max}=\frac{A}{4 G_N}\sim \frac{L^3}{4 G_N} \frac{\tilde{u}^3}{(1-\tilde{u}^2)^3}\bigg|_{\tilde{u}^2=1-\e}\sim \frac{L^3}{4 G_N \e^3}
\ee

The cutoff $\tilde{u}^2=1-\e$ correspond to a UV cuttoff in the CFT, therefore we may view $L\e<<1$ as a small distance cutoff in the field theory. We can therefore divide the boundary in$\sim 1/\e$ fundamental cells and since it is known that sYM theory has $N^2$ degrees of freedom, we can estimate the entropy as
\be
S_{FT}\sim \frac{N^2}{\e^3} \sim \frac{L^3}{G_N \e^3} \sim S_{max}
\ee
Where we have used (eq. \ref{sphere.red}) and $2 \kappa^2= 16 \pi G_N$. We conclude that in the context of AdS/CFT, the holographic bound is compatible with QFT based considerations of entropy.

As we have argued, the AdS/CFT is an example of strong/weak coupling correspondence. In simpler words a duality of a strongly coupled CFT that is inaccessible by perturbative methods and a weakly coupled gravity theory that is a priori solvable (and vice versa). One would expect that such a theory will have many applications and this is indeed the case. The most important areas that the AdS/CFT correspondence could be applied are Quantum chromodynamics and Condensed matter physics:

Quantum chromodynamics (QCD) is the theory of strong interactions. Holography provides a new method to study the low energy spectrum, where the theory becomes strongly coupled. For more information on this topic see \cite{Erlich}.

Condensed matter physics (CMP) is the branch of physics that deals with the physical properties of condensed phases of matter at finite density. Theories describing interesting states of matter such as superconductors, superfluids, Bose-Einstein condensates as well as theories describing transitions at zero temperature (quantum critical points) may be strongly coupled. Most of these phenomena are well studied by experiments. They are, however, very difficult to explain theoretically using standard techniques from QFT. As a consequence, holography can be invoked as a way to describe those strongly coupled theories with weakly coupled gravity ones.

\chapter{Holographic flows with broken Lorentz symmetry} \label{cpt:setup}

\indent As already stated, the scope of this work is to generalize the notion of holographic $\beta$-function in the case of non-relativistic holography. In holography we are mostly interested in the strongly coupled limit of the dual field theory, which according to AdS/CFT corresponds to low energy gravity. The simplest field theory in this case will be described by the expectation values of a single scalar operator $\mathcal{O}^{FT}$, a conserved current $J^{FT}_\mu$, which ensures the conservation of particles at finite density and a stress energy tensor $T^{FT}_{\mu \nu}$ which encodes the energy and momentum distributions. Those operators are mapped by the AdS/CFT correspondence to a scalar field, $\phi$, a $U(1)$ gauge field, $A_\mu$ and the metric $g_{\mu \nu}$ of the dual field theory, respectively. The Einstein-Maxwell-Dilaton action is the most general gravity action at the two-derivative level (low-energy limit) that contains those fields. This action reads
\be
S=\int d^{d+1} x \sqrt{-g} \left(R -\frac{1}{2}\pa_\mu \varphi \pa^\mu \varphi+ V(\varphi)-\frac{1}{4} Z(\varphi) F^{\mu \nu}F_{\mu \nu}\right). \label{EMDaction}
\ee
Note that the bulk dimension is $d+1$ and the boundary dimension $d$. The dimension of the boundary space is $d-1$.

In order to define the holographic $\beta$-functions, in analogy to the relativistic case we should use the coordinate frame where the component $g_{rr}=1$. We define the following ansatz, which satisfies this criterion
\be
ds^2=-e^{2 A} dt^2+dr^2+e^{2 B} \left( dx^i dx_i \right),~~ A_t=A_t(r),~~ A_i=0,~~ \phi=\phi(r). \label{Ansatz}
\ee
The functions $A$ and $B$ in contrast to the relativistic case are not fixed to the same value, $A=B$. We can therefore identify two RG scales in the dual field theory described by this metric the ``energy'' scale $M_E \sim \ln A$ and the ``momentum'' scale $M_P \sim \ln B$. We can define the corresponding $\beta$-functions as:
\be
\beta_P(\varphi)\equiv \frac{\pa \varphi}{\pa B} \sp \beta_E(\varphi)\equiv \frac{\pa \varphi}{\pa A}. \label{betafunctions}
\ee

\section{The equations of motion}

The covariant equations of motion for the EMD action (eq. \ref{EMDaction}) are well known in the literature. We present them below
\bs \be
R_{\mu \nu}-\frac{1}{2} g_{\mu \nu} R=T_{\mu \nu},
\ee
\be
\na_\mu \left( Z(\phi) F^{\mu \nu} \right)=0, \label{GF.coveq}
\ee
\be
\na_\mu \na^\mu \varphi + \frac{d V_\text{eff}}{d \varphi}=0,
\ee \label{covariant.eqmot} \es
where the stress energy tensor $T_{\mu \nu}$ of the fields $A_\mu$ and $\phi$ is defined as
\be
T_{\mu \nu}=\frac{g_{\mu \nu}}{2} V(\phi)+\frac{1}{2} (\pa_\mu \varphi)(\pa_\nu \varphi) -\frac{g_{\mu \nu}}{4} (\pa \varphi)^2+\frac{1}{2} Z(\varphi) \left( F_\mu^\rho F_{\nu \rho} - \frac{g_{\mu \nu}}{4} F^2 \right)
\ee
and the effective potential $V_\text{eff}$ of the scalar field reads
\be
V_\text{eff}=V(\varphi)-\frac{1}{4} Z(\varphi) F^2.
\ee

We can simplify the covariant equations of motion (eq. \ref{covariant.eqmot}) by introducing the ansatz (eq. \ref{Ansatz}). The equation stemming from the $A_t$ variation (eq. \ref{GF.coveq}) simplifies to
\be
A_t'(r)=\frac{q}{Z(\varphi)} e^{A -(d-1) B}, \label{emgf}
\ee
where we identify $q$ with the charge density.
The rest of the equations of motion (eq. \ref{covariant.eqmot}) with the substitution of (eq. \ref{emgf}) reduce to
\bs
\be
(d-2)(d-1)(B')^2+2(d-1) A' B' -\frac{1}{2}(\phi')^2=V(\phi)-\frac{q^2 e^{-2(d-1) B}}{2 Z(\phi)}, \label{eqmot1}
\ee
\be
(A')^2+(d-2)^2(B')^2-\frac{d-2}{2(d-1)}(\phi')^2+(3d -5) A' B' +A''=V(\phi), \label{eqmot2}
\ee
\be
A' B' -B''-(B')^2=\frac{1}{2(d-1)}(\phi')^2, \label{eqmot3}
\ee \label{eqmot}
\es
where prime denotes derivatives over the $r$ coordinate. The general solutions of the action (eq. \ref{EMDaction}) have been studied in the ``domain wall frame coordinates'' which can be mapped to the current ones. In the sequel, we will show that this mapping implies that both coordinate frames yield the same solutions.

\section{Domain wall frame}
\indent The general solutions of the action (eq. \ref{EMDaction}) have been studied by the use of the following ansatz in \cite{Kiritsis:2012ma}
\be
ds^2=e^{2\tilde{A}(u)} \left( -f(u) dt^2 + dx^i dx_i \right) +\frac{du^2}{f(u)},~~ A_t=A_t(r),~~ A_i=0,~~ \phi=\phi(r). \label{ansatz2}
\ee

In the ansatz (eq. \ref{ansatz2}) the metric is expressed in the so called ``domain wall frame'' coordinates which can be mapped to the ansatz (eq. \ref{Ansatz}) via a diffeomorphism. The functions $\tilde{A}$ and $f$ can be expressed in terms of $A$ and $B$ as
\be
\tilde{A}=B \sp f(u)=e^{2(A-B)} \sp \frac{dr}{du}=\frac{1}{\sqrt{f(u)}}=e^{B-A}. \label{ans.cor}
\ee
The mapping (eq. \ref{ans.cor}) holds only for $f(u) \geq 0$. According to the domain wall frame ansatz $f(u) < 0$ corresponds to the interior of a black hole horizon, as the radial $g_{rr}$ and time $g_{tt}$ components of the metric shift sign. We are interested in RG flows which terminate on IR and UV fixed points which according to the standard AdS/CFT prescription correspond to the black hole horizon, $g_{tt} \to 0$ and the spacetime boundary, $g_{tt} \to \infty$. Therefore, this mapping covers the part of the spacetime which is interesting for us.

The equations of motion for the domain wall frame ansatz (eq. \ref{ansatz2}) are:
\bs \be
\frac{d}{du}\left( e^{(d-2) \tilde{A}} Z \dot{A}_t \right)=0,\label{em1.a2} \ee
\be
2(d-1)\ddot{\tilde{A}}+\dot{\phi}^2=0,\label{em2.a2} \ee
\be
\ddot{f}+d \dot{\tilde{A}} \dot{f}- e^{-2 \tilde{A}} Z \dot{A}_0^2=0,\label{em3.a2} \ee
\be
(d-1) \dot{\tilde{A}} \dot{f}+ \left( d(d-1)\dot{\tilde{A}}^2 -\frac{1}{2} \dot{\phi}^2 \right) f -V+ \frac{1}{2} e^{-2 \tilde{A}} Z \dot{A}_0^2=0,\label{em4.a2}
\ee \label{eq.mot.an2} \es
where $\dot{} \equiv d/du$.

It is straight forward to show that the equations of motion (eq. \ref{eq.mot.an2}) reduce to the equations (\ref{eqmot}), by using (eq. \ref{ans.cor}). This showcases that the equations of motion (eq. \ref{eq.mot.an2}) yield the same solutions as the (eq. \ref{eqmot}).

The integral of equation (\ref{em1.a2}), with the introduction of (eq. \ref{ans.cor}) reduces to
\be
A'_t=\frac{q}{Z} e^{A-(d-1)B},
\ee
($' \equiv d/dr$), where $q$ is an integration constant we define as the charge density. This equation is identical to (eq. \ref{emgf}).

The equations (\ref{em2.a2}) to (\ref{em4.a2}) with the introduction of (eq. \ref{ans.cor}) reduce to
\bs
\be
A'B'-B''-(B')^2=\frac{1}{2(d-1)} (\phi')^2,\label{eq2.a2.new} \ee
\be
A''+\frac{1}{2(d-1)}(\phi')^2+(A')^2-(3-d)A'B'+(2-d)(B')^2=\frac{q^2}{2Z}e^{-2(d-1)B}, \label{eq3.a2.new} \ee
\be
(d-2)(d-1)(B')^2+2(d-1)A' B'-\frac{1}{2} (\phi')^2=V-\frac{q^2}{2Z}e^{-2(d-1)B}. \label{eq4.a2.new}
\ee \es
The (eq. \ref{eq2.a2.new}) can be immediately identified as the (eq. \ref{eqmot3}). The same is true for (eq. \ref{eq4.a2.new}) which can be identified as (eq. \ref{eqmot1}).

Finally, by adding (eq. \ref{eq3.a2.new}) with (eq. \ref{eq4.a2.new}) we obtain
\be
A''-\frac{1}{2}\frac{d-2}{d-1}(\phi')^2+(A')^2+(d-2)^2 (B')^2+(3d-5)A'B'=V,
\ee
which is identified as the (eq. \ref{eqmot2}).
This concludes the proof that the equations of motion in both frames are identical.

\section{Solutions for constant $\phi=\phi^*$} \label{sec:constant_phi}

According to the holographic prescription $\phi$ is dual to the coupling of a scalar operator in the dual field theory. By having $\phi$ equal to a constant $\phi=\phi^*$, the RG equations become trivial as the $\beta$-functions vanish in which case the scalar operator coupling is independent of the energy scale, such finite couplings are a rare occurrence in field theory. In the following we will extract the solutions of (eq. \ref{eq.mot.an2}) in this particular case.

The first equation (eq. \ref{em1.a2}) can be solved for $\dot{A}_t$ yielding
\be
\dot{A}_t=\frac{q}{Z(\phi)} e^{-(d-2)\tilde{A}}, \label{At_res1}
\ee
where $q$ is an integration constant which is interpreted as the charge density. The second equation, (eq. \ref{em2.a2}) dictates that
\be
\tilde{A}(u)=\tilde{A}_0 +\tilde{A}_1 u, \label{A.tilde_sol}
\ee
where $A_0$, $A_1$ are integration constants. The $A_0$ can be absorbed into a rescaling of the $x_i$ and $t$ coordinates and therefore we can safely set it to zero. By using (eq. \ref{A.tilde_sol}), as well as, (eq. \ref{At_res1}) we can verify that (eq. \ref{em3.a2}) is the derivative of (eq. \ref{em4.a2}). Therefore, the final equation that needs to be solved, (eq. \ref{em4.a2}), is a linear differential equation for $f$ that reads
\be
\dot{f}+ d \tilde{A}_1 f = \frac{1}{(d-1) \tilde{A}_1 } \left[V(\phi^*)+ \frac{q^2}{2 Z(\phi^*)} e^{-2(d-1)\tilde{A}_1 u} \right].
\ee
Which yields the solution
\be
f(u)=\frac{e^{-d \tilde{A}_1 u}}{(d-1) \tilde{A}_1^2} \left[ m + \frac{1}{d} V(\phi^*) e^{d \tilde{A}_1 u} - \frac{q^2 e^{-(d-2) \tilde{A}_1 u}}{2 (d-2) Z(\phi^*)} \right], \label{f.tilde_sol}
\ee
The solution found (eq. \ref{A.tilde_sol}) and (eq. \ref{f.tilde_sol}) is the AdS-Reissner-Nordstr\" om black hole, where $m$ is an integration constant which is the mass parameter of the black hole. This black hole depending on the values of $Z(\phi^*)$ and $V(\phi^*)$ may have no, one or two horizons, as well as, an AdS boundary. Its holographic properties have been investigated in \cite{Peca:1998cs,Konoplya:2008rq,Chamblin:1999hg,Chamblin:1999tk}. The fact that this black hole is a solution of the EMD action (eq. \ref{EMDaction}) for constant scalar field $\phi=\phi^*$ means that in the dual description the scalar coupling is independent of the renormalization scales.

\section{Reduction of equations of motion for running $\phi$} \label{sec:running_phi}

Having verified that the equations of motion of the domain wall frame ansatz (eq. \ref{ansatz2}) map to the corresponding ones for (eq. \ref{Ansatz}) and discussed the case of constant $\phi$ we can get occupied with the solution of the equations of motion (eq. \ref{eq.mot.an2}) for running $\phi$. The solution of those equations was first done in \cite{Kiritsis:2012ma}. In the sequel we will reformulate this procedure in order to cast them in a useful form for the extraction of the RG equations.

The first equation (eq. \ref{em1.a2}) can be solved for $\dot{A}_t$ yielding
\be
\dot{A}_t=\frac{q}{Z(\phi)} e^{-(d-2)\tilde{A}}, \label{At_res}
\ee
where $q$ is an integration constant which is interpreted as the charge density.

The second equation (eq. \ref{em2.a2}) can also be integrated. To show this, we introduce the superpotential, $W$
\be
\dot{\phi}=W', \label{superpotential}
\ee
where $' \equiv d/d\phi$. Then by a straightforward application of the chain rule and an integration of the $\phi$ variable the (eq. \ref{em2.a2}) is reduced to
\be
\frac{d \phi}{d \tilde{A}}=-2(d-1) \frac{W'}{W} . \label{beta_res}
\ee
As we will claim in the following this will be one of the RG equations, the function $\tilde{A}(u)$ which according to (eq. \ref{ans.cor}) is mapped to the function $B(r)$ is one of the RG scale factors of theory.
In order to have a well defined renormalization scheme we should define a reference RG scale, we choose $\ln M=\tilde{A}(\phi_0)$, where $\phi_0$ is an arbitrary value of the scalar field $\phi$, which is dual to source of a scalar operator in the dual field theory.

The third equation with the substitution of (eq. \ref{At_res}), (eq. \ref{beta_res}) and (eq. \ref{superpotential}) reduces to
\be
W' \left(W' f' \right)' - \frac{d}{2(d-1)} W W' f' = \frac{q^2}{Z} e^{-2(d-1)\tilde{A}}.
\ee
with the use of (eq. \ref{beta_res}) the equation above can be integrated once to obtain
\be
W' f'= e^{- d \tilde{A}} \left[D + q^2 \int_{\phi_0}^{\phi} d \tilde \phi ~\frac{e^{-(d-2)\tilde{A}}}{ Z W' } \right], \label{W.eq2}
\ee
where $D$ is an integration constant which depends on the temperature $T$, entropy $S$ and charge density, $q$ as \cite{Kiritsis:2012ma}
\be
D= - 4 \pi T S M^{(d-1)} - q^2 \int_{\phi_0}^{\phi_h} d \tilde \phi ~\frac{e^{-(d-2)\tilde{A}}}{ Z W' }, \label{D_defin}
\ee
where $\phi_h$ is the value of the scalar field at the black hole horizon and $\phi_0$ is the value of scalar field in the UV. Since we are interested in $T=0$ field theory for finite charge density there should not be a horizon and consequently we can set $T=0$ and $S=0$ to the equation (eq. \ref{D_defin}). Furthermore, the value of the scalar field in the IR will saturate to a value $\phi_{IR}$ and consequently $D$ is expressed as
\be
D=  - q^2 \int_{\phi_0}^{\phi_{IR}} d \tilde \phi ~\frac{e^{-(d-2)\tilde{A}}}{ Z W' }. \label{D_tzero}
\ee
For compactness we will nullify $D$ from our expressions by combining it with the integral in expression (\ref{W.eq2}).

The last equation (eq. \ref{em2.a2}) with the introduction of the previous results (eq. \ref{At_res}), (eq. \ref{beta_res}), (eq. \ref{superpotential}) and (eq. \ref{W.eq2}) reduces to an algebraic equation for $f$ that reads
\be
f= \frac{2 V - \frac{q^2}{Z} e^{-2(d-1)\tilde{A}} + q^2 W e^{- d \tilde{A}} \int_{\phi_{IR}}^{\phi} d \tilde \phi ~ e^{-(d-2)\tilde{A}} \left( Z W' \right)^{-1}}{  \frac{d~ W^2}{2(d-1)} - \left( W' \right)^2 }. \label{f.eq}
\ee
It is also instructive to write the metric component $\tilde A$  in terms of the superpotential (see eq. \ref{beta_res})
\be
\tilde A=\frac{-1}{2(d-1)} \int_{\phi_0}^{\phi} d\tilde\phi \frac{W}{W'}. \label{alpha_tilde}
\ee

Therefore the solution of the equations of motion (eq. \ref{eq.mot.an2}) have been reduced to the solution of an non-linear integro-differential equation for the superpotential, $W$, (eq. \ref{W.eq2}), via which all the metric and field components are expressed (see eq. \ref{f.eq}, \ref{superpotential}, \ref{alpha_tilde}).

Collecting all of the above and turning our attention to the initial problem we can express the momentum scale $\beta$-function as
\be
\beta_P(\phi)=- {2(d-1)}\frac{W'}{W}, \label{RG.eq1}
\ee
and the energy scale $\beta$-function as
\be
\beta_E(\phi)=\frac{\beta_P}{1+ \beta_P \frac{f'}{2 f}}. \label{RG.eq2}
\ee
And the solution of the equations of motion (eq. \ref{eq.mot.an2}) have been reduced to the evaluation of the superpotential, $W$, which satisfies the integro-differential equation

\be
f'= q^2 \frac{e^{\frac{d}{2(d-1)} \int_{\phi_0}^{\phi} d\tilde\phi \frac{W}{W'} }}{M^{2(d-1)}~W'}  \int_{\phi_{IR}}^{\phi} \frac{d \tilde\phi}{Z~W' } ~ e^{\frac{d-2}{2(d-1)} \int_{\phi_0}^{\phi} d\tilde\phi \frac{W}{W'}}. \label{integrodiff}
\ee
In the above expressions (eq. \ref{RG.eq2}) and (eq. \ref{integrodiff}) the function $f$ which can be expressed in terms of $W$ by the algebraic equation
\be
\begin{split}
f=&\frac{1}{  \frac{d~ W^2(\phi)}{2(d-1)} - \left( W' \right)^2 } \bigg[ 2 V -\frac{q^2}{ M^{2(d-1)} Z} e^{\int_{\phi_0}^{\phi} d\tilde \phi \frac{W}{W' }} +\\
&+ q^2 \frac{W e^{\frac{d}{2(d-1)} \int_{\phi_0}^{\phi} d\tilde \phi \frac{W}{W'} }}{M^{2(d-1)}}  \int_{\phi_{IR}}^{\phi} \frac{d \tilde \phi}{Z W'} ~ e^{\frac{d-2}{2(d-1)} \int_{\phi_0}^{\phi} d \tilde{\tilde{\phi}} \frac{W}{W'} } \bigg] ,
\end{split} \label{f_algeq}
\ee
was not substituted for the sake of compactness.

From the  above mentioned results one can conclude that the non-relativistic $\beta_P$-function (eq. \ref{RG.eq1}) retains the same dependence on the superpotential as its relativistic counterpart $\beta(\phi)$ (see \cite{Bourdier:2013axa}), while the beta function $\beta_E$ (eq. \ref{RG.eq2}) shifts away from its relativistic value $\beta_E=\beta_P=\beta$, as a function of $f(W)$ (eq. \ref{f_algeq}) which is, in turn, determined by the charge density $q$, the scalar potential $V(\phi)$ and the coupling between the scalar and gauge fields, $Z(\phi)$. The two $\beta$-functions are not independent from one another which is an indication that the two considered scales are also non-independent.

\section{Case study: The relativistic $\beta$-function} \label{sec:red_to_rel_b}

We devote this section in showing that the non-relativistic RG equations (eq. \ref{RG.eq1}, \ref{RG.eq2}, \ref{integrodiff}), worked out in section \ref{sec:running_phi}, reduce to the relativistic ones in the case of zero charge density and temperature. With the assumption of zero temperature and charge density ie. $q=0$ and $D=0$ (see eq. \ref{D_defin}) the equations for $f$ (eq. \ref{f_algeq}) and $W$ (eq. \ref{W.eq2}) reduce to
\be
f(\phi) = \frac{2 V}{  \frac{d~ W^2}{2(d-1)} - \left( W' \right)^2 }, ~~~ W' f' = 0.
\ee
We have already considered the case that $W' = \dot{\phi}=0$ in section \ref{sec:constant_phi}, where we have obtained a solution with $\beta(\phi)=0$. In this section we will consider $f' = 0$. In this case the value of $f(\phi)$ is constant, the value of this constant can be absorbed by a rescaling of the time coordinate $t$ and in that sense we are allowed to fix its value to $f(\phi)=1$. With this choice the equations (\ref{RG.eq1}, \ref{RG.eq2}, \ref{integrodiff}) simplify to
\be
\beta_P=- \frac{2(d-1)}{W} W', \label{NR.eq1}\ee
\be
\beta_E=\beta_P=\beta, \label{NR.eq2} \ee
\be
2 V=\frac{d~ W^2}{2(d-1)} - \left( W' \right)^2 \label{NR.eq3}
\ee
By differentiating (eq. \ref{NR.eq3}) and using (eq. \ref{NR.eq1}) we can reduce this equations to
\be
\beta(\phi)=-(d-1)\frac{d}{d\phi} \left(\frac{\beta^2(\phi) - 2d(d-1)}{V(\phi)} \right),
\ee
which is the non-linear differential equation that the relativistic holographic $\beta$-function satisfies, that was first found in \cite{Bourdier:2013axa}.

\section{Relevant asymptotic EMD gravity solutions} \label{sec:asympotic_EMD}

Asymptotic solutions for the EMD action (eq. \ref{EMDaction}) have been already examined. The most notable ones are the hyperscaling violating solutions and the Lifshitz invariant ones. In order to obtain those solutions one considers that the  couplings scale with $\phi$ as
\be
V(\phi)=2 \Lambda ~e^{-\delta \varphi},\sp Z(\varphi)=e^{\gamma \varphi}, \label{scaling}
\ee
in the IR asymptotic limit. Those solutions do not correspond to a complete RG flow as they have no AdS boundary that we could identify as the UV fixed point of the dual field theory description. However these kind of solutions can provide us with some insight on how the beta functions and the couplings behave in the dual field theory.

The properties of hyperscaling violating solutions are discussed in appendix \ref{sec:Hyperscaling_violation}. The derivation of the general hyperscaling violating solution that EMD gravity admits \cite{cgkkm,gk1} is presented in appendix \ref{sec:Hyperscaling_solutions}.

The Lifshitz solution, stemming from the general hyperscaling violating one (eq. \ref{sol_hv_ex.1} to \ref{sol_hv_ex.5}) for $\theta=0$, is written in the metric coordinates (eq. \ref{Ansatz}) as
\be
\begin{split}
ds^2&=-e^{-\frac{2 z r}{\sqrt{B_0}}} dt^2+dr^2+e^{-\frac{2 r}{\sqrt{B_0}}} \left( dx^i dx_i \right),\\
A_0=& \frac{(z-1)^{3/2}}{z+d-1} \sqrt{\frac{\Lambda}{z+d-2}} e^{\frac{z+d-1}{(z-1)\sqrt{B_0}}r \mp \phi_0 \sqrt{\frac{d-1}{2(z-1)}}},\;\;\; A_i=0,\\
\phi_\pm(r)&=\phi_0 \pm \sqrt{\frac{2(d-1)(z-1)}{B_0}} r, \; B_0=\frac{(z+d-1)(z+d-2)}{2 \Lambda},\\
&z=1+\frac{2(d-1)}{\gamma^2},\;\; \delta=0.
\end{split} \label{Lifshitz}
\ee
where $z$ is the Lifshitz exponent. The $\beta$-functions in this asymptotic limit are
\be
\beta_E(\phi)=\mp \frac{\sqrt{2(d-1)(z-1)}}{z}, ~~ \beta_P(\phi)=\mp \sqrt{2(d-1)(z-1)},~~ d \ln M_E = z ~ d \ln M_p,
\ee
where $M_E$ and $M_E$ the energy and momentum scale of the dual field theory, keep in mind that only $z \geq 1$ Lifshitz space times are well behaving (see criteria \ref{cond1} to \ref{cond6} in section \ref{sec:Hyperscaling_violation}). This means that the UV asymptotics are obtained for $r \to -\infty$ and the IR asymptotics are obtained for $r \to +\infty$. The UV asymptotic values of couplings $\varphi_\pm$ can be obtained from the UV value of the scalar fields by ignoring the divergent linear part. In the case $\phi_+$ the UV value of the coupling $\varphi_+ \xrightarrow{UV} +\sqrt{\frac{2(d-1)(z-1)}{B_0}}$ is repulsive ($\beta_E,\beta_p<0$). In the case $\phi_-$ the UV value $\varphi_- \xrightarrow{UV} +\sqrt{\frac{2(d-1)(z-1)}{B_0}}$ is attractive ($\beta_E,\beta_p>0$).

The Hyperscaling violating solution, can be written in the coordinates (eq. \ref{Ansatz}) as
\be
\phi_{\pm}=\phi_0 \pm \frac{1}{\theta} \sqrt{2(d-1)(\theta+d-1)\left[ (d-1)(z-1)-\theta \right]} \ln \left( \frac{\left| \theta \right| r}{(d-1)\sqrt{B_0}}  \right), \label{scalar_hv_rsol}
\ee
\be
A_t=q_{\pm} \left( \frac{\left| \theta \right| r}{(d-1)\sqrt{B_0}} \right)^{-\frac{(d-1)^2(z+(d-1)-\theta)}{\theta \left[ \theta-(d-1)z+(d-1) \right]}} \sp A_r=A_i=0,
\ee
\be
q_{\pm}=\frac{(d-1)(z-1)-\theta}{(d-1) \left[ z+(d-1)-\theta \right]} \sqrt{\frac{\Lambda \ell^2 (z-1)}{z+(d-2)-\theta}} e^{\pm \phi_0 \frac{(d-3) \theta -(d-1)^2}{\sqrt{2 (d-1) (d-\theta -1) \left[ (d-1) (z-1)-\theta \right]}}}.
\ee
\be
ds^2=-\left( \frac{\left| \theta \right| r}{(d-1)\sqrt{B_0}} \right)^{2-\frac{2z (d-1)}{\theta}} dt^2 +dr^2 +\left( \frac{\left| \theta \right| r}{(d-1)\sqrt{B_0}} \right)^{2-\frac{2 (d-1)}{\theta}} dx_i dx^i, \label{metric_hv_rsol}
\ee
\be
B_{0,\pm}=\frac{\left[ z+(d-1)-\theta \right] \left[ z+(d-2)-\theta \right]}{2 \Lambda \ell^2} e^{\pm \phi_0 \frac{\sqrt{2} \theta}{\sqrt{2 (d-1) (d-\theta -1) \left[ (d-1) (z-1)-\theta \right] }}}.
\ee
\be
\theta=\frac{\delta (d-1)^2}{\gamma +(d-2) \delta} \sp z= 1 + \frac{(d-1)\left[2 -(\delta-\gamma)\delta \right]}{(\gamma-\delta)\left[ \gamma+(d-2)\delta \right]}.
\label{exponents_rsol}
\ee
where $z$ the Lifshitz exponent, $\theta$ is the hyperscaling violation exponent and $r \geq 0$. The beta functions in this asymptotic case are more involved than before. Their form is presented below
\be
\begin{split}
\beta_E(\phi)=\pm \frac{\sqrt{2(d-1)(\theta+d-1)\left[(d-1)(z-1)-\theta \right]}}{\theta-z(d-1)},\\
\beta_P(\phi)=\pm \frac{\sqrt{2(d-1)(\theta+d-1)\left[(d-1)(z-1)-\theta \right]}}{1-d},\\
d \ln M_E = \left( z- \frac{\theta}{d-1} \right) ~ d \ln M_p.
\end{split}\label{beta.hyper}
\ee
The running of the couplings $\varphi_\pm$ is more involved than before. The IR and UV limits are identified by identifying the boundary and horizon of the metric (eq. \ref{metric_hv_rsol}). Then the asymptotic value of couplings $\varphi_\pm$ can be derived for the scalar field near the boundary $\phi_\pm$ by using (eq. \ref{scalar_hv_rsol}) and ignoring the divergent logarithmic part. Finally, the behavior of the fixed points (attractive or repulsive) can be identified by evaluating the sign of the $\beta_E$ and $\beta_P$-function (eq. \ref{beta.hyper}). We observe that for $\theta -z(d-1)>0$ the energy and momentum scales are inversely proportional. This poses the question on whether ``saddle points'' appear in the RG flow that is asymptotically hyperscaling violating. The saddle points are attractive when the momentum dependence, $M_p$, is examined and repulsive when the energy dependence, $M_E$, is examined or vice-versa. We identify four different possible behaviors of the couplings in the UV limits, based on (eq. \ref{metric_hv_rsol}, \ref{scalar_hv_rsol} and \ref{beta.hyper}). Those are summarized in the following table
\be
\begin{array}{| c | c | c | c | c | c | c |}
\hline
\# & \theta & \theta -z (d-1) & r & \lim \varphi_\pm & M_E & M_P \\ \hline
a &+	& + 	& \infty & \pm  & \text{att.} & \text{rep.}\\
b &+ 	& - 	& 0	     & \mp  & \text{att.} & \text{att.}\\
c &- 	& + 	& \infty & \mp  & \text{rep.} & \text{att.}\\
d &- 	& - 	& 0      & \pm  & \text{rep.} & \text{rep.}\\
\hline
\end{array} \label{hv_asympt_cases}
\ee
The first column assigns a label to each of the cases. The following two columns show the signs of $\theta$ and $\theta -z (d-1)$ respectively for each of the cases. Finally, in the last four columns we show the position of the UV asymptotics (column $r$), the sign of the asymptotic value of the couplings $\varphi_\pm$ (column $\lim \varphi_\pm$) , and whether the point is attractive or repulsive for both scales $M_e$ and $M_p$ (columns $M_e$ and $M_p$ respectively). Positive signs are denoted by $+$ and negative by $-$. Attractive behavior is denoted by ``att'' and repulsive by ``rep''.

By examining the ``saddle point'' cases, $a$ and $c$ (eq. \ref{hv_asympt_cases}), we can verify that they are excluded as unphysical because they violate the constraints (eq. \ref{constraint235}).

We conclude that, in the case of hyperscaling violating solutions we can realize both attractive or repulsive, UV fixed points where the dual field theory has positive or negative couplings by appropriately selecting the parameters $\delta$ and $\gamma$. Those can be mapped to the corresponding hyperscaling violating parameters $z$ and $\theta$ via (eq. \ref{exponents_rsol}) and the increasing or decreasing case for the scalar field ($\phi_+$ and $\phi_-$ respectively) is selected by the sign of $\delta-\gamma$. We summarize those findings in the following equation
\be
\begin{split}
\delta \left[ \gamma+(d-2)\delta \right]>0, \delta-\gamma>0 : \text{ attractive }  \varphi \xrightarrow{UV} \varphi<0, \\
\delta \left[ \gamma+(d-2)\delta \right]<0, \delta-\gamma>0 : \text{ repulsive }  \varphi \xrightarrow{UV} \varphi<0, \\
\delta \left[ \gamma+(d-2)\delta \right]>0, \delta-\gamma<0 : \text{ attractive }  \varphi \xrightarrow{UV} \varphi>0, \\
\delta \left[ \gamma+(d-2)\delta \right]<0, \delta-\gamma<0 : \text{ repulsive }  \varphi \xrightarrow{UV} \varphi>0.
\end{split}
\ee

\chapter{RG flow near a  holographic Lifshitz fixed point.} \label{hp}

\indent We intend to study the backreaction of the Lifshitz metric when perturbed by a relevant operator. We can obtain a model for this by considering Einstein gravity coupled with a massive gauge field which is perturbed by a dilaton field. The action for this model reads
\be
S=\int dt d^{d} x \sqrt{-g} \left[R -\frac{1}{2}\pa_\mu \phi \pa^\mu \phi +V(\phi)-\frac{1}{4} \left(F^{\mu \nu}F_{\mu \nu}+m^2 A^2\right) \right], \label{massive.gf}
\ee
where the potential of the dilaton field $V(\phi)$ can be expanded in even powers of $\phi$ as
\be
V(\phi)= \frac{d(d-1)}{\ell^2} -\frac{M^2}{2} \phi^2 + \frac{g}{4!} \phi^4 + \mathcal{O}\left( \phi^6 \right). \label{dil_pot}
\ee

For vanishing $\phi$ there is a well known Lifshitz solution first studied in \cite{Taylor:2008tg}
\be
ds^2=-{dt^2\over r^{2z}}+B_0{dr^2\over r^2}+{dx^idx^i\over r^2}, ~~~ A_0(r)=\frac{1}{r^z}\sqrt{\frac{2(z-1)}{z}},
\ee
where the $B_0$ and $z$ are given by the action parameters (eq. \ref{massive.gf}) as
\be
B_0=\frac{2 z (d-1)}{m^2}~~\text{and}~~m^2 \ell^2= \frac{2 z d (d-1)^2}{z^2+(d-2)z+(d-1)^2} ,
\ee
where the cases $d=1$ and $d=2 \&\& z=1$ are excluded.

We introduce a perturbation to that solution by the introduction of a non-trivial scalar field that scales as $\phi\sim \sqrt{\epsilon}  r^{\zeta}$, near the Lifshitz boundary. The solution for the scalar field to leading order provokes a backreaction to order $\e$ to the metric and gauge field due to the kinetic $-\frac{1}{2}\pa_\mu \phi \pa^\mu \phi$ and mass $\frac{M^2}{2} \phi^2$ terms that contribute to the action (eq. \ref{massive.gf}). The correction of the metric and gauge field in turn provokes a backreaction to the scalar field of order $\e^\frac{3}{2}$, which is of the same order as the effect of the interaction $g \phi^4$ and so on. This process defines a self-consistent perturbation scheme for the complete theory around the Lifshitz background.
The ansatz for this perturbative expansion of the metric and fields to leading order reads
\be
\begin{split}
ds^2=-{dt^2\over r^{2z}} \left[ 1+\epsilon ~\tilde{g}_{tt}(r) \right]+B_0{dr^2\over r^2} \left[ 1+\epsilon ~\tilde{g}_{rr}(r) \right]+{dx^idx^i\over r^2} +\mathcal{O}\left( \epsilon^2 \right),\\
A_0(r)=\frac{1}{r^z}\sqrt{\frac{2(z-1)}{z}} \left[ 1+\epsilon ~\tilde{A}_0(r) \right] +\mathcal{O}\left( \epsilon^2 \right), ~~~ \phi= \sqrt{\epsilon} ~\tilde{\phi} (r) +\mathcal{O}\left( \epsilon^\frac{3}{2} \right).
\end{split} \label{metric.pert}
\ee

The equations of motion stemming from the variational principle (eq. \ref{massive.gf}) read
\be
\begin{split}
2 r (z -1) \tilde{A}_0' -(d-1) r \tilde{g}_{rr}' +(z-1) (z+d-1) \left( \tilde{g}_{tt} -2 \tilde{A}_0 \right)\\
+(z(z-1)+d(d-1)) \tilde{g}_{rr} -\frac{1}{2} \left( B_0 M^2 \tilde{\phi}^2 +r^2 \tilde{\phi} '^2 \right)=0 \label{pert_eq1}
\end{split}
\ee
\be
\begin{split}
2 r (z -1) \tilde{A}_0' +(d-1) r \tilde{g}_{tt}' +(z-1)(z-d+1) \left( \tilde{g}_{tt} -2 \tilde{A}_0 \right)\\
+ (z+d-1)(z+d-2) \tilde{g}_{rr} -\frac{1}{2} \left( B_0 M^2 \tilde{\phi}^2-r^2 \tilde{\phi} '^2 \right)=0 \label{pert_eq2}
\end{split}
\ee
\be
2 r^2 \tilde{A}_0'' -2 (d+z-2) r \tilde{A}_0' +z \left[ r \tilde{g}_{tt}'+ r \tilde{g}_{rr}' -2 (d-1) \tilde{g}_{rr} \right]=0 \label{pert_gauge}
\ee
\be
r^2 \tilde{\phi}''-(d+z-2) r \tilde{\phi}'-B_0 M^2 \tilde{\phi}=0. \label{eq_scalar_field}
\ee

Before we continue with the presentation of the solutions to the equations of motion (eq. \ref{pert_eq1}) to (eq. \ref{eq_scalar_field}) we briefly sketch the method we used in order to solve them. The system of equations (eq. \ref{pert_eq1}) to (eq. \ref{eq_scalar_field}) is invariant under a rescaling $r \to \lambda r$ and therefore they  can be reduced to a system of differential equations with constant coefficients by the substitution $r= \ln(u)$. The equation of motion for the scalar field (eq. \ref{eq_scalar_field}) is independent from the rest of equations of motion. This allows us to solve (eq. \ref{eq_scalar_field}) separately and in the following substitute the solution to (eq. \ref{pert_eq1}) and (eq. \ref{pert_eq2}), this effectively cancels the non linear way that the latter equations depend on (eq. \ref{eq_scalar_field}). This process yields a non-homogeneous system of linear differential equations which can be solved with standard methods. Details about the solution of this system are presented in the appendix \ref{sec:Linearization}.

For $M^2 > - \frac{(z+(d-1))^2}{4 B_0}$ the solution for the dilaton field satisfies the BF bound and it reads
\be
\phi(r)=\phi_- r^{ \frac{1}{2} \alpha_-} + \phi_+ r^{ \frac{1}{2} \alpha_+} \; \text{where}\;  \alpha_\pm= z+(d-1) \pm \sqrt{\left( z+(d-1) \right)^2+ 4 B_0 M^2}.
\ee
Note above that $\alpha_{-}>\alpha_{+}$. Therefore, the leading behavior near the boundary is
\be
\phi(r)=\phi_- r^{ \frac{1}{2} \alpha_-}+\cdots
\label{p0}\ee
Consequently the coefficient $\phi_-$ is a source that corresponds to the appropriate coupling constant of the dual field theory that multiplies the operator ${\cal O}_{\phi}$ which dual to the scalar field $\phi$ in the bulk
\be
S_{QFT}=S_{*}+\int dt ~d^{d-1}x~ \phi_-~{\cal O}_{\phi}(t,\vec x).
\ee
If the (mass) scaling dimension of ${\cal O}_{\phi}$ is $\Delta_{\phi}$, then the dimension of $\phi_-$ is
\be
[\phi_-]=z+d-1-\Delta_{\phi}.
\ee
From (eq. \ref{p0}) we obtain
\be
\Delta_{\phi}=z+d-1-{1\over 2}\alpha_{-}={1\over 2}\alpha_{+},
\ee
where we have used the fact that the bulk scalar field $\phi$ is dimensionless.

The corresponding $\beta$ functions can be obtained by coordinate transforming to the frame $g_{rr}=1$ (eq. \ref{Ansatz}) and using the definitions for the $\beta$ functions (eq. \ref{betafunctions}) the datails of this process are provided in appendix \ref{sec:der_hol_beta}.
Our calculation of the leading backreaction to the gravitational part of the action (eq. \ref{massive.gf}) allows for the determination of the $\beta$ functions up to order $\mathcal{O} (e^2)$ which according to (eq. \ref{metric.pert}) corresponds to $\mathcal{O} (\phi^4)$.
By following that prescription we are able to extract the $\beta$-functions
\be
\begin{split}
\beta_E(\phi)\equiv &\frac{d \phi}{d A}=\frac{(z+d-1)-\Delta_\phi}{z} \phi \\
&- \frac{(z-1)(\Delta_\phi-z)\left[ \Delta_\phi -(d-1) \right] \left[\Delta_\phi-(z+d-1) \right] \phi^3}{4z(d-1) \left[ 2 \Delta_\phi ^2 -3 \Delta_\phi  (d+z-1) +z (3d-2)+(d-1)(d-2) \right]} +\mathcal{O}\left( \phi^4 \right), \label{betae}
\end{split}
\ee
\be
\beta_P(\phi)=\frac{d \phi}{d B}=\left[(z+d-1)-\Delta_\phi \right] \phi +\mathcal{O}\left( \phi^4 \right).
\label{betap}\ee
The first term that appears in the holographic $\beta$ functions is the tree-level contribution. This term appears due to the fact that the operator $\mathcal{O}_\phi$ is dimensionfull. The second term in (eq. \ref{betae}) is the one corresponding to quantum corrections.

\chapter{The perturbative $\beta$-functions of the interacting Lifshitz scalar field theory.} \label{sec:field_theory_treatment}

\section{General considerations}
\indent The minimal example of a (free) quantum field theory that exhibits Lifshitz scaling symmetry is the free Lifshitz scalar field. The action for such a field reads
\be
S_{QFT}=\frac{1}{2}\int dt d^{d-1}x\left[\dot\phi^2- \phi\square ^z \phi + m^2 \phi^2 \right],
\label{Lifshitz.free}\ee
where $\dot{\;}=\partial_0$. The mass dimension of each of the components of the above action are
\be
\left[ x \right]=-1 \sp \left[ t \right]=-z \sp \left[ \phi \right]=\frac{d-z-1}{2} \sp \left[ m \right]= z. \label{Lifshitz.symmetry}
\ee

In order to extract the $N$-point functions we introduce the sources $J(x,t)$ with vanishing boundary conditions, $J(x,t_i)=0=J(x,t_f)$. We also introduce the generating functional of the field theory
\be
Z[J]=\int \mathcal{D} \phi~ e^{i S_{QFT}[J]}. \label{gen.func}
\ee
We can simplify this equation by performing the Gaussian integral in the exponent of (eq. \ref{gen.func}). The propagator for the free theory is
\be
\Delta_F \left(x,t \right)=\frac{1}{\left(2 \pi \right)^{d}} \int d\omega d^{d-1}k \frac{e^{-i \left(\mathbf{k} \cdot \mathbf{x} + \omega t \right)}}{ k ^{2z} - \omega^2 +m^2} .
\ee

We now consider the case of interacting theory
\be
S_{QFT}={1\over 2} \int dt d^{d-1}x\left[\dot\phi^2+\left( \pa_i^z\phi \right)^2 +m^2 \phi^2- \frac{g}{n!} \phi^n \right].
\label{Lifshitz.int}\ee
The interaction adds the additional Feynman rule that each vertex is weighted by $g$, the mass dimension of which is \be
[g]=\frac{n+2}{2}z-\frac{n-2}{2}\left( d-1 \right)\;.
 \ee
 Now we consider amputated 1 particle irreducible (1-PI) Feynman diagrams with $N$ external legs $I$ internal propagators, $L$ number of loops and $V$ vertices. In the interacting theory each internal line has two vertices at its endpoints and each external line has only one. The fact that each vertex has $n$ legs implies that
\be
n V= N + 2 I \label{vertices}
\ee
The total momentum (or for this case energy) that goes into a vertex is zero and thus introduces one constraint equation for the momenta (energies). Each internal lines carries momentum and energy and there is an additional momentum (energy) that comes from the external legs. For each loop  there is a momentum (energy) that is not constrained (and therefore integrated over). Therefore, the number of loops is equal to the number of unconstrained energies or momenta:
\be
L=I-V+1 \label{loops}
\ee
In the standard prescription of renormalization (momentum space cutoff method), we introduce UV cutoffs to the relativistic ($d$)-momentum in order to evaluate the diagrams. Here however, we are interested in the case of non-relativistic QFTs which are not Lorentz symmetric. In that context we introduce two cutoffs the energy cutoff $\Lambda$ and the momentum cutoff $\Lambda_p$, being the upper bounds of the energy and spatial momentum integrations  respectively. In the UV a particular divergent Feynman diagram scales with the cutoffs as $\sim \Lambda^{D_E} \Lambda_p^{D_p}$, where $D_E$ and $D_p$ are the energy and momentum superficial degrees of divergence. The momentum superficial degree of divergence is
\be
D_p=\left( d-1 \right) L -2 z I
\ee
and the superficial degree of divergence for the energy is
\be
D_E=L -2 I.
\ee
We can express the aforementioned degrees of divergence in terms of loops, $L$ and the parameters of the theory $z$, $d$ and $n$ as follows
\be
D_p=-2 z \frac{N-n}{n-2} + \left(d -1 - \frac{2 z n}{n-2} \right)L \sp D_E=-2 \frac{N-n}{n-2} + \left(1 - \frac{2 n}{n-2} \right)L.
\ee
We can obtain a criterion for renormalizability by utilizing $D_E$ and $D_p$ and demanding that their value is decreasing with the number of loops. This is always true for $D_E$. While for $D_p$ we obtain the condition
\be
\text{Condition $D_{p}$:~~~~ } n\leq \frac{2(d-1)}{d-z-1},
\ee
Because of (eq. \ref{Lifshitz.symmetry}) the dimensions of energy and momentum are connected by the relation $[E]=z[P]$. Therefore, we can obtain a power-counting superficial degree of divergence, $D_{PC}$, by subtracting the scaling dimension of the denominator from the scaling dimension of the numerator in a particular 1PI amputated diagram.
\be
D_{PC}=
\begin{cases}
(d+z)L-2z I & \text{for} ~ z \geq 1 \\
(d+z)L-2 I & \text{for} ~ z<1
\end{cases}.
\ee
We express the superficial degrees of divergence with respect to the number of loops $L$ and the external lines $E$ by using (eq. \ref{vertices}) and (eq. \ref{loops})
\be
D_{PC} = -2z \frac{N-n}{n-2} +\left(d -1 +z -2 z \frac{n}{n-2} \right) L.
\ee
Therefore there is a finite number of graphs that are superficially divergent and therefore the theory is said to be power-counting renormalizable in the case that
\be
\text{Condition $D_{PC}$: }\frac{d-1}{2z} \leq \frac{1}{2} \frac{n+2}{n-2}, \label{power_counting}
\ee
this condition could be also be obtained by demanding that the momentum dimension of the coupling constant is positive, $[g] \geq 0$.

In the following we examine the 1-loop renormalization of power-counting renormalizable scalar Lifshitz $\phi^4$ theories. First, though, we will review the well known example of the relativistic $\phi^4$ theory.

\section{Standard renormalization of $3+1$ dimensional $z=1$, $\phi^4$ theory}

Before continuing with the examination of non-relativistic Lifshitz theories, we will briefly examine the Lorentz invariant $z=1$, $d=3$, $g \phi^4$ theory in order to illustrate the approach to evaluating $\beta$ functions that we will use in the treatment of non-Lorentz invariant theories.

\begin{figure}[ht]
\begin{center}
\includegraphics[width=0.7\textwidth]{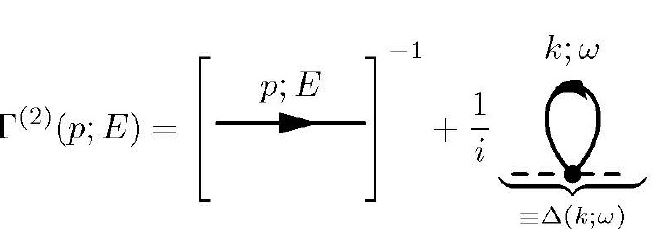}
\end{center}
\caption{The expression of the two-point proper vertex $\Gamma^{(2)}$ in terms of 1PI amputated diagrams up to one-loop order. Internal lines are represented as solid lines, the momentum and energy carried by each is given by the corresponding label. Interaction vertices are represented by dots and amputated propagators by dashed lines. The 1PI diagram referred to as $\Delta(k;\omega)$ in the main text is also defined.}
\label{fig:two_point_vertex}
\end{figure}

\begin{figure}[ht]
\begin{center}
\includegraphics[width=\textwidth,height=4cm]{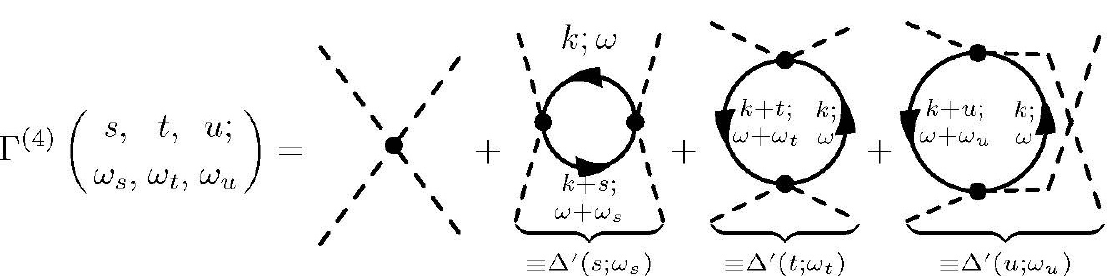}
\end{center}
\caption{The expression of the four-point proper vertex $\Gamma^{(2)}$ in terms of 1PI amputated diagrams up to one-loop order. Internal lines are represented as solid lines, the momentum and energy carried by each is given by the corresponding label. Interaction vertices are represented by dots and amputated propagators by dashed lines. The 1PI diagrams referred to as $\Delta'(k;\omega)$ in the main text are also defined.}
\label{fig:four_point_vertex}
\end{figure}

By employing power-counting arguments we can infer that up to one loop the divergent proper vertices are the $\Gamma^{(2)}$ and the $\Gamma^{(4)}$. The equations for those quantities are presented pictorially, via Feynman diagrams, in Fig. \ref{fig:two_point_vertex} and Fig. \ref{fig:four_point_vertex} respectively.

The diagram, $\Delta$ (see Fig. \ref{fig:two_point_vertex}) is expected to be quadratically divergent. Its amputated amplitude is equal to
\be
\Delta= \frac{g}{2} \int \frac{d^3k d\omega}{(2\pi)^4} \frac{1}{k^2 -\omega^2 +m^2}.
\ee
First we transform to Euclidean time by setting $t=i \tau$, that is equivalent to the substitution of $\omega \to i \omega$ in the previous expression.
\be
\Delta= \frac{i g}{2} \int \frac{d^3k d\omega}{(2\pi)^4} \frac{1}{k^2 +\omega^2 +m^2}.
\ee
In the standard one-scale renormalization we define the $3+1$ dimensional measure as $k_0=k^2 +\omega^2$ and the cutoff is introduced by imposing $k_0 < \Lambda$. This process applied to $\Delta$ yields
\be
\Delta= \frac{i g}{2 (2\pi)^4} \int_{S^3} d\Omega_3 \int^\Lambda_0 dk_0 \frac{k_0^3}{k_0^2 +m^2}=\frac{i g}{16 \pi^2} \int^\Lambda_0 dk_0 \frac{k_0^3}{k_0^2 +m^2},
\ee
where the purpose for not evaluating the $k$ integral will be made clear in the following. The proper vertex $\Gamma^{(2)}(p)$ for $\Lambda \gg m^2$ reads
\be
\Gamma^{(2)}(p;\omega)=p^2 -\frac{\omega^2}{2} +m ^2 + \frac{g}{16 \pi^2} \int^\Lambda_0 dk_0 \frac{k_0^3}{k_0^2 +m^2}. \label{gamma2}
\ee
The renormalized proper vertex $\Gamma_r^{(2)}(p;\omega)$ should not depend on the value of the cutoff. In order to achieve that the couplings $g$ and $m$ should depend on the cutoff (run) and cancel the dependence of the renormalized proper vertex. By the differentiation of the (eq. \ref{gamma2}) and by demanding $\frac{d}{d \Lambda} \Gamma_r^{(2)}(p;\omega)=0$  we obtain the following equation
\be
\frac{d m^2}{d \Lambda}+\frac{1}{32 \pi^2} \frac{d g}{d \Lambda} \int^\Lambda_0 dk_0 \frac{k_0^3}{k_0^2 +m^2} + \frac{g}{16 \pi^2} \frac{\Lambda^3}{\Lambda^2 +m^2} =0. \label{consistency1}
\ee
By inspecting the (eq. \ref{consistency1}) we can conclude that the second term is quadratic in the loop expansion, because the bare couplings $g$ and $m$ do not depend on the cutoff $\Lambda$ in the absence of loop corrections. We are therefore allowed to drop that term. Since we are interested in the asymptotic behavior for $\Lambda \gg m^2$ we can expand the last term in (eq. \ref{consistency1}) in powers of $m^2/\Lambda^2$ obtaining
\be
\frac{d m^2}{d \Lambda}=- \frac{g}{16 \pi^2} \Lambda \left[1 - \frac{m^2}{\Lambda^2} + \mathcal{O} \left( \frac{m^4}{\Lambda^4} \right) \right].
\ee
by multiplying this expression by $\Lambda/m^2$ we obtain the well known result for the $\gamma_m$ function
\be
\gamma_m = \frac{\Lambda}{m} \frac{d m}{d \Lambda}= - \frac{g}{32 \pi^2 m^2} \left(\Lambda^2 - m^2 \right). \label{gamma_phi_4}
\ee

We can renormalize the four-point proper vertex $\Gamma^{(4)}$, similarly to the aforementioned case. In order to achieve that, we need to regularize the diagram $\Delta'$, defined in Fig. \ref{fig:four_point_vertex} which is expressed as
\be
\Delta'= \frac{g^2}{2(2\pi)^4}\int \frac{d^3k d\omega}{(k^2 -\omega^2 +m^2)[(k+p)^2 -(\omega+E)^2 +m^2]}.
\ee
By using the Feynman integration formula
\be
\frac{1}{A B}=\int_0^1 dx~ \frac{1}{\left( x A+(1-x) B \right)^2}
\ee
and $k_0=k^2 +\omega^2$, $\Delta'$ reduces to
\be
\Delta'= \frac{i g^2}{2(2\pi)^4} \int_0^1 d x~\int_{S^3} d\Omega_3~ \int_0^{\Lambda} \frac{d k_0 k_0^3}{(k_0^2 +M^2)^2}=\frac{i g^2}{16 \pi^2}\int_0^1 dx \int_0^{\Lambda} \frac{d k_0 k_0^3}{(k_0^2 +M^2(p))^2},
\ee
where $M^2(p) \equiv m^2+x(1-x)p^2$. According to Fig. \ref{fig:four_point_vertex} the four-point vertex function reads
\be
\begin{split}
\Gamma^{(4)}(s,t,u;\omega_s,\omega_t,\omega_u)=-i g +\frac{i g^2}{16 \pi^2} \int_0^1 dx  \bigg\{ \int_0^{\Lambda} \frac{d k_0 k_0^3}{(k_0^2 +M^2(s))^2}+\\
\int_0^{\Lambda} \frac{d k_0 k_0^3}{(k_0^2 +M^2(t))^2} + \int_0^{\Lambda} \frac{d k_0 k_0^3}{(k_0^2 +M^2(u))^2} \bigg\},
\end{split}
\ee
where $s$, $t$, $u$ denote the Mandelstam variables, defined as $s=(p_1+p_2)^2 $, $t=(p_1+p_3)^2 $ and $u=(p_1+p_4)^2 $. Similarly as before the renormalized four-point proper vertex does not depend on the value of the cutoff, $\Lambda$. Therefore, by demanding $\frac{d}{d \Lambda} \Gamma_r^{(4)}(s,t,u;\omega_s,\omega_t,\omega_u)=0$ we obtain the equation
\be
-i \frac{d g}{d \Lambda} +\frac{i g^2}{16 \pi^2} \int_0^1 dx  \bigg\{ \frac{\Lambda^3}{(\Lambda^2 +M^2(s))^2}+
+\frac{\Lambda^3}{(\Lambda^2 +M^2(t))^2} + \frac{\Lambda^3}{(\Lambda^2 +M^2(u))^2} \bigg\}=0, \label{consistency2}
\ee
where we have omitted the quadratic term to the loop expansion by using the same argumentation as in the two-point vertex case. By multiplying the whole expression by $\Lambda$ and assuming $\Lambda \gg M^2$ we obtain the beta function of the theory
\be
\beta=\Lambda \frac{d g}{d \Lambda}= \frac{3 g^2}{16 \pi^2} + \mathcal{O} \left( \frac{M^2(s,t,u)}{\Lambda^2} \right). \label{beta_phi_4}
\ee
The equations (\ref{beta_phi_4}) and (\ref{gamma_phi_4}), conclude the textbook results for the $\beta$ and $\gamma_m$ of the relativistic $d=3$, $\phi^4$ theory \cite{Ryder,Peskin,Zee}.

\section{1-loop two-scale renormalization of the  Lifshitz $\phi ^4$ theories}

By power-counting we can conclude that the divergent proper vertices are the $\Gamma^{(2)}$ and $\Gamma^{(4)}$, being represented in Fig. \ref{fig:two_point_vertex} and Fig. \ref{fig:four_point_vertex} respectively. To begin with, we focus to the evaluation of the diagram $\Delta$
\be
\Delta= \frac{i g}{2} \int_0^{\Lambda_p} d^{d-1}k \int_0^\Lambda \frac{ d\omega}{(2\pi)^{d}} \frac{1}{k^{2 z} +\omega^2 +m^2}. \label{sunset_z}
\ee
The integral for $\omega$ is elementary and as such the diagram reduces to
\be
\Delta= \frac{i g}{(2 \sqrt{\pi})^{d} \Gamma \left( \frac{d}{2} \right)} \int_0^{\Lambda_p} dk \frac{k^{d-2}}{\sqrt{k^{2z}+m^2}} \arctan \left( \frac{\Lambda}{\sqrt{k^{2z}+m^2}} \right).
\ee
It turns out that we can nullify the $\Lambda$ dependence of the couplings by taking the well defined limit $\Lambda \to \infty$. We postpone the proof of this statement until after we obtain the $\Lambda_p$ dependence of $m^2$. The vertex function $\Gamma^{(2)}$ reads
\be
\Gamma^{(2)}(p;\omega)=p^2 -\frac{\omega^2}{2} +m ^2 + \frac{g \pi}{(2 \sqrt{\pi})^{d} \Gamma \left( \frac{d}{2} \right)} \int_0^{\Lambda_p} dk \frac{k^{d-2}}{\sqrt{k^{2z}+m^2}}.
\ee
Similarly as in the case examined before we demand that the two-point vertex function does not depend on the cutoff, $\Lambda_p$. Thus we obtain
\be
\begin{split}
\gamma^m_p=\frac{\Lambda_p}{m}\frac{d m}{d \Lambda_p}&= -\frac{g \pi}{2(2 \sqrt{\pi})^{d} \Gamma \left( \frac{d}{2} \right)} \sum_{n=0}^\infty \binom{-1/2}{n} m^{2(n-1)} \Lambda_p^{d-(2n+1)z-1}\\
&= -\frac{g \pi}{2(2 \sqrt{\pi})^{d} \Gamma \left( \frac{d}{2} \right)} \left( \frac{\Lambda_p^{d-z-1}}{m^2}+\frac{\Lambda_p^{d-3z-1}}{2}+\frac{3 m^2 \Lambda_p^{d-5z-1}}{8}+\dots \right). \label{gamma_phi_z}
\end{split}
\ee
It can be checked that, for $(d-1)=3$ and $z=1$, (eq. \ref{gamma_phi_z}) reproduces the results of (eq. \ref{gamma_phi_4}). In order to explicitly show that the mass coupling $m^2$ is independent of the energy cutoff scale, $\Lambda$, we perform the integral over $k$ in (eq. \ref{sunset_z}) and express the vertex function as
\be
\Gamma^{(2)}(p;\omega)=p^2 -\frac{\omega^2}{2} +m ^2 + \frac{g \pi}{(2 \sqrt{\pi})^{d} \Gamma \left( \frac{d}{2} \right)} \int_0^{\Lambda} d\omega \frac{\Lambda_p^{d-1} \;_2F_1 \left[1,\frac{d-1}{2z},1+\frac{d-1}{2z}; -\frac{\Lambda_p^{2z}}{\omega^2 +m^2} \right]}{(d -1) \left( \omega^2 +m^2 \right)}.
\ee
The two scales are independent in the sense that $\frac{d \Lambda}{d \Lambda_p}=0$ and therefore the differentiation over the energy cutoff, $\Lambda$, to one-loop order yields
\be
\begin{split}
\frac{d m^2}{d \Lambda}&= -\frac{g \pi}{(2 \sqrt{\pi})^{d} \Gamma \left( \frac{d}{2} \right)} \frac{\Lambda_p^{d-1} \;_2F_1 \left[1,\frac{d-1}{2z},1+\frac{d-1}{2z}; -\frac{\Lambda_p^{2z}}{\Lambda^2 +m^2} \right]}{(d -1) \left( \Lambda^2 +m^2 \right)}\\
&=-\frac{g \pi \Lambda_p^d }{d (2 \sqrt{\pi})^{d} \Gamma \left( \frac{d}{2} \right)} \Lambda^{-2} \left( 1 -\frac{m^2}{\Lambda^2} + \dots \right) \left( 1 - \frac{\left(\frac{d-1}{2z} \right)_1}{\left(1+\frac{d-1}{2z} \right)_1} \frac{\Lambda_p^{2 z}}{\Lambda^2} + \dots \right),
\end{split}
\ee
where $(a)_n$ is the Pochhammer symbol. For large $\Lambda$ the dependence of $m$ on the energy cutoff, $\Lambda$, is suppressed at least quadratically and, as such, the coupling $m^2$ does not depend on $m^2$ ie. $\gamma_E^m=0$. This argument justifies why we can can safely neglect the $\Lambda$ dependence of the vertex-function by taking the limit $\Lambda \to \infty$.

We proceed by evaluating the diagram, $\Delta'$, which is expressed as
\be
\Delta'= \frac{i g^2}{2(2\pi)^4} \int_0^{\Lambda_p} d^{d-1}k \int_0^\Lambda \frac{d\omega}{(k^2 +\omega^2 +m^2)[(k+p)^2 +(\omega+E)^2+m^2]}
\ee
By using the Feynmann integration formula and integrating over $\omega$ the diagram reduces to
\be
\Delta'= \frac{i g^2}{(2 \sqrt{\pi})^{d} \Gamma\left(\frac{d}{2} \right)} \int_0^1 dx \int_0^{\Lambda_p} \frac{dk k^{d-1}}{2 M^3(k,p)} \left[ \arctan\left(\frac{\Lambda}{M(k,p)} \right) + \frac{\Lambda M(k,p)}{M^2(k,p)+\Lambda^2} \right],
\ee
where $M^2(k,p)=m^2+x(1-x) E^2 +x (k+p)^{2 z} +(1-x) k^{2 z}$. As previously, we can nullify the $\Lambda$ dependence of the couplings by taking the well defined limit $\Lambda \to \infty$. The fact that this limit is well behaved means that the leading order dependence of the diagram $\Delta'$ on the scale $\Lambda$ satisfies $\Delta' \propto 1/\Lambda^n$, with $n>1$. The four point vertex function in this case reads
\be
\begin{split}
\Gamma^{(4)}(s,t,u;\omega_s,\omega_t,\omega_u)=-i g + \frac{i \pi g^2}{4 (2 \sqrt{\pi})^{d} \Gamma\left(\frac{d}{2} \right)}  \int_0^1 dx  \bigg\{ \int_0^{\Lambda_p} \frac{dk k^{d-2}}{M^3(k,s)}+\\
+\int_0^{\Lambda_p} \frac{dk k^{d-2}}{M^3(k,t)} + \int_0^{\Lambda_p} \frac{dk k^{d-2}}{M^3(k,u)} \bigg\},
\end{split} \label{4point_vertex}
\ee
where $s$, $t$, $u$ the Mandelstam variables. In the case that the coupling $g$ is dimensionfull, $d-3z-1 \neq 0$, the dimensionless coupling $\tilde{g}= g \Lambda_p^{d-3z-1}$ depends on the cutoff. To determine the terms in (eq. \ref{4point_vertex}) that contribute in the case $\Lambda_p \gg p$ we expanding the  appearing integrand in Taylor series with respect to $k$
\be
\begin{split}
\frac{k^{d-2}}{M^3(k,p)}&=k^{d-3z-2} \sum_{m=0}^{\infty} \binom{-3/2}{m} \left[ x \sum_{n=1}^{2z} \binom{2 z}{n} \left( \frac{p}{k} \right)^n +x(1-x) \frac{E^2}{k^{2z}} + \frac{m^2}{k^{2z}} \right]^m\\
&=k^{d-3z-2} -3 z x p k^{d-3z-3} + \frac{15 z^2 x^2 -3 z x (2z -1)}{2} p^2 k^{d-3z-4} + \mathcal{O} \left( k^{d-3z-5} \right)\\
&~~~-\frac{3}{2} \left(m^2 +x(1-x)E^2 \right) k^{d-5z-2} + \frac{15 z}{4} x^2 p \left(m^2 +x(1-x)E^2 \right) k^{d-5z-3} \\
&~~~ + \mathcal{O} \left( k^{d-5z-4} \right).
\end{split} \label{integrand}
\ee
In the case of power-counting renormalizable $\phi^4$ theories $d-3z-1 \leq 0$. By substituting the expansion (eq. \ref{integrand}) in (eq. \ref{4point_vertex}) and working with the dimensionless coupling $g$ we can verify that the subleading terms in the Taylor expansion are suppressed by additional negative powers of $\Lambda_p$. By demanding $\Lambda_p \frac{d}{d \Lambda_p} \Gamma^{(4)}(s,t,u;\omega_s,\omega_t,\omega_u)=0$ we obtain the corresponding $\beta_p$ function that reads
\be
\beta_P=\Lambda_p \frac{d \tilde g}{d \Lambda_p}=(d-3z-1) \tilde g - \frac{3 \pi \tilde g^2}{4(2\sqrt{\pi})^{d} \Gamma\left( \frac{d}{2} \right)} + \mathcal{O} \left( \frac{p}{\Lambda_p} \right) . \label{beta_p_z1}
\ee
we can obtain the $\beta_E$ function by defining the dimensionless coupling as
$\tilde{g}= g \Lambda^{\frac{d-3z-1}{z}}$ and following a similar process
\be
\beta_E=\Lambda \frac{d \tilde g}{d \Lambda}=\frac{1}{z} (d-3z-1) \tilde g - \frac{3 \pi \tilde g^2}{2 z (2\sqrt{\pi})^{d} \Gamma\left( \frac{d}{2} \right)} \left( \frac{\Lambda_p^z}{\Lambda} \right)^{\frac{d-3z-1}{z}}. \label{beta_e_z}
\ee
The first term in (eq. \ref{beta_p_z1}) and (eq. \ref{beta_e_z}) refers to the tree-level result for the beta function that stems form the fact that the corresponding operator $\phi^4$ is dimensionfull. The second terms provide the quantum correction obtained by perturbation theory.
In the dimensionless case $\tilde{g}= g$, $d-3z-1=0$ the $\beta$ functions read
\be
\beta_P=\frac{3 \pi g^2}{4(2\sqrt{\pi})^{d} \Gamma\left( \frac{d}{2} \right)} \sp \beta_E=0. \label{beta_phi_z}
\ee
It can be verified that for $z=1$ and $(d-1)=3$ the result of (eq. \ref{beta_phi_z}) is equivalent to the one obtained for (eq. \ref{beta_phi_4}). In that case the coupling $g(\Lambda)$ runs as
\be
g(\Lambda_p)=\frac{g(\Lambda_{p,0})}{1-\frac{3 \pi g \left( \Lambda_{p,0} \right)}{4(2\sqrt{\pi})^{d} \Gamma\left( \frac{d}{2} \right)} \ln \left( \frac{\Lambda_p}{\Lambda_{p,0}} \right)}.
\ee
Hence this class of theories are free in the IR, as $g \to 0$ for $\Lambda \to 0$, but exhibit a Landau pole (ie. become strongly coupled, $g \to \infty$ for finite cutoff values) at $\Lambda_p=\Lambda_{p,0} \exp\left( \frac{4(2\sqrt{\pi})^{d} \Gamma\left( \frac{d}{2} \right)}{3 \pi g \left( \Lambda_{p,0} \right)} \right)$. Such theories are considered trivial in the sense that in order to take the limit $\Lambda \to \infty$ we should accept $g \to 0$.

\section{Comparison with holography}
In order to compare with holography we consider the scale invariant case of the previously analyzed Lifshitz $\phi^4$ field theories
\be
S_{QFT}={1\over 2} \int dt d^{d-1}x\left[\dot\phi^2+\left( \pa_i^z\phi \right)^2 - \frac{g}{n!} : \phi^4 : \right],
\ee
where $:\phi^4:$ denotes that the interaction term is normal ordered. Here by normal ordering we impose the condition that all contractions of the interaction vertex do not contribute to the proper vertices. As a consequence, the diagram $\Delta$ does not contribute to $\Gamma^{(2)}(p;E)$ (see Fig. \ref{fig:two_point_vertex}) and thus the running of scalar mass with the renormalization scale is avoided up to 1-loop order. The $\beta$-functions (eq. \ref{beta_p_z1}) and (eq. \ref{beta_e_z}) do not depend on $m$ and thus retain the same dependence with the renormalization scale
\be
\begin{split}
\beta_P&=\Lambda_p \frac{d \tilde g}{d \Lambda_p}=(d-3z-1) \tilde g - \frac{3 \pi \tilde g^2}{4(2\sqrt{\pi})^{d} \Gamma\left( \frac{d}{2} \right)}, \\
\beta_E&=\Lambda \frac{d \tilde g}{d \Lambda}=\frac{1}{z} (d-3z-1) \tilde g - \frac{3 \pi \tilde g^2}{2 z (2\sqrt{\pi})^{d} \Gamma\left( \frac{d}{2} \right)} \left( \frac{\Lambda_p^z}{\Lambda} \right)^{\frac{d-3z-1}{z}}.
\end{split}
\ee

The holographically obtained beta functions in the case of a UV  Lifshitz invariant fixed point (eq. \ref{betap}) and (eq. \ref{betae}) read
\be
\beta_P(g)=\left[(z+d-1)-\Delta_g \right] g +\mathcal{O}\left( g^4 \right). \label{betaP}
\ee
\be
\begin{split}
\beta_E(g)= &\frac{(z+d-1)-\Delta_g}{z} g \\
&- \frac{(z-1)(\Delta_g-z)\left[ \Delta_g -(d-1) \right] \left[\Delta_g-(z+d-1) \right] g^3}{4z(d-1) \left[ 2 \Delta_g ^2 -3 \Delta_g  (d+z-1) +z (3d-2)+(d-1)(d-2) \right]} +\mathcal{O}\left( g^4 \right),
\end{split} \label{betaE}
\ee

where we have denoted the coupling constant as $g$. The dimension of the $:
\phi^4 :$ operator is $\Delta_g=2(d-z-1)$, consequently the classical
contributions to the $\beta$ function agree for both cases. The quantum
correction terms in the UV (eq. \ref{betae}) are different from the
corresponding IR ones (eq. \ref{beta_p_z1}) and (eq. \ref{beta_e_z}).
Furthermore, the dependence of the quantum correction on the coupling in the UV
case (eq. \ref{betae}) is $\propto g^3$, such terms can be obtained by
perturbation theory when diagrams with three vertices are divergent. The
corresponding diagrams with three vertices are a 2-loop correction to the
four-point vertex and a 3-loop correction to the two-point vertex (eq.
\ref{vertices}, eq. \ref{loops})  and consequently do not enter the 1-loop
calculation performed here.  The absence of quadratic corrections in (eq.
\ref{betaP}) and (eq. \ref{betaE}), can be explained by the absence of a
term $\frac{g}{4!} \phi^4$ in the dilaton potential (eq.  \ref{dil_pot}).  The
inclusion of such terms is expected to yield corrections quadratic in the
coupling $g$.

\chapter{Conclusions} \label{cpt:conclusions}

In the present work we have examined a formalism that utilizes two distinct UV cutoff scales for energy and momentum. In this framework the Renormalization of the couplings is described by two different $\beta$ functions, arising from independent variations of the energy ($\beta_E$) or momentum ($\beta_P$) cutoffs ($\Lambda_E$ and $\Lambda_P$ respectively). Such a formalism is desired because in the case of non-Lorentz symmetric field theories the energy and momentum scales are allowed to vary in different ways.

The evaluation of the $\beta$ functions can be performed within holography by evaluating the metric of the dual field theory in domain wall frame coordinates.
More specifically, we have analyzed holographic non-relativistic theories with rotational invariance, described by the Einstein Maxwell Dilaton action. In this case we have developed a technique to calculate such $\beta$-functions using a generalization of the superpotential formalism developed in \cite{Kiritsis:2012ma}. In this formalism the complete solution of the system is encoded in a non-linear integro-differential equation for the superpotential, $W$. If $W$ is known the $\beta$ functions are obtained by algebraic equations. Within this analysis the $\beta_P$ function is shown to retain the same dependence on the superpotential $W$ as in the relativistic case while the $\beta_E$ shifts away from the relativistic value $\beta_P=\beta_E$ as a function of the EMD couplings. Furthermore, the property of the $\beta_P$ and $\beta_E$ functions to obtain their corresponding relativistic values $\beta=\beta_P=\beta_E$ in the case of Lorentz symmetry is manifested.

We then proceed and examine the properties of a renormalization group flow near the Lifshitz and hyperscaling violating UV asymptotics of the EMD gravity. We can realize different types of non-trivial fixed points depending on the couplings of gravity theory $V$ and $Z$ in the vicinity of the fixed points. Such fixed points can be either attractive or repulsive for both renormalization scales and they are non-trivial in the sense that the value of the coupling is finite. Saddle points i.e. points attractive in one of the scales and repulsive in the other cannot be realized in the endpoints of an RG flow because in that case the Gubser bound is violated. An intriguing question, requiring further study is whether such saddle points can be realized away from the fixed points of the RG flow and whether they can be utilized in order to lead the system to different IR or UV fixed points.

We subsequently investigate the RG flow that emerges when a Lifshitz critical point is perturbed by a relevant operator. In this case the fixed point at $\phi_{UV}=\phi_-$ is repulsive and quantum corrections are incorporated into the $\beta$ functions by using perturbation theory for the dual gravity. We do a similar computation by introducing an interaction term to a Lifshitz-invariant free QFT. In this case, we calculate the quantum corrections by explicitly introducing momentum and energy cutoffs and treating the corresponding renormalization scales separately. In this case the IR fixed point for the interaction $g=0$ is attractive for both scales and termd depending on the fraction of the scales appear the energy $\beta_E$ function.

Finally, we compare the holographic approach in the evaluation of the RG flow, as well as, the field theory one and conclude that both approaches yield the same tree-level dependence for the corresponding $\beta$ functions, but the calculated quantum corrections for the UV and IR case are different.

\newpage
\appendix
\addcontentsline{toc}{chapter}{Appendices\label{app}}
\chapter*{Appendices}
\renewcommand{\theequation}{\Alph{chapter}.\arabic{equation}}

\chapter{Anti-de Sitter Space} \label{subsec:AdS}

\indent $AdS_n$ space is the $n$ dimensional hyperboloid embedded in a $n+1$ dimensional flat space, $R^{2,n-1}$. If we suppose the coordinate system $\{X_0,...,X_n \}$, the metric for $R^{2,n-1}$ is presented below
\be
ds^2=-dX_0^{\;2} -dX_n^{\;2} +\sum\limits_{i=1}^{n-1}dX_i^{\;2}. \label{AdS.1}
\ee
The equation of the hyperboloid is
\be
dX_0^{\;2} +dX_n^{\;2} -\sum\limits_{i=1}^{n-1}dX_i^{\;2}=L^2. \label{AdS.2}
\ee
As a sidenote, we observe that the $AdS_n$ space has the symmetry $O(2,n-1)$ by construction. This symmetry is of great importance in the context of AdS/CFT. This symmetry of $AdS_n$, corresponds to the conformal symmetry of the dual field theory, that lives in $n-1$ dimensions.

Therefore the metric of the hyperboloid and consequently $AdS_n$ space, is obtained by solving (eq. \ref{AdS.1}) and (eq. \ref{AdS.2}). We define $\rho$, $\tau$ and $\Omega_i$, with $i=1,...,n-1$ and $ \sum\limits_{i=1}^{n-1}\Omega_i^{\;2} =1 $, as
\be
X_0\equiv L \cosh \rho \cos \tau \sp
X_n\equiv L \cosh \rho \sin \tau \sp
X_i\equiv L \sinh \rho \; \Omega_i,
\ee
$\Omega_i$ are the standard coordinates of the $n-1$ sphere, $S^{n-1}$. This parametrization yields the following metric for the $AdS_n$ space
\be
ds^2=L^2 \left( -\cosh^2 \rho \; d\tau^2+d\rho^2 +\sinh^2 \rho \; d\Omega_{n-2}^{\;2} \right), \label{AdS.metric}
\ee
where $\rho$, $\tau$ and $\Omega_i$, with $i=1,...,n-1$ are the global coordinates of the $AdS_n$ space. In order to cover the hyperboloid once we should  take $\rho \in \{0,\infty \}$ and $\tau \in [0, 2\pi )$. In the limit $\rho \to 0$ the $AdS_n$ metric (eq. \ref{AdS.metric}) asymptotes
\be
ds^2=L^2 \left( - d\tau^2+d\rho^2 + \rho^2 \; d\Omega_{n-2}^{\;2} \right)
\ee
This is recognized as $S^1 \times R^{n-1}$, where the $S^1$ coordinate is $\tau$. This indicates that this definition of $AdS_n$ space leads to a spacetime has closed timelike curves and therefore theories that live on it cannot be causal. In order to re-establish causality we consider the universal cover of the $\tau$ coordinate, allowing it to take values in region $\tau \in (-\infty,\infty)$. Most of the time the notion $AdS_n$ space in literature, stands for this universal cover.

Poincar\' e coordinates $(t, u, \vec{x})$ are useful in the study of $AdS_n$. Those are defined as
\be
\begin{split}
X_0 \equiv \frac{u}{2} \left[1+\frac{1}{u^2} \left( L^2+ \vec{x}^2-t^2 \right) \right]& \sp
X_{n-1} \equiv \frac{u}{2} \left[1-\frac{1}{u^2} \left( L^2- \vec{x}^2+t^2 \right) \right] \\
X_n \equiv &\frac{L t}{u} \sp X_i \equiv \frac{L x^i}{u},
\end{split}
\ee
These coordinates cover half of the hyperboloid defined by (eq. \ref{AdS.2}). The metric in Poincar\' e coordinates is
\be
ds^2=\left(\frac{L}{u}\right)^2 \left( - dt^2+du^2 + dx^i dx_i \right).
\ee
In this coordinates the Poincar\' e symmetry acting on $(t,\vec{x})$ coordinates is obvious. Lorentz symmetry, SO(1,1), acts as a dilation on coordinates $(t, \vec{x})$,
\be (t, u, \vec{x}) \to (at,a u,a \vec{x}), \;\;\; a>0. \ee
The Riemann tensor for $AdS_n$ and the scalar curvature are
\be
R_{\mu \mu \rho \sigma}=- \frac{1}{L^2} \left( g_{\mu \rho} g{\nu \sigma}- g_{\mu \sigma} g{\nu \rho} \right) \Rightarrow R^{(n)}=- \frac{n(n-1)}{L^2} \ee
The scalar curvature of $AdS_n$ is a constant and negative quantity.

The considerations above, prove the claim that ``The $AdS_n$ spacetime is maximally symmetric and is characterized by constant and negative scalar curvature, $R^{(n)}$".

\chapter{Fields in AdS space}

\indent In this section we consider the dynamics of the fields that live in an $AdS_n$ spacetime. For a massive scalar field $\phi(u,t,\vec{x})$, the Klein-Gordon equation is written in Poincar\' e coordinates as:
\be
(\square -m^2)\phi=0 \Leftrightarrow \frac{u^2}{L^2}\left[ \pa_u^2 - \frac{n-2}{u} \pa_u -\pa_t^2+ (\vec{\pa} \cdot \vec{\pa}) \right] \phi(u,t,\vec{x})=m^2 \phi(u,t,\vec{x}).
\ee
By Fourier transforming the $(t,\vec{x})$ coordinates
\be
\phi(u,t,\vec{x})= \int \frac{d\omega d^{n-2}q}{(2\pi)^{n-1}} e^{i(\vec{q} \cdot \vec{x} - \omega t)} \phi(u,\omega,\vec{q}),
\ee
we obtain the following equation
\be
\left[\pa_u^2-\frac{n-2}{u} \pa_u -(\vec{q}^2-\omega^2)- \frac{m^2 L^2}{u^2} \right]\phi(u,\omega,\vec{q})=0.
\ee
The solution is then given in terms of Bessel functions
\be
\phi(u,\omega,\vec{q}) \sim u^{(n-1)/2} Z_\nu(\sqrt{\vec{q}^2-\omega^2\;} u), ~~~~ \nu=\frac{1}{2}\sqrt{(n-1)^2+4 m^2 L^2},
\ee
where $Z_\nu$ stands for one of the two linearly independent solutions of the Bessel equation, $I_\nu$ and $K_\nu$. If the so called Breitenlohner-Freedman (BF) bound holds, $\nu$ is real and positive. The BF bound is
\be
m^2 \geq - \left( \frac{n-1}{2L} \right)^2. \label{BF.bound}
\ee
An observation that will be useful is that this bound allows $m^2$ to be negative, and the corresponding field to be tachyonic. We define
\be
\Delta_\pm=\frac{1}{2} (n-1) \pm \nu \longrightarrow \nu=2\Delta_+ -d \geq 0 \Leftrightarrow  \Delta_-=d-\Delta_+ \leq \Delta_+,
\ee
the asymptotic behavior near the boundary ($u \to 0$) is dominated by $\phi_- \propto u^{\Delta_-}$. While near the horizon ($u \to \infty$), the asymptotic behavior is dominated by $\phi_+ \propto u^{\Delta_+}$.
\be
\phi_\pm(u,\omega,\vec{q}) \sim u^{\Delta_\pm} \phi_\pm(\omega,\vec{q}), ~~~~~ \Delta_\pm=\frac{1}{2} (n-1) \pm \frac{1}{2}\sqrt{(n-1)^2+4 m^2 L^2}. \label{basy.scal}
\ee
This result can be generalized to contain the case of any p-form field, i.e. to any antisymmetric tensor $A_{\mu_1,...,\mu_p}$ with p indices. The asymptotic behavior, is dominated by the $u^{\Delta_-}$ term near the boundary and the $u^{\Delta_+}$ term near the horizon.
\be
A_{\mu_1,...,\mu_p}^\pm(u,\omega,\vec{q}) \sim u^{\Delta_\pm} A_{\mu_1,...,\mu_p}^\pm(\omega,\vec{q}),~~ \Delta_\pm=\frac{1}{2} (n-1) \pm \frac{1}{2}\sqrt{(n-1-2p)^2+4 m^2 L^2}.
\ee
Finally, for a fermionic field with spin equal to $1/2$ the asymptotic behavior is
\be
v_\a^\pm(u,\omega,\vec{q}) \sim u^{\Delta_\pm} v_\a^\pm(\omega,\vec{q}), ~~~~~ \Delta_\pm=\frac{1}{2} (n-1) \pm \left|m L \right|,
\ee
where $v_\a \to v_\a^-$ near to the boundary and $v_\a \to v_\a^+$ near to the horizon.

\chapter{Conformal Transformations} \label{subsec:ConfTrns}

\indent Under a general coordinate transformation, $x \to \tilde{x}(x)$, the metric $g_{\mu \nu}$ transforms as
\be
g_{\mu \nu}(x) \to \tilde{g}_{\mu \nu}(\tilde{x})=\frac{\pa x^\a}{\pa \tilde{x}^\mu} \frac{\pa x^\b}{\pa \tilde{x}^\nu} g_{\a \b}(x)
\ee
The group of conformal transformations is the subgroup of these coordinate transformations that leave the metric invariant up to a rescaling, such as
\be
g_{\mu \nu}(x) \to \tilde{g}_{\mu \nu}(\tilde{x})=\Omega(x) g_{\a \b}(x)
\ee
Consider a point in spacetime $x=P$, and a set of curves through that point. These transformations preserve oriented angles between curves through P with respect to their orientation, hence the name conformal transformations. For Minkowski space the Poincar\' e group is a subgroup of conformal transformations, with $\Omega(x)=1$.

We will examine the infinitesimal conformal transformations, $x^\mu \to \tilde{x}^\mu=x^\mu+\e ^mu$. Under such transformations the metric should shift as
\be
\delta g_{\mu \nu}=-g_{\lambda \nu} \pa^\lambda \e_\mu -g_{\lambda \mu} \pa^\lambda \e_\nu - \e^\lambda \pa_\lambda \delta g_{\mu \nu}= \a(\e) g_{\mu \nu}. \label{conf1}
\ee
By supposing Minkowski space and contracting on both sides with $\delta^{\mu \nu}$ the factor $\a(\e)$ is equal to
\be
\a(\e)=-\left( 1+ \frac{2}{d} \right) \pa^\mu \e_\nu, \label{conf2}
\ee
where $d$ the dimension of the spacetime we are considering. We, then, substitute (eq. \ref{conf2}) into (eq. \ref{conf1}) and act on both sides of the equation with $\pa^\mu$ and $\square \equiv \pa^\mu \pa_\mu$, this process yields
\be
\begin{split}
\square \e_\nu+\left( 1-\frac{2}{d} \right) \pa_\nu (\pa^\sigma \e_\sigma)=0\\
\pa_\mu \square \e_\nu + \pa_\nu \square \e_\mu - \frac{2}{d} \delta_{\mu \nu} \square (\pa^\sigma \e_\sigma)=0\\
\left[ \delta_{\mu \nu} \square + (d-2) \pa_\mu \pa_\nu \right](\pa^\sigma \e_\sigma)=0
\end{split}\label{conf3}
\ee
It is easy to verify from (eq. \ref{conf3}) that for $d=2$ we obtain a special case of conformal transformations. We will not expand in the properties of this class of transformations, as there is a topic about these in any textbook about the calculus of complex variables. For $d \neq 0$, we can identify the following possibilities for $\e^\mu$:
\be
\begin{split}
\e^\mu=a^\mu \sp &\text{Translations.}\\
\e^\mu=\tensor{\omega}{^\mu_\nu} x^\nu \sp &\text{Rotations ($\omega_{\mu \nu}= -\omega_{\nu \mu}$).}\\
\e^\mu=\lambda x^\mu \sp &\text{Scale transformations.} \\
\e^\mu=b^\mu (x^\sigma x_\sigma) - 2 x^\mu (b^\sigma x_\sigma) \sp &\text{Special conformal transformations.}
\end{split}
\ee
Finite transformations can be obtained by exponentiation of infinitesimal ones. The generators of conformal transformations are presented below
\be
\begin{split}
&\text{Translations:} ~~~~~~~~~~~~~~ P_m= -i \pa_\mu \\
&\text{Rotations: } ~~~~~~~~~~~~~~~~ J_{\mu \nu}=i(x_\mu \pa_\nu-x_\nu \pa_\mu) \\
&\text{Scale Transformations:} \;  D= -i x^\sigma \pa_\sigma \\
&\text{Special Conf. Transf. :} \; K_\mu=-i \left[ (x^\sigma x_\sigma) \pa_\mu -2 x_\mu (x^\sigma \pa_\sigma) \right]
\end{split}
\ee
And they obey the following commutation relations
\be
\begin{split}
[J_{\mu \nu}, P_\rho]= -i( \eta_{\mu \rho} P_\nu - \eta_{\nu \rho} P_\mu ),\;\;\; [P_\mu, K_\nu]= 2i J_{\mu \nu} - 2 i \eta_{\mu \nu} D, \\ [J_{\mu \nu},K_\rho]= -i( \eta_{\mu \rho} K_\nu - \eta_{\nu \rho} K_\mu ), \;\;\;
[J_{\mu \nu},J_{\rho \sigma}]=-i (\eta_{\mu \rho} J_{\nu \sigma} -\eta_{\mu \sigma} J_{\nu \rho} -\eta_{\nu \rho} J_{\mu \sigma} +\eta_{\nu \sigma} J_{\mu \rho} ), \\ [D,K_\mu]= i K_\mu, \;\;\; [D,P_\mu]=-i P_\mu, \;\;\; [J_{\nu \nu},D]=0
\end{split}
\ee
A general conformal transformation has $\frac{1}{2}(d+2)(d+1)$ parameters. In a space of signature (p,q)\footnote{If we are allowed to think about (p,q) signature in context of relativity, p is the number of timelike coordinates and q the number of spacelike ones.} The conformal group is $O(p+1,d+1)$ and the generators can be written as the components of an antisymmetric $(d+2) \times (d+2)$ matrix as
\be
M_{\mu \nu}=J_{\mu \nu}, \;\;\;\; M_{\mu, d}=\frac{1}{2}(K_\mu -P_\mu),\;\;\;\; M_{\mu, d+1}=\frac{1}{2}(K_\mu +P_\mu), \;\;\;\; M_{d,d+1}=D.
\ee

\chapter{The electric hyperscaling violating solutions in EMD}

\section{Properties of hyperscaling violating solutions} \label{sec:Hyperscaling_violation}

\indent An important class of asymptotic metric solutions for the EMD gravity (eq. \ref{EMDaction}) is the Hyperscaling violating solutions. The behavior of such $d+1$ dimensional space-times depends on the Lifshitz exponent, $z$, which introduces an anisotropy in the scaling of space and time and the hyperscaling violation exponent $\theta$, which control the scaling of the proper distance. The scaling properties of hyperscaling violating space-times are summarized in the following expression
\be
t \to \lambda^z t \sp x_i \to \lambda x_i \sp s \to \lambda^{\frac{\theta}{d-1}} s, \label{hyperscaling_sym}
\ee
where $t$, $x_i$ the time-like and space-like coordinates respectively and $s$ the proper distance.
If we assume Poincar\'e coordinates the corresponding metric of the Hyperscaling violating space-time with parameters $(d,z,\theta)$ reads
\be
ds^2=r^{\frac{2 \theta}{d-1}}\left(-\frac{d t^2}{r^{2z}} +\frac{B_0 dr^2 + dx_i dx^i}{r^2} \right),
\ee
where $r$ denotes the anisotropic spatial coordinate and $i \in \{2,3,\dots,d\}$. In fact this metric is the most general one that satisfies the scaling property (eq.\ref{hyperscaling_sym}) and has a homogeneous $d-1$ dimensional spatial part. For $\theta = 0$, the proper distance, $s$, is not affected by scale transformations and thus the scaling property (eq.\ref{hyperscaling_sym}) is elevated to scaling invariance. The corresponding space-time is defined by the parameters $(d,z)$ and is called the Lifshitz space-time. In the case that both $z=\theta=0$, both the spatial and temporal part of the space-time scale uniformly under scale transformations and the corresponding space-time is the AdS$_{d+1}$.

Space-times exhibiting those properties are an acceptable solution of the EMD equations of motion (eq. \ref{covariant.eqmot}) if an amount of conditions (involving both the geometry and the auxiliary fields) is satisfied. Those conditions are summarized below
\begin{enumerate}
	\item Metric and field components ought to be real numbers, $g_{\mu \nu}(x), A_{\mu}(x), \phi(x) \in \mathbb{R}$ and only one coordinate should be timelike.\label{cond1}
	\item The metric ought to have regular asymptotic behaviour in the IR and UV
	$$g_{t t} \xrightarrow{x \to x_{IR}} 0 \Rightarrow \lim\limits_{x \to x_{IR}} g_{i i} \neq \infty, ~~
	g_{t t} \xrightarrow{x \to x_{UV}} \infty  \Rightarrow \lim\limits_{x \to x_{UV}} g_{i i} \neq 0.$$ \label{cond2}
	\item \textbf{Linearized spectrum (Temperature modes):} The finite temperature mode should be irrelevant in the UV.\\
	The finite temperature mode is given by the first correction to the asymptotic expansion of the blackness function\footnote{The blackness function $f(r)$ is defined as $\tilde g_{tt} = f(r) g_{tt}$ and $\tilde g_{rr} = \frac{g_{rr}}{f(r)}$, where $\tilde g_{\mu \nu}$ describes a space-time containing a black hole and $g_{\mu \nu}$ the background space-time.}. This mode should be irrelevant in the UV in order to recover the background geometry in regions of the space-time away from the event horizon. In holographic language this criterion ensures that arbitrarily small temperatures do not affect the physics of the dual field theory at arbitrarily high energies. \label{cond3}

	\item \textbf{Linearized spectrum (Dynamical Instabilities):} The metric ought to be free of dynamical instabilities .\\
	Dynamical instabilities are manifested as complex frequencies $\sim e^{\pm i k r}$ in the spectrum of linearized perturbations of the hyperscaling violating background. This can be seen by Wick rotating, $r \to i \tilde t$, in this case, those complex frequencies in the space-like coordinate, $r$ are mapped to positive frequencies in the time-like coordinate $\tilde t$. \label{cond4}
	\item \textbf{Null-energy condition:} The contraction of a null vector, $N^\mu : N_\mu N^\mu=0$, with the stress energy tensor, $T_{\mu \nu}$, is non-negative, $T_{\mu \nu} N^\mu N^\nu \geq 0$.\\
	This condition ensures that fields propagating null-like, such as the electromagnetic field, have non-negative energy. \label{cond5}
	\item \textbf{Gubser bound:}\cite{Gubser_bound}, The scalar potential evaluated at the scalar field solution, $V(\phi(x))$, ought to be bounded from below when $x$ approaches a naked singularity at $x=x_0$, $\lim\limits_{x \to x_0} V(\phi(x)) > - \infty$. \footnote{Note that in this work we use the opposite sign convention (see eq. \ref{scaling}) than in the case of \cite{Gubser_bound}, where this criterion was first discussed. By using the convention of (eq. \ref{scaling}) we obtain AdS-like solutions for $\Lambda>0$.} \\
	A naked singularity at $x=x_0$, is identified by the divergence of scalar curvature related quantities, such as the Ricci scalar, $R$, and the  Kretschmann scalar, $R_{\mu \nu \rho \sigma} R^{\mu \nu \rho \sigma}$, as $x \to x_0$, while an event horizon enclosing the point $x_0$ is absent. The Gubser bound is a necessary condition for a repulsive singularity. \label{cond6}
\end{enumerate}

From the conditions \ref{cond1} and \ref{cond2}, we straightforwardly obtain the constraints
\be
B_0>0 \sp (\theta-d +1) \left[ \theta -(d-1)z \right] >0, \label{constraint1}
\ee
plus any constraints arising from the solutions of the scalar and gauge field.
It had been shown that for quite generic cases\cite{Dong_hv_sol} that the temperature mode of the hyperscaling violating solution scales with $r$ as $r^{d-1+z-\theta}$. Therefore, according to condition \ref{cond3}, in order to accept an endpoint of the hyperscaling violating metric as the IR or UV fixed point the following conditions should apply
\be
\begin{array}{l l l l l}
\theta < d-1, & \theta < (d-1) z, & d-1+z-\theta > 0 \Rightarrow & r \xrightarrow{IR} \infty, & r \xrightarrow{UV} 0,\\
\theta > d-1, & \theta > (d-1) z, & d-1+z-\theta < 0 \Rightarrow & r \xrightarrow{IR} 0, & r \xrightarrow{UV} \infty.
\end{array} \label{constraint2}
\ee
In order to enforce the null-energy condition\cite{Dong_hv_sol} (condition \ref{cond5}) we contract the null vectors
\be
N^t=\sqrt{B_0} r^{z-\frac{\theta}{d-1}} \sp N^r=\frac{1}{\sqrt{B_0}} r^{1-\frac{\theta}{d-1}} \cos \psi \sp N^i=\frac{1}{\sqrt{B_0}} r^{1-\frac{\theta}{d-1}} \sin \psi,
\ee
for $\psi=\{0,\pi/2\}$ with the Einstein tensor $G_{\mu \nu}=R_{\mu \nu}-\frac{1}{2}R g_{\mu \nu}$ (which in view of the Einstein equation $G_{\mu \nu}=T_{\mu \nu}$ is equal to the stress energy tensor). Those conditions yield the constraints\cite{Dong_hv_sol}
\be
(d-1-\theta)\left[ (d-1)(z-1)-\theta \right] \geq 0 \sp (z-1) \left[ d-1 +z -\theta \right] \geq 0. \label{constraint3}
\ee
The conditions \ref{cond1}, \ref{cond4} and \ref{cond6}, depend on the scaling of the scalar and gauge fields as they asymptote the endpoints of hyperscaling violating space-time. As such the corresponding constraints cannot be inferred based solely on the parameters of the hyperscaling violating metric $(d,z,\theta)$. As a last remark we note that the Ricci and Kretschmann curvature scalars exhibit the following dependence on the $r$ coordinate near the endpoints
\be
R \propto r^{- \frac{2\theta}{d-1}} \sp R_{\mu \nu \rho \sigma} R^{\mu \nu \rho \sigma} \propto r^{- \frac{4\theta}{d-1}}. \label{curvature_scalars}
\ee
The (eq. \ref{curvature_scalars}) manifests the existence of a naked singularity at $r \to 0$ ($r \to \infty$) for $\theta > 0$ ($\theta < 0$) which is relevant for the enforcement of the Gubser bound (condition \ref{cond6}).

In order to recapitulate, we combine (eq. \ref{constraint1}, \ref{constraint2}, \ref{constraint3})\cite{Kiritsis_hv_conditions}, and obtain the following constraint
\be
\begin{array}{l l l l}
	r \xrightarrow{IR} \infty, & r \xrightarrow{UV} 0 : & \bigg\{
	\begin {array}{l}
		z \geq 1,\\
		z \geq 1 + \frac{\theta}{d-1},
	\end{array} &
	\begin {array}{l}
		\text{if } \theta \leq  0,\\
		\text{if }  0 < \theta < d -1,
	\end{array}\\
	r \xrightarrow{IR} 0, & r \xrightarrow{UV} \infty : & \bigg\{
	\begin {array}{l}
		z < \theta -(d -1),\\
		z \leq 1,
	\end{array} &
	\begin {array}{l}
		\text{if } d-1 < \theta  \leq d, \\
		\text{if } d < \theta.
	\end{array}
\end{array} \label{constraint235}
\ee
which ensure that the asymptotics are well behaved in the corresponding regions according to the criteria \ref{cond2}, \ref{cond3} and \ref{cond5}.

\section{The general hyperscaling violating solution} \label{sec:Hyperscaling_solutions}

The Hyperscaling violating solutions have been generalized to arbitrary number of dimensions in \cite{cgkkm,gk1}, where they have been formulated by using the domain wall frame ansatz (\ref{ansatz2}). The couplings are assumed to scale exponentially with the dilaton field in the scaling regime as $V(\phi)=2 \Lambda e^{-\delta \phi}$ and $Z(\phi)=e^{\gamma \phi}$. By defining
\be
e^{\tilde A(u)}=\alpha(u) ~~~ \text{and} ~~~ e^{\phi(u)}= e^{\phi_0} \varphi(u),
\ee
the equations of motion (\ref{em2.a2}, \ref{em3.a2}, \ref{em4.a2}) read
\be
\left( \frac{\dot{\varphi}}{\varphi} \right)^2 =2 (d-1) \left[\left( \frac{\dot{\alpha}}{\alpha} \right)^2 - \frac{\ddot{\alpha}}{\alpha} \right]
\ee
\be
\ddot f + d \dot f \frac{\dot{\alpha}}{\alpha} =\frac{q^2 e^{- \gamma \phi_0}}{\alpha^{2 (d-1)} \varphi^{\gamma}}
\ee
\be
2(d-1) \dot f \frac{\dot{\alpha}}{\alpha} +\frac{q^2 e^{- \gamma \phi_0}}{\alpha^{2 (d-1)} \varphi^{\gamma}} -4 \frac{\Lambda e^{- \delta \phi_0}}{\varphi^{\delta}}=\left[ \left( \frac{\dot{\varphi}}{\varphi} \right)^2 - 2d(d-1) \left( \frac{\dot{\alpha}}{\alpha} \right)^2 \right] f
\ee
and the gauge field is expressed as follows
\be
A_t=q e^{- \gamma \phi_0} \int du~\alpha^{-(d-2)} \varphi^{-\gamma}.
\ee
The solution of those equations of motion is summarized below
\be
\phi(u)=\phi_0+(\delta - \gamma) \ln \left( \frac{u}{\ell} \right), \label{sol_hv.1}
\ee
\be
ds^2=-\left( \frac{u}{\ell} \right)^{1 -\frac{d-2}{2(d-1)}(\delta -\gamma)^2} M(u) dt^2+ \left( \frac{u}{\ell} \right)^{\frac{d(\delta -\gamma)^2}{2(d-1)}-1} \frac{dr^2}{M(u)} + \left( \frac{u}{\ell} \right)^{\frac{(\delta -\gamma)^2}{d-1}} dx_i dx^i, \label{sol_hv.2}
\ee
\be
M(u)=\frac{8(d-1) \Lambda \ell^2 e^{-\delta \phi_0}}{(\gamma^2-\gamma \delta +2) \left[ (\gamma -\delta)^2 +(d-1)(\gamma^2-\delta^2+2) \right]}
\left[\left( \frac{u}{\ell} \right)^{1+ \frac{\gamma^2-\delta^2}{2} +\frac{(\delta-\gamma)^2}{2(d-1)}} -m \right], \label{sol_hv.3}
\ee
\be
A_t=\frac{4(d-1)\Lambda^{1/2} \ell e^{-\frac{\gamma+\delta}{2} \phi_0}}{(\gamma -\delta)^2 +(d-1)(\gamma^2-\delta^2+2)} \sqrt{-\frac{\delta^2-\gamma \delta -2}{\gamma^2-\gamma \delta +2}} \left[\left( \frac{u}{\ell} \right)^{1+ \frac{\gamma^2-\delta^2}{2} +\frac{(\delta-\gamma)^2}{2(d-1)}} -m \right]. \label{sol_hv.4}
\ee
this solution is manifestly black hole - like and $m$ is an integration constant which we identify with the mass of the black hole. The event horizon is situated at
\be
u_0= \ell m^{\frac{2(d-1)}{ (\gamma -\delta)^2 +(d-1)(\gamma^2-\delta^2+2)}},
\ee
and whenever $u-u_0$ changes sign the $u$ and $t$ coordinates interchange their roles as space-like and time-like.

The relevant asymptotic behaviour is the behaviour of the solution in the regions $u=0$, $u=u_0$ and $u \to \infty$. We begin with the asymptotics near the event horizon $u=u_0$, those asymptotics will make manifest where the interior and the exterior of the black hole lies. The expansion of $M(u)$ in powers of $u/u_0$ reads
\be
M(u) = \frac{4 u_0 \Lambda \ell^2 e^{- \delta \phi_0}}{\gamma^2- \gamma \delta +2} \left[-1 + \frac{u}{u_0} + \mathcal{O} \left( \frac{u}{u_0} \right) \right].
\ee
Observe that the sign of the metric components depend solely on the sign of $M$ since we are working in the range $u>0$. More specifically, the coordinate $t$ has the opposite sign than $M$ and $u$ has the same one. The sign of $M$ depends only on the sign of $\gamma^2- \gamma \delta +2$ and on whether we approach the event horizon at $u_0$ from higher or lower $u$ values. By using the above statements we obtain the following table
\be
\begin{array}{c | c c}
 \text{Case:} & u < u_0 & u > u_0 \\ \hline
 \gamma^2- \gamma \delta +2 > 0 & {\scriptstyle r \to \text{ time-like,} \atop \scriptstyle t \to \text{ space-like}} & {\scriptstyle t \to \text{ time-like,} \atop \scriptstyle r \to \text{ space-like}} \\[0.25cm]
 \gamma^2- \gamma \delta +2 < 0 & {\scriptstyle t \to \text{ time-like,} \atop \scriptstyle r \to \text{ space-like}} & {\scriptstyle r \to \text{ time-like,} \atop \scriptstyle t \to \text{ space-like}}
\end{array} \label{horizon}
\ee
The metric components of the solution (eq. \ref{sol_hv.1}, \ref{sol_hv.2}, \ref{sol_hv.3}, \ref{sol_hv.4}) depend solely on $r$ and since we are interested in static solutions, we conclude that the exterior of the black hole is located at $u>u_0$ for $\gamma^2- \gamma \delta +2 > 0$ and at $u<u_0$ for $\gamma^2- \gamma \delta +2 < 0$. In any of those cases the event horizon is regular and it can be casted in a Rindler form\footnote{We remind that Rindler metric, $ds^2=-\kappa^2 \rho^2 dt^2 +d \rho^2 +dx_i dx^i$, is the proper description for flat space with a homogeneous gravity field and uniform gravitational acceleration $\kappa \hat e_\rho$.}, like the horizon of a Schwartzchild black hole. Similarly, with the aforementioned black hole the temperature, $T$, is obtained by the gravitational acceleration (surface gravity) at the horizon, $\kappa$, as $T= \kappa/{2 \pi}$.

The asymptotic form of the metric at the endpoints (ie. $u \to 0$, $u \to \infty$) can be obtained by considering that either the constant term $m$ or the power scaling term dominate $M$ at the corresponding endpoint region. We are interested in the asymptotic behavior outside of the event horizon, ie. the static regions where $r$ ($t$) is space-like (time-like) according to (eq. \ref{horizon}).

\subsubsection*{Case $\mathbf{\gamma^2- \gamma \delta +2 > 0}$}
In the case $\gamma^2- \gamma \delta +2 > 0 $ the power scaling term dominates the asymptotic form of the metric as $u \to \infty$. The metric of the solution reduces to
\be
ds^2=\left( \frac{u}{\ell} \right)^{-\delta (\gamma-\delta)} \left[-M_0 \left( \frac{u}{\ell} \right)^{\frac{(\gamma-\delta)\left[ \gamma +(d-2) \delta \right]}{d-1} +2 -\delta(\delta-\gamma)} dt^2 +\frac{\ell^2}{M_0} \frac{du^2}{u^2} + \left( \frac{u}{\ell} \right)^{\frac{(\gamma-\delta)\left[ \gamma +(d-2) \delta \right]}{d-1}} dx_i dx^i \right],
\ee
where we defined $M_0$ as
\be
M_0 \equiv \frac{8(d-1) \Lambda \ell^2 e^{-\delta \phi_0}}{(\gamma^2-\gamma \delta +2) \left[ (\gamma -\delta)^2 +(d-1)(\gamma^2-\delta^2+2) \right]}.
\ee
 After the change of variables
\be
r \equiv \left( \frac{u}{\ell} \right)^{-\frac{(\gamma-\delta)\left[ \gamma +(d-2) \delta \right]}{2(d-1)}} \sp \tilde{t} \equiv M_0 t,
\label{change_variables}
\ee
it is manifested that metric solution is asymptotically hyperscaling violating
\be
ds^2=r^{\frac{2 \theta}{d-1}}\left(-\frac{d\tilde {t}^2}{r^{2z}} +\frac{B_0 dr^2 + dx_i dx^i}{r^2} \right).
\ee
The parameters of the asymptotic hyperscaling violating metric are
\be
\theta=\frac{\delta (d-1)^2}{\gamma +(d-2) \delta} \sp z= 1 + \frac{(d-1)\left[2 -(\delta-\gamma)\delta \right]}{(\gamma-\delta)\left[ \gamma+(d-2)\delta \right]}.
\label{exponents}
\ee
\be
B_0=-\frac{(d-1)(\gamma^2-\gamma \delta +2) \left[ (\gamma -\delta)^2 +(d-1)(\gamma^2-\delta^2+2) \right]}{2 (\gamma - \delta)^2 (\gamma + (d-2) \delta)^2 \Lambda \ell^2 e^{-\delta \phi_0}}.
\ee
We find the subset of acceptable solutions that comply with the criteria presented in section \ref{sec:Hyperscaling_violation}. We can readily see that since we are working in the range $u > 0$ and $\Lambda>0$ the constraint \ref{cond1} is satisfied if
\be
(\delta^2-\gamma \delta -2)(\gamma^2-\gamma \delta +2)<0. \label{hv_sol_con1}
\ee
The constraints to the $\theta$ and $z$ in order to ensure well defined asymptotics was presented in (eq. \ref{constraint235}). It was also shown that the endpoints of the hyperscaling violating metric interchange their role as the IR and UV asymptotics depending on the sign of $\theta-(d-1)$. In view of (eq. \ref{exponents}). This condition can be expressed in terms of $\gamma$ and $\delta$ as
\be
\theta-(d-1)=\frac{(d-1)(\delta-\gamma)}{\gamma+(d-2)\delta}
\label{asympt_1}
\ee
Therefore the sign of (eq. \ref{asympt_1}) is the same as the sign of the exponent in the definition of the $r$ coordinate (eq. \ref{change_variables}). We can conclude that the limit $u \to \infty$ of the solution (eq. \ref{sol_hv.1}, \ref{sol_hv.2}, \ref{sol_hv.3}, \ref{sol_hv.4}) always maps to the UV asymptotic endpoint of the corresponding hyperscaling violating metric.
The scalar potential reads
\be
V(\phi(r))=2 \Lambda e^{- \delta \phi_0} r^{-\frac{2 \theta}{(d-1)}}.
\ee
Therefore the Gubser criterion is always satisfied as $V(\phi(r))>0$ and consequently it does not introduce further constraints.

\begin{figure}[ht]
\begin{center}
\includegraphics[width=\textwidth]{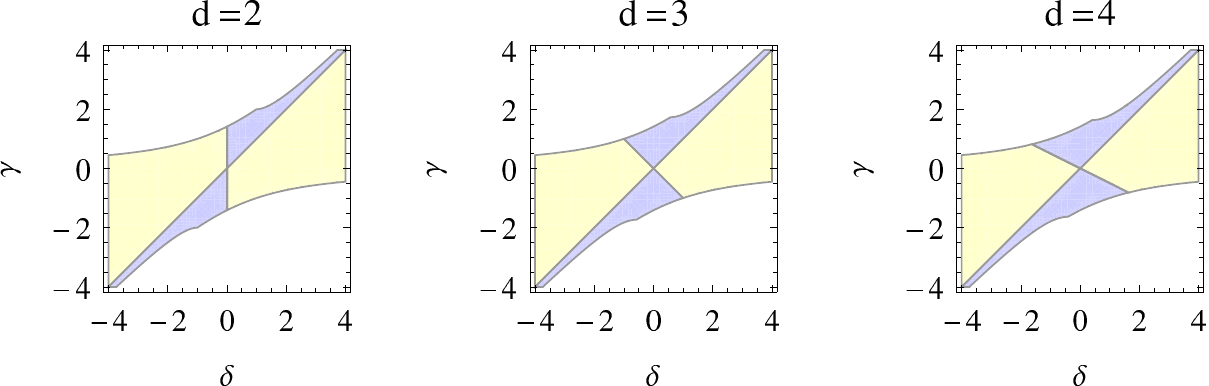}
\end{center}
\caption{The validity regions for the massive (eq. \ref{sol_hv.1} to \ref{sol_hv.4}) and the extremal (eq. \ref{sol_hv_ex.1} to \ref{sol_hv_ex.5}) solution in the ($\gamma$, $\delta$) plane, as given by the constraints (eq. \ref{bound1} to \ref{bound3}). The cases $d=2$, $d=3$ and $d=4$ are shown. The color denotes the position of the IR asymptotics in Poincar\'e coordinates, yellow regions denote that the IR is located at $r\to \infty$ and blue regions denote that the IR is located at $r \to 0$.}
\label{fig:validity_regions}
\end{figure}

If we do not discriminate between the case that $u \to \infty$ maps to $r \to 0$ or $r \to \infty$ the above mentioned constraints (eq. \ref{hv_sol_con1} and \ref{constraint235}) can be summarized as
\be
\delta^2-\gamma \delta -2 < 0, \label{bound1}
\ee \be
\gamma^2-\gamma \delta +2 > 0, \label{bound2}
\ee \be
(\gamma -\delta)^2 +(d-1)(\gamma^2-\delta^2+2)>0. \label{bound3}
\ee
In Fig. \ref{fig:validity_regions} we present the aforementioned validity regions for $d=2$, $d=3$ and $d=4$. In the same figure we also present where the IR asymptotics lie in Poincar\'e coordinates, a remark that is useful for the discussion of the extremal case.

We conclude that massive ($m>0$) solutions (eq. \ref{sol_hv.1}, \ref{sol_hv.2}, \ref{sol_hv.3}, \ref{sol_hv.4}) that satisfy the constraints (eq. \ref{bound1}, \ref{bound2} and \ref{bound3}) describe RG flows in the dual field theory from a hyperscaling violating UV fixed point ($u \to \infty$) to an IR thermal fixed point ($u=u_0$).

\subsubsection*{Case $\mathbf{\gamma^2- \gamma \delta +2 < 0}$}

We continue by examining the case $\gamma^2- \gamma \delta +2 < 0$ and $r \to 0$. In this case the constant term in $M(u)$ dominates the asymptotic behavior
\be
ds^2=\left( \frac{u}{\ell} \right)^{1+\frac{d (\gamma-\delta)^2}{2(d-1)}} \left[m M_0 \left( \frac{u}{\ell} \right)^{ -(\delta -\gamma)^2} dt^2 -\frac{\ell^2}{m M_0} \frac{du^2}{u^2} + \left( \frac{u}{\ell} \right)^{\frac{(2-d)(\gamma-\delta)^2}{2(d-1)}-1} dx_i dx^i \right],
\ee
the shifting of the signs of $t$ and $r$ might seem worrying, however, in the regime of parameters we are working, $M_0<0$ since  $\gamma^2- \gamma \delta +2 < 0$(see eq. \ref{horizon}). After the change of variables
\be
r \equiv \left( \frac{u}{\ell} \right)^{\frac{2(d-2)(\gamma-\delta)^2 -d(d-1)}{2(d-1)}} \sp \tilde{t} \equiv -m M_0 t
\ee
it is manifested that metric solution is asymptotically hyperscaling violating
\be
ds^2=r^{\frac{2 \theta}{d-1}}\left(-\frac{d\tilde {t}^2}{r^{2z}} +\frac{B_0 dr^2 + dx_i dx^i}{r^2} \right).
\ee
The parameters of the asymptotic hyperscaling violating metric are
\be
\theta=\frac{(d-1) \left[d (\gamma -\delta )^2+2 (d-1)\right]}{2 \left[2 (d-2) (\gamma -\delta )^2+d-1\right]} \sp
z= \frac{(d-1) (\gamma -\delta )^2}{2 (d-2) (\gamma -\delta )^2+d-1}.
\ee
\be
B_0=-\frac{(\gamma^2-\gamma \delta +2) \left[ (\gamma -\delta)^2 +(d-1)(\gamma^2-\delta^2+2) \right]}{4(d-1) (\gamma - \delta) (\gamma + (d-2) \delta) \Lambda e^{-\delta \phi_0}}.
\ee
By imposing the constraints (eq. \ref{constraint235}) it can be verified that those values of $z$ and $\theta$ do not yield well-defined asymptotics for any values of $\gamma$ and $\delta$. Therefore we have to reject all solutions with $\gamma^2- \gamma \delta +2 < 0$ as unphysical.

\section{Extremal case}

As we have remarked previously, solutions (eq. \ref{sol_hv.1}, \ref{sol_hv.2}, \ref{sol_hv.3}, \ref{sol_hv.4}) that satisfy the constraints (eq. \ref{bound1}, \ref{bound2} and \ref{bound3}) are acceptable. In this section we examine the extremal case, $m=0$. In this case the space-time is hyperscaling violating for all values of $u$. The extremal solution can be casted in Poincarr\'e coordinates as follows
\be
\phi_{\pm}=\phi_0 \pm \sqrt{\frac{2(\theta+d-1)\left[ (d-1)(z-1)-\theta \right]}{d-1}} \ln r, \label{sol_hv_ex.1}
\ee
\be
A_t=q_{\pm} r^{-\frac{(d-1)(z+(d-1)-\theta}{\theta-(d-1)z+(d-1)}} \sp A_r=A_i=0, \label{sol_hv_ex.2}
\ee
\be
q_{\pm}=\frac{(d-1)(z-1)-\theta}{(d-1) \left[ z+(d-1)-\theta \right]} \sqrt{\frac{\Lambda \ell^2 (z-1)}{z+(d-2)-\theta}} e^{\pm \phi_0 \frac{(d-3) \theta -(d-1)^2}{\sqrt{2 (d-1) (d-\theta -1) \left[ (d-1) (z-1)-\theta \right]}}}. \label{sol_hv_ex.3}
\ee
\be
ds^2=r^{\frac{2 \theta}{d-1}}\left(-\frac{d t^2}{r^{2z}} +\frac{B_{0,\pm} dr^2 + dx_i dx^i}{r^2} \right), \label{sol_hv_ex.4}
\ee
\be
B_{0,\pm}=\frac{\left[ z+(d-1)-\theta \right] \left[ z+(d-2)-\theta \right]}{2 \Lambda \ell^2} e^{\pm \phi_0 \frac{\sqrt{2} \theta}{\sqrt{2 (d-1) (d-\theta -1) \left[ (d-1) (z-1)-\theta \right] }}}. \label{sol_hv_ex.5}
\ee
For a given set of $(d,z,\theta)$ there are two solutions depending on whether the scalar field increases or decreases with $r$. Since the scalar potential is always positive we can accept both solutions as neither of those violates the Gubser bound (condition \ref{cond6}). In the previous expression we discriminate between the case of increasing and decreasing scalar field by the addition of the subscript $\pm$. We remind that the exponents $z$ and $\theta$ are determined from the action parameters $\gamma$ and $\delta$ via (eq. \ref{exponents}), which we rewrite here for completeness
\be
\theta=\frac{\delta (d-1)^2}{\gamma +(d-2) \delta} \sp z= 1 + \frac{(d-1)\left[2 -(\delta-\gamma)\delta \right]}{(\gamma-\delta)\left[ \gamma+(d-2)\delta \right]}.
\ee
The validity regions of the extremal solution presented in this section, as well as, the limit it asymptotes the IR of the dual QFT are depicted in Fig \ref{fig:validity_regions}. One very important property of the validity regions is that if the point $(\delta,\gamma)$ is in the validity region then the symmetric point with respect to the origin $(-\delta,-\gamma)$ also belongs to the validity region, that allows for both increasing and decreasing running of the scalar field.

\chapter{Solution of the linearized equations equations of motion} \label{sec:Linearization}

\indent The equations of motion (eq. \ref{pert_eq1}) to (eq. \ref{pert_gauge}) can be casted in linear form by substituting $r=e^u$ and defining $B'_0 \equiv A_0$. The system reduces to a 4-by-4 first order linear differential equation system of the form $\bar x'=A \bar x + \bar b$
\be
\begin{split}
\left[
\begin{array}{c}
\tilde g'_{tt}(u) \\
\tilde g'_{rr}(u) \\
\tilde A'_0(u) \\
\tilde B'_0(u) \\
\end{array}
\right]&=
\left[
\begin{array}{cccc}
 -\frac{z^2-d z+d-1}{d-1} & -\frac{(d+z-2) (d+z-1)}{d-1} & -\frac{2 (d-z-1) (z-1)}{d-1} & -\frac{2 (z-1)}{d-1} \\
 \frac{(z-1) (d+z-1)}{d-1} & -\frac{-d^2+d-z^2+z}{d-1} & -\frac{2 (z-1) (d+z-1)}{d-1} & \frac{2 (z-1)}{d-1} \\
 0 & 0 & 0 & 1 \\
 (1-z) z & z (d+z-2) & 2 (z-1) z & d+z-1 \\
\end{array}
\right]
\left[
\begin{array}{c}
\tilde g_{tt}(u) \\
\tilde g_{rr}(u) \\
\tilde A_0(u) \\
\tilde B_0(u) \\
\end{array}
\right] \\
&+\left[
\begin{array}{c}
 +\frac{B_0 M^2 \phi^2 (u)-\left( \phi '(u) \right)^2}{2 (d-1)} \\
 -\frac{B_0 M^2 \phi^2 (u)+\left( \phi '(u) \right)^2}{2 (d-1)} \\
 0 \\
 \frac{z \left( \phi '(u) \right)^2}{2 (d-1)} \\
\end{array}
\right]. \label{Linear_system}
\end{split}
\ee
The non homogeneous term is non-linearly dependent on $\phi(r)$. The scalar field is given by the solution of the linear equation (eq. \ref{eq_scalar_field})
\be
r^2 \tilde{\phi}''-(d+z-2) r \tilde{\phi}'-B_0 M^2 \tilde{\phi}=0,
\ee
which does not depend on $\tilde g_{tt}$, $\tilde g_{rr}$, $\tilde A_0$ and $\tilde B_0$ and thus can be solved independently from the linear system (eq. \ref{Linear_system}).

To begin with, we focus on solving the homogeneous equation. We evaluate the eigenvectors and eigenvalues of the matrix $A$. The first eigenvalue and eigenvector are
\be
\lambda_1=0 \sp \bar v_1=
\left[
\begin{array}{c}
 2 \\
 0 \\
 1 \\
 0 \\
\end{array}
\right],
\ee
this eigenstate simply expresses the gauge invariance of the field $A_0$ and is a marginal perturbation mode of the Taylor solution. More specifically it shows the invariance of the theory under the transformation $A_0 \to A_0 + \text{constant}$ and thus it is insignificant for our analysis.
The second eigenvector (eigenvalue) is
\be
\lambda_2=d+z-1 \sp \bar v_2=
\left[
\begin{array}{c}
 \frac{2 \left(-d z+d+z^2-1\right)}{(d+z-1) \left((d-3) z+(d-1)^2+2 z^2\right)} \\
 \frac{2-2 z}{(d-3) z+(d-1)^2+2 z^2} \\
 \frac{1}{d+z-1} \\
 1 \\
\end{array}
\right],
\ee
this mode corresponds to a mode that is irrelevant in the UV and corresponds to a flow to solutions with finite temperature.
The third pair is
\be
\begin{split}
\lambda_3=\frac{1}{2} \left[ (d+z-1)- \Delta \right ] \sp \bar v_3=
\left[
\begin{array}{c}
 \frac{d^2+d \left(\Delta -2 z+2\right)+(z-1) \left(\Delta +5 z+3\right)}{4 (z-1) z (d-z-1)} \\
 -\frac{\Delta +d+5 z-5}{4 z (d+z-2)} \\
 -\frac{2}{\Delta -d-z+1} \\
 1 \\
\end{array}
\right],
\end{split}
\ee
where $\Delta \equiv \sqrt{-6 d z+d (d+6)+9 z^2-2 z-7}$. And the last ones reads
\be
\begin{split}
\lambda_4=\frac{1}{2} \left[ (d+z-1)+\Delta \right] \sp \bar v_4=
\left[
\begin{array}{c}
 \frac{d^2-d \left(\Delta +2 z-2\right)-(z-1) \left(\Delta -5 z-3\right)}{4 (z-1) z (d-z-1)} \\
 \frac{\Delta -d-5 z+5}{4 z (d+z-2)} \\
 \frac{2}{\Delta +d+z-1} \\
 1 \\
\end{array}
\right].
\end{split}
\ee
By having the eigenvectors of the matrix $A$ we can readily obtain information about the stability of the Lifshitz background under perturbations. By imposing the constraint $z \geq 1$ we avoid dynamical instabilities caused by an imaginary value of $\Delta$ (see also condition \ref{cond4}). We examine whether the amplitude of each those modes diverge for $r \to 0$ (or equivalently $u \to -\infty$), we can conclude that this is not the case for $d > 2$ and $1 < z < d-1$, under this constraint both of the modes $\left( \lambda_3, \bar v_3 \right)$ and $\left( \lambda_4, \bar v_4 \right)$ are irrelevant in the UV. In the case $z > d-1$, however, the mode $\left( \lambda_3, \bar v_3 \right)$ becomes relevant as its amplitude diverges for $r \to 0$, while the mode $\left( \lambda_4, \bar v_4 \right)$ remains irrelevant. In the relativistic case $z=1$ both of those modes are marginal. We need also to consider whether it is possible that $\lambda_3$ and $\lambda_4$ become degenerate, because in such a case we would obtain further solutions. We are able to verify that there is no such degenerate case within the stability region, $z>1$.

The solution of the homogeneous equation can be expressed as the Wronskian, $W$ times a vector of constant coefficients which are to be determined by the boundary conditions. We remind that the Wronskian is defined as follows
\be
W=\left[ \bar v_1 e^{\lambda_1 u} ~~ \bar v_2 e^{\lambda_2 u}~~ \bar v_3 e^{\lambda_3 u}~~ \bar v_4 e^{\lambda_4 u} \right].
\ee
Since we have the general solution of the homogeneous differential equation we can formulate a particular solution. This solution is to be added to the one for the homogeneous equation in order to obtain the solution of the inhomogeneous system. That solution is given by
\be
\bar x_{sp}=W. \int du~W^{-1}. \bar b, \label{specific}
\ee
where $\bar b$ is the vector that contains the inhomogeneous terms, the exponent $-1$ implies matrix inversion and dots imply matrix multiplication.

The solution of the equation of motion for the scalar field (eq. \ref{eq_scalar_field}) in the case $M^2 \neq -\frac{(d+z-1)^2}{4 B_0}$ is
\be
\phi(u)= \phi_+  e^{ \frac{u}{2} \left(d+z-1 +\delta\right) } + \phi_- e^{\frac{u}{2} \left(d+z-1 -\delta \right) },
\ee
where $\delta \equiv \sqrt{(d+z-1)^2+4 B_0^2 M^2}$. In order to avoid dynamical instabilities we demand $\delta \in \mathbb{R}$, this defines the Breitenlohner-Freedman (BF) bound which for the case studied reads
\be
M^2 \geq -\frac{(d+z-1)^2}{4 B_0}.
\ee
We consider the behaviour of the scalar field as $r \to 0$. For a non-tachyonic field (ie. $M^2>0$), $\delta > d+z-1$ and consequently, $d+z-1 -\delta<0$, this means that the coupling $\phi_-$ becomes dominant near $r \to 0$ and thus the dual operator of that field is relevant in the UV. In the case of a tachyonic field (ie. $M^2<0$) the operator dual to $\phi_-$ is irrelevant as $d+z-1 -\delta>0$. For all cases the operator dual to $\phi_+$ is irrelevant in the UV.

The vector $b$ that is carrying the backreaction of the dilaton field to the metric components and the gauge field can be expanded in three modes.
\be
b=b_+ + b_- +b_c.
\ee
The irrelevant mode $b_+$ reads
\be
b_+=\phi_+^2 e^{ u \left(d+z-1 +\delta\right) }
\left[
\begin{array}{c}
 -\frac{\frac{1}{4} \left(d+z-1+\delta\right)^2-B^2_0 M^2}{2 (d-1)} \\
 \frac{-B^2_0 M^2-\frac{1}{4} \left(d+z-1+\delta\right)^2}{2 (d-1)} \\
 0 \\
 \frac{z \left(d+z-1+\delta\right)^2}{8 (d-1)} \\
\end{array}
\right].
\ee
The relevant mode $b_-$ reads
\be
b_-=\phi_-^2 e^{ u \left(d+z-1 -\delta\right) }
\left[
\begin{array}{c}
 -\frac{(d+z-1) \left(d+z-1-\delta\right)}{4 (d-1)} \\
 \frac{-B^2_0 M^2-\frac{1}{4} \left(d+z-1 -\delta\right)^2}{2 (d-1)} \\
 0 \\
 \frac{z \left(d+z-1-\delta\right)^2}{8 (d-1)} \\
\end{array}
\right]
\ee
and the irrelevant mode $b_c$ reads
\be
b_c=\phi_+ \phi_- e^{ u \left(d+z-1 \right) }
\left[
\begin{array}{c}
 -\frac{-2 B^2_0 M^2}{d-1} \\
 0 \\
 0 \\
 \frac{-B^2_0 M^2 z}{d-1} \\
\end{array}
\right].
\ee

We can evaluate the contribution of each of those modes by evaluating the corresponding particular solutions. Note that $W$ is a linear operator and as such we have
\be
x_{sp}=\sum_{i=\{+,-,c\}}x_{sp,i} \sp x_{sp,i}=W.\int W^{-1}. b_i.
\ee

The particular solution for the $+$ and $-$ modes is expressed as
\be
x_{sp,\pm}=\phi_\pm^2 r^{\alpha_\pm}
\left[
\begin{array}{c}
 \frac{(z-1) (\alpha_\pm -2 d+2) (\alpha_\pm -2 z)}{4 (d-1) \left(\alpha_\pm ^2+\alpha_\pm -\alpha_\pm  (d+z)+2 d (z-1)-2 z^2+2\right)} \\
 \frac{-\alpha_\pm \left(\alpha_\pm ^2-\alpha_\pm  (d+2 z-2)+2 (d-1) (z-1)\right)}{4 (d-1) \left(\alpha_\pm ^2+\alpha_\pm -\alpha_\pm  (d+z)+2 d (z-1)-2
   z^2+2\right)} \\
 \frac{z (\alpha_\pm -2 d+2) (\alpha_\pm -2 z+2)}{8 (d-1) \left(\alpha_\pm ^2+\alpha_\pm -\alpha_\pm  (d+z)+2 d (z-1)-2 z^2+2\right)} \\
 \frac{\alpha_\pm  z (\alpha_\pm -2 d+2) (\alpha_\pm -2 z+2)}{8 (d-1) \left(\alpha_\pm ^2+\alpha_\pm -\alpha_\pm  (d+z)+2 d (z-1)-2 z^2+2\right)} \\
\end{array}
\right]. \label{leading_backreaction}
\ee

While the coupling term yields
\be
x_{sp,c}=-\phi_+ \phi_- B_0 M^2 r^{d+z-1}
\left[
\begin{array}{c}
 \frac{-1}{(d-1) (d+z-1)} \\
 \frac{-1}{(d-1) (d-z-1)} \\
 \frac{z (d+3 z-3)}{2 (d-1) (z-1) \left((d-1)^2-z^2\right)} \\
 \frac{z (d+3 z-3)}{2 (d-1) (z-1) (d-z-1)} \\
\end{array}
\right].
\ee

Therefore, the solution for $c_i=0$ (case of no excited homogeneous modes) is summarized by the following expressions (where we have substituted $u=\ln(r)$)
\be
\begin{split}
\tilde g_{tt}= r^{d+z-1} \bigg[
\phi_+^2 r^{+\delta} \frac{(z-1) (\alpha_+ -2 d+2) (\alpha_+ -2 z)}{4 (d-1) \left(\alpha_+ ^2+\alpha_+ -\alpha_+  (d+z)+2 d (z-1)-2 z^2+2\right)}\\
+\phi_-^2 r^{-\delta} \frac{(z-1) (\alpha_- -2 d+2) (\alpha_- -2 z)}{4 (d-1) \left(\alpha_- ^2+\alpha_- -\alpha_-  (d+z)+2 d (z-1)-2 z^2+2\right)}\\
+\frac{\phi_+ \phi_- B_0 M^2}{(d-1) (d+z-1)}
\bigg],
\end{split}
\ee

\be
\begin{split}
\tilde g_{rr}= r^{d+z-1} \bigg[
\phi_+^2 r^{+\delta} \frac{- \alpha_+ \left(\alpha_+ ^2-\alpha_+ (d+2 z-2)+2 (d-1) (z-1)\right)}{4 (d-1) \left(\alpha_+ ^2+\alpha_+ -\alpha_+  (d+z)+2 d (z-1)-2 z^2+2\right)}\\
+\phi_-^2 r^{-\delta} \frac{- \alpha_- \left(\alpha_- ^2-\alpha_- (d+2 z-2)+2 (d-1) (z-1)\right)}{4 (d-1) \left(\alpha_- ^2+\alpha_- -\alpha_- (d+z)+2 d (z-1)-2 z^2+2\right)}\\
+\frac{\phi_+ \phi_- B_0 M^2}{(d-1) (d+z-1)}
\bigg],
\end{split}
\ee

\be
\begin{split}
\tilde A_0= r^{d+z-1} \bigg[
\phi_+^2 r^{+\delta} \frac{z (\alpha_+ -2 d+2) (\alpha_+ -2 z+2)}{8 (d-1) \left(\alpha_+ ^2+\alpha_+ -\alpha_+  (d+z)+2 d (z-1)-2 z^2+2\right)}\\
+\phi_-^2 r^{-\delta} \frac{z (\alpha_- -2 d+2) (\alpha_- -2 z+2)}{8 (d-1) \left(\alpha_- ^2+\alpha_- -\alpha_-  (d+z)+2 d (z-1)-2 z^2+2\right)}\\
-\frac{z (d+3 z-3) \phi_+ \phi_- B_0 M^2}{2 (d-1) (z-1) \left((d-1)^2-z^2\right)}
\bigg].
\end{split}
\ee

In the case that $M^2 =- \frac{(d+z-1)^2}{4 B_0}$ the solution of the equation for the scalar field is
\be
\phi(u)= \phi_0  e^{ \frac{u}{2} \left(d+z-1\right) } + \phi_1 u e^{\frac{u}{2} \left(d+z-1 \right) },
\ee
by using the same arguments and methodology as in the $M^2 \neq - \frac{(d+z-1)^2}{4 B_0}$  case we obtain the following solution for the inhomogeneous problem
\be
\begin{split}
\tilde g_{tt}&= r^{d+z-1} \bigg[
-\frac{ (d+z-1) \phi_0^2 }{8 (d-1)} +\phi_1^2 \frac{(-d+z-1) \left(2 (d-3) z+(d-1)^2+5 z^2\right)}{16 (z-1)^2 \left((d-1)^3-(d-1) z^2\right)}\\
&+\phi_1^2 \frac{2 (z-1) (d-z-1) (d+z-1) \ln (r) ((z-1) (-d+z+1) \ln (r)+d+z-1)}{16 (z-1)^2 \left((d-1)^3-(d-1) z^2\right)}\\
&+\phi_0 \phi_1 \frac{2 (z-1) \left(z^2-(d-1)^2\right) \ln (r)+2 (d+1) z+(d-1)^2-3 z^2}{8 (d-1) (z-1) (d+z-1)}
\bigg],
\end{split}
\ee

\be
\begin{split}
\tilde g_{rr}&= r^{d+z-1} \bigg[
-\frac{(d+z-1)^2 \phi_0^2 }{8 (d-1) (d-z-1)} +\phi_1^2 \frac{(d+z-1) \left(-4 d z+d (d+4)+3 z^2+2 z-5\right)}{16 (d-1) (z-1)^2 (-d+z+1)^2}\\
&+\phi_1^2 \frac{\ln (r) \left((-d+z+1) \left(2 (d+1) z+(d-1)^2-3 z^2\right)-(z-1) (-d+z+1)^2 (d+z-1) \ln (r)\right)}{8 (d-1) (z-1) (-d+z+1)^2}\\
&-\phi_0 \phi_1 \frac{(d+z-1) (2 (z-1) (d-z-1) \ln (r)+d+z-1)}{8 (d-1) (z-1) (d-z-1)}
\bigg],
\end{split}
\ee

\be
\begin{split}
\tilde A_0&= r^{d+z-1} \bigg[
\phi_0^2 \frac{z (d+z-1) (d+3 z-3)}{16 (d-1) (z-1) (d-z-1)} \\
&-\phi_1^2 \frac{z \left(d^4-2 d^2 (z-1) (3 z-7)-8 d (z-4) (z-1)^2+(z-1)^3 (13 z+19)\right)}{32 (d-1) (z-1)^3 (-d+z+1)^2 (d+z-1)}\\
&-\phi_1^2 \frac{z \left((d-1)^2-z^2\right) \ln (r) \left(-d^2+(z-1) \left((d-z)^2-1\right) \ln (r)+z^2-6 z+5\right)}{16 (d-1) (z-1)^2 (-d+z+1)^2 (d+z-1)}\\
&+\phi_0 \phi_1 \frac{z \left(d^3+5 d^2 (z-1)+3 d (z-1)^2+(z-1)^2 (7 z+1)\right)}{16 (z-1)^2 \left((d-1)^3-(d-1) z^2\right)}\\
&-\phi_0 \phi_1 \frac{z (d-z-1) (d-z+1) (d+z-1) \ln (r)}{8 (z-1) \left((d-1)^3-(d-1) z^2\right)}.
\bigg].
\end{split}
\ee

\chapter{Derivation of the holographic $\beta$ functions} \label{sec:der_hol_beta}

Near the boundary $r \to 0$, the leading correction to the gauge field and metric components (eq. \ref{leading_backreaction}) due to the UV relevant coupling $\phi_-$ reads
\bs \be
\tilde g_{tt} \simeq \frac{(z-1) (\Delta_\phi -z) \left[ \Delta_\phi -(d-1) \right]}{2 (d-1) C_0} \phi_-^2 r^{-2\left[ \Delta_\phi -(z+d-1) \right]}+\cdots,
\ee
\be
\tilde g_{rr} \simeq \frac{\left[ \Delta_\phi -(z+d-1) \right] (\Delta_\phi- \Delta_+) (\Delta_\phi- \Delta_-)}{(d-1) C_0} \phi_-^2 r^{-2\left[ \Delta_\phi -(z+d-1) \right]}+\cdots,
\ee
\be
\tilde A_{0} \simeq \frac{z (\Delta_\phi -d) (\Delta_\phi -z)}{4 (d-1) C_0} \phi_-^2 r^{-2\left[ \Delta_\phi -(z+d-1) \right]}+\cdots, \label{coeff.gf}
\ee \label{per.solution1} \es
where
\be
\begin{split}
C_0=2 \Delta_\phi ^2 -3 \Delta_\phi  (d+z-1) +z (3d-2)+(d-1)(d-2),\\
\Delta_\pm=\frac{1}{4} \left[ (3 d+2 z-2)\pm\sqrt{4 z^2-4 d z+d (d+4)-4} \right].
\end{split}
\ee

In order to obtain the leading correction to the $\beta$ functions of the dual theory due to the scalar perturbation above we have to recast the metric (eq. \ref{metric.pert}) in the frame that $g_{uu}=1$ (eq. \ref{Ansatz}) and use (eq. \ref{betafunctions}). We use the coordinate transformation
\be
r=\exp\left[ \frac{m u}{\sqrt{2z(d-1)}} + \epsilon \phi_-^2 \frac{(\Delta_\phi -\Delta_-)(\Delta_\phi -\Delta_+)}{4(d-1) C_0} e^{\frac{2 m u\left[ (z+d-1)-\Delta_\phi \right]}{\sqrt{2z(d-1)}}} +\mathcal{O}\left( \epsilon^2 \right)  \right].
\ee

The metric and the dilaton field are expressed in the new coordinates as
\bs
\be
ds^2=-e^{2A(u)}dt^2 +dr^2 + e^{2B(u)} dx_i dx^i,
\ee
\be
\begin{split}
A(u)=- \frac{z m u}{\sqrt{2z(d-1)}} -\epsilon \phi_-^2 \frac{z(\Delta_\phi-\Delta_-)(\Delta_\phi-\Delta_+)}{4(d-1) C_0}
e^{\frac{2 m \left[ (z+d-1)-\Delta_\phi \right] u}{\sqrt{2z(d-1)}}}\\
+\epsilon \phi_-^2 \frac{(z-1)(\Delta_\phi-z)\left[\Delta_\phi-(d-1)\right]}{4(d-1) C_0}
e^{\frac{2 m \left[ (z+d-1)-\Delta_\phi \right] u}{\sqrt{2z(d-1)}}},
\end{split}
\ee
\be
B(u)=- \frac{m u}{\sqrt{2z(d-1)}} -\epsilon \phi_-^2 \frac{(\Delta_\phi -\Delta_-)(\Delta_\phi -\Delta_+)}{4(d-1) C_0}
e^{\frac{2 m \left[ (z+d-1)-\Delta_\phi \right] u}{\sqrt{2z(d-1)}}},
\ee
\be
\begin{split}
\ln\frac{\phi}{\phi_-}=&\frac{m u \left[ (z+d-1)-\Delta_\phi \right]}{\sqrt{2z(d-1)}} \\
&+\epsilon \phi_-^2 \frac{\left[ (z+d-1)-\Delta_\phi \right](\Delta_\phi -\Delta_-)(\Delta_\phi -\Delta_+)}{4(d-1) C_0}
e^{\frac{2 m \left[ (z+d-1)-\Delta_\phi \right] u}{\sqrt{2z(d-1)}}}.
\end{split}
\ee \label{comp.transf}
\es
We express the functions $A$ and $B$ in terms of the dilaton field as
\be
A(\phi)=\frac{z \ln\frac{\phi}{\phi_0}}{(z+d-1)-\Delta_\phi}+\frac{(z-1)(\Delta_\phi-z)\left[\Delta_\phi-(d-1)\right]}{4(d-1)C_0}\phi^2,
\ee
\be
B(\phi)=\frac{1}{(z+d-1)-\Delta_\phi} \ln\frac{\phi}{\phi_0}.
\ee
Finally, by using (eq.\ref{betafunctions}) we extract the $\beta$-functions to third order in $\phi$,
\be
\begin{split}
\beta_E(\phi)\equiv &\frac{d \phi}{d A}=\frac{(z+d-1)-\Delta_\phi}{z} \phi \\
&- \frac{(z-1)(\Delta_\phi-z)\left[ \Delta_\phi -(d-1) \right] \left[\Delta_\phi-(z+d-1) \right] \phi^3}{4z(d-1) \left[ 2 \Delta_\phi ^2 -3 \Delta_\phi  (d+z-1) +z (3d-2)+(d-1)(d-2) \right]} +\mathcal{O}\left( \phi^4 \right), \label{betae1}
\end{split}
\ee
\be
\beta_P(\phi)=\frac{d \phi}{d B}=\left[(z+d-1)-\Delta_\phi \right] \phi +\mathcal{O}\left( \phi^4 \right).
\label{betap1}\ee

\end{document}